\documentclass{aa}  

\usepackage{graphicx}
\usepackage[varg]{txfonts}
\usepackage{longtable,lscape}
\usepackage{natbib}
\bibpunct{(}{)}{;}{a}{}{,}

\begin{document}

   \title{Chemical tagging of the Ursa Major moving group}

   \subtitle{A northern selection of FGK stars \thanks{ Based on observations obtained with the HERMES spectrograph at the 
		   \emph{ Observatorio del Roque de los Muchachos} (La Palma), the FOCES spectrograph
   at Calar Alto, and with the Coud\'e-\'Echelle spectrograph of the Alfred-Jensch-Teleskop at the Th\"uringer Landessternwarte Tautenburg.}}

   \titlerunning{Ursa Major MG}
   \author{
           H.M. Tabernero\inst{1},
	   D. Montes\inst{1},
	J. I. Gonz\'alez Hern\'andez\inst{1,2,3},
	\and 
	M. Ammler-von Eiff \inst{4,5}}
   \institute{
Dpto. Astrof\'isica, Facultad de CC. F\'isicas, Universidad Complutense de Madrid, E-28040 Madrid, Spain.\\
\email{htabernero@ucm.es} 
         \and
Instituto de Astrof\'isica de Canarias, E-38205 La Laguna, Tenerife, Spain.
        \and
Universidad de La Laguna, Dept. Astrofísica, E-38206 La Laguna, Tenerife, Spain.
	\and
Thüringer Landessternwarte Tautenburg, Sternwarte 5, 07778, Tautenburg, Germany.  \and
Max-Planck-Institut f\"ur Sonnensystemforschung, Justus-von-Liebig-Weg 3, 37077 G\"ottingen, Germany             }
   \authorrunning{H. M. Tabernero et al.}	     
   \date{Received 22 August, 2013; accepted 04 September, 2014}

\abstract
{Stellar kinematic groups are kinematical coherent groups of stars which might share a common origin. 
These groups spread through the Galaxy over time due to tidal effects caused by Galactic rotation and disc 
heating. However, the chemical information survives these processes.}
{The information provided by the analysis of chemical elements can
reveal the origin of these kinematic groups. Here we investigate the origin of 
the stars that belong to the Ursa Major (UMa) Moving Group (MG).}
{We present high-resolution spectroscopic observations obtained from three different spectrographs of kinematically 
selected FGK stars of the Ursa Major moving group. Stellar atmospheric parameters ($T_{\rm eff}$, $\log{g}$, $\xi$, and [Fe/H])
were determined using our own automatic code ({\scshape StePar}) which makes use of the sensitivity of 
iron equivalent widths (EWs) measured in the spectra. We critically compare
the {\scshape StePar} results with other methods ($T_{\rm eff}$ values derived using the infrared flux 
method (IRFM) and $\log{g}$ values based on {\sc Hipparcos} parallaxes). We derived the 
chemical abundances of 20 elements, and their [X/Fe] ratios of all stars in the sample. We perform a 
differential abundance analysis with respect to a reference star of the UMa MG (HD~115043). We have 
also carried out a systematic comparison of the abundance pattern of the Ursa Major MG and the Hyades SC 
with the thin disc stellar abundances.} 
{Our chemical tagging analysis indicates that the Ursa Major MG is less affected by field 
star contamination than other moving groups (such as the Hyades SC). We find a roughly solar 
iron composition [Fe/H]~=~0.03 $\pm$ 0.07 dex for the finally selected stars, whereas the [X/Fe] ratios are 
roughly sub-solar except for super-solar Barium abundance.
}
{We conclude that 29 out of 44 (i.e. 66\%) candidate stars share a similar chemical 
	composition. In addition, we find that the abundance pattern of the Ursa Major MG might be marginally different from that
	of the Hyades SC.
} 


\maketitle 

\section{Introduction}

Stellar  kinematic groups (SKGs) --superclusters (SCs) and moving groups (MGs)-- are kinematic coherent groups 
of stars \citep{egg94} that might share a common origin. Among them, the youngest SKGs are  \citep[see][]{mon01a}: the Hyades 
SC (600 Myr), the Ursa Major MG (Sirius SC, 300 Myr), the Local Association or Pleiades MG (20 to 150 Myr), the 
IC 2391 SC (35-55 Myr), and the Castor MG (200 Myr).

Since Olin Eggen introduced the concept of MGs and the fact that stars can maintain a kinematic signature 
over long periods of time, their existence (mainly in the case of the old MGs) has been disputed. The disruption 
of MGs is caused by the Galactic differential rotation. Furthermore, disc heating causes the velocity dispersion of disc 
stars to increase gradually with age \citep[][]{wie71}.

The over density of stars in some regions of the Galactic velocity UV-plane may be the result of global dynamical 
mechanisms linked with the non-axisymmetry of the Galaxy \citep{fae05}, namely the presence of 
a rotating central bar \citep[e.g.][]{deh98, fux01, min10}, and spiral arms \citep[e.g.][]{qui05, ant09, ant11}, or both \citep[see][]{qui03, mfa10}.

Previous works show that different age sub-groups are located in the same region of the velocity plane as the 
classical MGs \citep{asi99} suggesting that both field stars and young coeval sub-groups can coexist 
in MGs \citep{fae07, fae08, ant08, kle08, sil08, fra09a, fra09b, Zhao09}.

Using different age indicators (e.g. the lithium line \ion{Li}{i} at 6707.8 {\AA}, chromospheric activity) it is possible
to quantify the contamination by younger or older field stars among late-type candidate members of 
a SKG \citep[e.g.][]{mon01b, mar10, lop06, lop09, lop10, mal10}.

However, the detailed analysis of the chemical content (\textit{chemical tagging}) is another powerful and 
complementary approach that provides clear constraints on  the membership \citep{fre02,mit13}. Unfortunately, chemical 
composition alone cannot provide the answers to the common origin unless previous information is available 
beforehand (i.e. information on kinematics). Regarding this approach, studies usually start from 
already known information \citep[see][and references therein]{mit13}, in order to fully exploit the {\it chemical tagging} 
approach.

Studies of open clusters such as the Hyades and Collinder 261 \citep{pau03, sil06, sil07a, sil09} found high levels
of chemical homogeneity, showing that chemical information is preserved within the stars, and that the possible 
effects of any external sources of pollution are negligible. Since chemical homogeneity is found among open 
clusters, it is possible to trace back dispersed clusters based on their chemical 
composition. In this sense {\it chemical tagging} was applied to the HR~1614 \citep{sil07b}, to the 
Hercules stream \citep{ben07}, Wolf 360 MG \citep{bub10}, and the Hyades SC \citep{pom11,sil11,tab12}. These 
studies proved or disproved the common origin of these structures by using chemical abundance information. In particular, the 
Hyades SC is an interesting case. This MG is supposed to originate from the Hyades 
cluster and was an excellent test since there is a whole cluster to choose a reference star 
for the differential analysis. \citet{tab12} found that 46 \% percent of Hyades SC members sharing similar 
abundances  to the original Hyades cluster. On the contrary \citet{pom11} and \citet{sil11} found that 
10-15 \% of the stars seem to originate from the Hyades cluster. These differences arise from the 
different sizes of the samples employed, $\approx$ 60 stars where analysed in \citet{tab12}, whereas 
in \citet{pom11} and \citet{sil11} analyse $\approx$ 20 stars. The comparison of these three studies 
shows that it is not possible to constrain the contamination level in moving groups until more complete 
samples are analysed. However, it would be still possible to find stars that may originate from 
a single cluster using the {\it chemical tagging} approach. The Hyades SC is not a unique case, \citet{sil13} 
linked the open cluster IC2391 and the Argus association using chemical analysis.

In this paper, we apply the \textit{chemical tagging} technique to a homogeneous sample of kinematically selected northern 
FGK Ursa Major MG candidates. This group has been previously investigated by \citet{sod93}, \citet{kin03}, \citet{kin05}, 
\citet{mor05}, \citet{amm09}, \citet{bia12}, \citet{dor12}. These studies demonstrate that their candidate members are 
consistent with a true MG of marginally sub-solar composition.  \citet{sod93} find [Fe/H]~$=$~$-0.08$~$\pm$~0.09 dex, whereas 
\citet{amm09}  get a slightly higher value [Fe/H]~$=$~$-0.03$~$\pm$~0.05 dex. Finally, \citet{bia12} 
obtain [Fe/H]~$=$~0.01~$\pm$~0.03, higher but consistent with previous measurements within the uncertainties. The study 
of individual abundances in \citet{amm09} covers Fe and Mg. \citet{bia12} analyse 11 different chemical elements, 
whereas \citet{dor12} also treat some s-process elements.  

More importantly, the age of the Ursa Major MG is close to the time scale of the dissolution of open 
clusters \citep{wie71}. Therefore, this  is an important case to study some aspects of the open cluster 
evolution and to apply the \textit{chemical tagging} approach. In Sect.~2, we give details on the 
sample selection. Observations and data reduction are described in Sect.~3. Descriptions for the 
derivation of the stellar parameters and chemical abundances are provided in Sect.~4. Chemical abundances are 
given in Sect.~5 together with the discussion of the results. Finally in Sect.~6, we summarize our 
conclusions about  UMa MG membership extracted 
from the \textit{chemical tagging} approach. 

\section{Sample Selection}

The sample analyzed in this paper (see Table~\ref{tablavel}) was selected using kinematical criteria based on $U$, $V$, and $W$ 
Galactic velocities of a given target being approximately within 10 km s$^{-1}$ of the mean velocity of the Ursa Major
nucleus \citep{kin03}. We selected our kinematic candidates from \citet{amm09}, \citet{hol09}, \citet{lop10}, \citet{mar10}, 
and \citet{mal10}. This candidate selection was later verified once more with the radial velocities coming 
from the spectroscopic data presented here (see Section~\ref{secobs}).

After the first stage of selection based on kinematical criteria, we then discarded those stars that were 
unsuitable for our standard abundance analysis, namely stars cooler than K4 and hotter than F6, because 
for these stars we would have been unable to measure 
the spectral lines required for our particular abundance analysis. Stars with high rotational
velocities (namely those greater than 15 km s$^{-1}$) were also discarded. In addition, we also removed spectroscopic
binaries (SB2) to avoid confusion between the spectral lines of the two components during the analysis. After these 
considerations, we were left with 45 stars suitable for the present analysis.

We recalculated the Galactic velocities of our selected targets by employing the radial velocities and uncertainties derived 
by the HERMES spectrograph automated pipeline \citep{ras11}. However, for stars observed with the FOCES 
and TLS spectrographs, we applied the cross-correlation technique using the routine {\sc fxcor} in IRAF \footnote{IRAF is distributed by 
the National Optical Observatory, which is operated by the Association of Universities for 
Research in Astronomy, Inc., under contract with the National Science Foundation.}, by adopting 
a solar spectrum as radial velocity template \citep[the Kurucz solar {\it ATLAS}][]{kur84}. Those radial 
velocities were derived after applying the heliocentric correction to the observed velocity.  Uncertainties were 
computed by {\sc fxcor} based on the fitted peak height and the antisymmetric noise, as 
described in \citet{td79}. The obtained radial velocities and their associated errors 
are given in Table~\ref{tablavel}. Proper motions and  parallaxes were taken from the {\sc Hipparcos} and 
Tycho catalogues \citep{esa97}, the Tycho-2 catalogue \citep{hog00}, and the new 
reduction of the {\sc Hipparcos} catalogue \citep{lee07}.

Following the method described in \citet{mon01a} we determine the $U$, $V$, and $W$ velocities. The Galactic 
velocities are in a right-handed coordinate system (positive in the directions of the Galactic centre, Galactic 
rotation, and the North Galactic Pole, respectively). \citet{mon01a} 
modified the procedures in \citet{jo87} to perform the velocity calculation and associated errors. 

\include{tabvel}
\include{tabpar}
This modified program uses 
coordinates adapted to the epoch J2000 in the International Celestial Reference System (ICRS). The calculated velocities 
are given in Table~\ref{tablavel}. As some stars are observed with two or more spectrographs, we decided to run
an internal consistency check to verify whether significant differences exist for different spectrographs. We find there 
is a small scatter of about 0.14 km s$^{-1}$. For these stars, final values of $U$, $V$, and $W$ velocities were 
derived from the weighted average of their radial velocities, since the parallaxes and proper motion
data are the same (see Table~\ref{tablavel}).

\section{Observations\label{secobs}}

\begin{figure}
\centering
\centerline{
\includegraphics[scale=0.55]{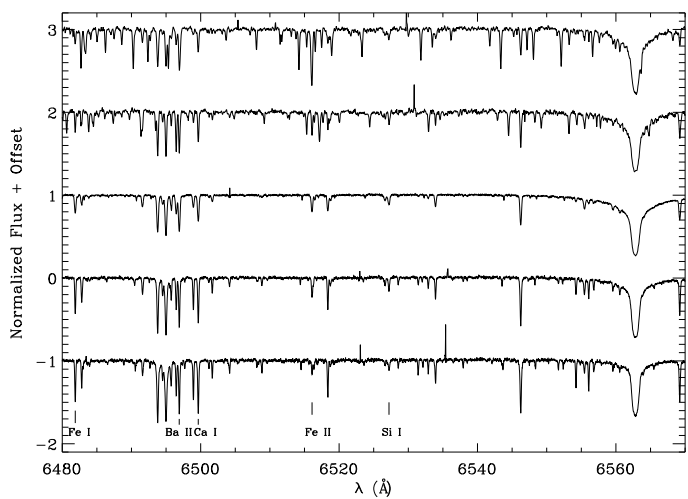}}
\caption{High-resolution spectra for some representative stars from our 
sample (from top to bottom): HD~4048 (F8~V), HD~13829 (F8~V), HD~115043 (a G2~V reference star
known to be a member of the Ursa Major nucleus), HD~76218 (G5~V), and HD~56168 (K0~V). Some lines used in 
the abundance analysis are highlighted in the bottom part of the diagram.}
\label{spec}
\end{figure}

Spectroscopic observations (see Fig.~\ref{spec}) were obtained at the 1.2-m Mercator 
Telescope\footnote{Supported by the Fund for Scientific Research of Flanders (FWO), Belgium, the Research 
Council of K.U. Leuven, Belgium, the Fonds National Recherches Scientific (FNRS), Belgium, the Royal Observatory of 
Belgium, the Observatoire de Genève, Switzerland, and the Thüringer Landessternwarte Tautenburg, Germany.} at 
the \emph{Observatorio del Roque de los Muchachos} (La Palma, Spain) in 2011-2012 with HERMES \citep[High Efficiency 
and Resolution Mercator Echelle Spectrograph,][]{ras11} with the high-resolution mode. Additional  spectra were taken 
in 2002–2004 with the 2.2-m telescope of the Centro Astron{\'o}mico Hispano Alem{\'a}n (CAHA) at Calar Alto with 
FOCES (operated by the Max-Planck-Institut für Astronomie Heidelberg and the Instituto de Astrofísica de Andalucía, CSIC), and 
the  Coud{\'e}-Echelle spectrograph at 2~m-{\it the Alfred-Jensch-Teleskop} at the Th{\"u}ringer Landessternwarte in Tautenburg
(TLS thereafter). Resolutions are 86,000 for HERMES, 40,000 for FOCES, and 67,000 for TLS. The wavelength range 
covered by the three spectrographs includes the range needed for our purposes: $\lambda$3600~{\AA} to $\lambda$9000~{\AA} 
approximately for HERMES and FOCES, $\lambda$4700~{\AA} to $\lambda$7400~{\AA} for TLS.

The typical signal-to-noise ratio ($S/N$) of the analyzed spectra is approximately 150 in the $V$ band (at $\lambda$6070~{\AA}). We 
analysed single main-sequence stars (from F6 to K4), being 45 candidates in total. Among them, there are 27 HERMES, 13 FOCES, 
and 17 TLS spectra (10  out of them are observed  with more than one spectrograph). Our observations also include the reference  star 
used in the differential abundance analysis (with respect to HD~115043). Additionally we  took 
three solar spectra, one  of the asteroid Vesta with HERMES, and two Moon spectra with FOCES and TLS.

The HERMES \emph{echelle} spectra were reduced with the automatic pipeline \citep{ras11} at the Mercator Telescope. 
Additionally, the FOCES and TLS data comprise spectroscopic observations presented in \citet{amm09}. The IDL - based
FOCES EDRS data reduction suite  was adapted by Klaus Fuhrmann for use with the Tautenburg Coudé-Echelle spectrograph. The
common steps of data reduction were followed \citep{hor86,mcl97} including bias subtraction, scattered light removal, order 
extraction, wavelength calibration using ThAr exposures, and division by flat-field exposures. 

We later used several IRAF tasks to transform the observed spectra into a unique  one-dimensional spectrum and applying 
the Doppler correction required to account for the radial velocity. In case several exposures were taken for the same
star, we combined all of the individual spectra to obtain a unique spectrum at higher $S/N$.

\section{Spectroscopic analysis}
\subsection{Stellar parameters}

Stellar atmospheric parameters ($T_{\rm eff}$, $\log{g}$, $\xi$, and [Fe/H]) were computed using the automatic 
code {\scshape StePar} \citep{tab12}. This automatic code employs a 2002 version of the MOOG code \citep{sne73} and 
a grid of Kurucz ATLAS9 plane-parallel model atmospheres \citep{kur93}. As damping 
prescription, we used the Uns$\ddot{{\rm o}}$ld approximation multiplied by a factor recommended by the Blackwell 
group (option 2 within MOOG). As line list we employed 300 $\ion{Fe}{i}$-$\ion{Fe}{ii}$  lines 
from \citet{sou08}. A typical star of our sample has 230-250 measurable 
$\ion{Fe}{i}$ and 20-25 $\ion{Fe}{ii}$ lines.   The {\scshape StePar} code  iterates within the 
parameter space until the slopes of $\chi$ versus (vs.) $\log{\epsilon(\textrm{\ion{Fe}{i}})}$ and 
$\log{(EW / \lambda)}$ vs. $\log{\epsilon(\textrm{\ion{Fe}{i}})}$ are zero (excitation equilibrium). In addition, 
it imposes the ionization equilibrium, such that $\log{\epsilon(\textrm{\ion{Fe}{i}})} = \log{\epsilon(\textrm{\ion{Fe}{ii}})}$. We 
also imposed that the [Fe/H] average of the MOOG output is equal to the metallicity of the atmospheric model.  Tolerance 
values for these conditions are needed, thus reasonable limits must be defined. In {\sc StePar}, we have chosen to 
iterate until the absolute value of the slope $\chi$ vs. $\log{\epsilon(\textrm{\ion{Fe}{i}})}$ was $\leq$~0.001 dex~eV$^{-1}$, 
whereas the absolute value of the slope of $\log{(EW / \lambda)}$ vs. $\log{\epsilon(\textrm{\ion{Fe}{i}})}$ 
was $\leq$~0.002. For the ionization balance we 
chose $|\log{\epsilon(\textrm{\ion{Fe}{i}})}$--$\log{\epsilon(\textrm{\ion{Fe}{ii}})}|$ $\leq$~0.005.  

{\scshape StePar}  employs a Downhill Simplex Method \citep{pre92}, and the problem function to minimize is a quadratic 
form composed of the excitation and ionization equilibrium conditions. Thus, {\scshape StePar} convergence towards the
best solution in the stellar parameter space takes only a few minutes. We have tested that the obtained solution for a 
given star is independent of the initial set of parameters employed. Hence,  we used the canonical solar values as initial 
input values ($T_{\rm eff}$~=~5777 K, $\log{g}$~=~4.44 dex, $\xi$~=~1~km~s$^{-1}$). In addition, we performed 
a 3-$\sigma$ rejection of the deviant $\ion{Fe}{i}$ and $\ion{Fe}{ii}$ lines after a first determination of 
the stellar parameters. Therefore, we re-run the {\scshape StePar} program again without the rejected lines.

The $EW$ determination of Fe lines was carried out with the ARES\footnote{The ARES code can be downloaded at http://www.astro.up.pt/} code \citep{sou07}. We followed
the approach of \citet{sou08} to adjust the $rejt$ parameter of ARES according to the $S/N$ of each spectrum  -- the $rejt$ parameter
allows ARES to determine the stellar pseudocontinuum to fit the aimed $EW$s. The other ARES parameters 
we employed are $smoother$ = 4 -- the recommended parameter for smoothing the derivatives used for line identification,  $space$ = 3  -- the 
wavelength interval (in \AA) from each side of the central line to perform the EW computation, $lineresol$ = 0.07 -- the minimum 
distance for ARES to resolve lines, and $miniline$ = 2 - minimum $EW$ that will be printed in the ARES output. 
Details regarding the ARES parameters can be found in \citet{sou07}. In addition, ARES is able to measure automatically 
weak gaussian lines giving  negligible systematic differences about 1-2~m$\AA$ when compared against ``manual'' EW 
measurements \citep[i.e. estimated with the IRAF {\scshape splot} task, see][]{sou07,ghe10}.\\

The uncertainties on the stellar parameters were computed taking into account one or more error sources for
	uncertainty for each parameter that will be added quadratically. The uncertainty on $\xi$ is obtained 
using the slope of $\log{\epsilon(\textrm{\ion{Fe}{i}})}$ vs. $\log{(EW / \lambda)}$. 
The uncertainty on $T_{\rm eff}$ is inferred by propagating two error sources added
in quadrature: the slope $\log{\epsilon(\textrm{\ion{Fe}{i}})}$ vs. $\chi$ and the variation introduced 
by the uncertainty of $\xi$.

We considered three error sources for $\log{g}$: the standard deviation of  $\ion{Fe}{ii}$ and the previous 
uncertainty on $\xi$ and $T_{\rm eff}$. 

Finally, to determine the error in the $\ion{Fe}{i, ii}$ abundance, we propagate the previously derived uncertainty on 
each stellar parameter plus the standard deviation of the $\ion{Fe}{i, ii}$ abundances.

\begin{figure}[ht]
\centering
\centerline{\includegraphics[scale=0.50]{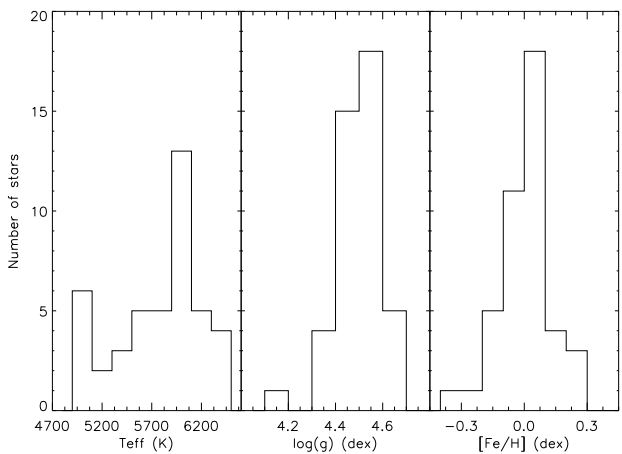}}
\caption{Histograms for the determined values of $T_{\rm eff}$, $\log{g}$, and [Fe/H] of the candidate stars.}
\label{histpar}
\end{figure}

We have also performed the parameter determination of the solar spectra taken with the three different instruments. We are able 
to reproduce the solar parameters (see Table~\ref{tablaparsol}). Solar values for each chemical abundance, also 
given in Table~\ref{tablaparsol}, represent the zero-point for the solar abundance values. Ideally, abundance measurements in 
each solar reference spectrum should provide the same solar photospheric abundances for each spectrograph. However, small differences 
are noticed probably due to systematic effects,  due to the different instrumental configurations, in the data taken with a specific instrument. These 
effects will likely apply to all candidate spectra of the UMa MG. Since our analysis is fully differential, the solar references are only used to convert 
the individual abundances (in a line-by-line basis) from $\log{\epsilon{(X)}}$ to [X/H]. Thus, the obtained chemical abundances 
will be referred to a solar spectrum corresponding to the instrument in which they were taken.

The obtained stellar parameters $T_{\rm eff}$, $\log{g}$, $\xi$, $\log{\epsilon(\ion{Fe}{i})}$, $\log{\epsilon(\ion{Fe}{ii})}$, and 
[Fe/H] (using our solar references) are given in Table~\ref{tablapar} (available online), together with the internal uncertainties in the stellar 
parameters.  In Fig.~\ref{histpar}, we show the histogram distributions of $T_{\rm eff}$, $\log{g}$, and [Fe/H] values. The effective temperature ranges approximately from
4800 K to 6500 K. The surface gravities of all stars in the sample are those typical of main sequence stars. 

We also verify that systematic errors of the stellar parameters are small when we use different spectrographs for the same object. 
We find that differences between $T_{\rm eff}$ are less than 100 K, with a dispersion of 30 K. $\log{g}$ and [Fe/H] show differences of 
less than 0.15 and 0.05 dex respectively. The dispersion is approximately 0.05 dex for surface gravity and 0.02 for [Fe/H]. These
differences are quite small and they do not represent  any significant difference when we derive stellar parameters 
from spectra taken with different {\it echelle} spectrographs. For these repeated spectra we 
employed an error-weighted average for their final stellar parameters. Then, the uncertainties
are given as the mean value of the individual ones.

\begin{table}
\small
\caption{Stellar Parameters for our solar spectra (Moon and Vesta).}
\label{tablaparsol}
\centering
\begin{tabular}{c c c c c c c}
\hline\hline
\noalign{\smallskip}
Spectrograph & HERMES  &  FOCES   &  TLS  & average\tablefootmark{a}   & $\sigma$\tablefootmark{b}  \\
\hline
\noalign{\smallskip}
{\it T}$_{\rm eff}$ (K) &  5776   &  5778 &  5789   & 5781 $\pm$ 7  &  15 \\
	$\log{g}$       &  4.48   & 4.43  & 4.45   & 4.45 $\pm$ 0.03   &  0.05\\  
  $\xi$ (km s$^{-1}$)   &   0.97   & 1.08   &  1.05  & 1.03 $\pm$ 0.06 & 0.03\\
\hline
\noalign{\smallskip}
Element     &         & $\log{\epsilon(X)}$      &     &      &   \\
\hline
\noalign{\smallskip}
Fe  &  7.46 & 7.47 & 7.48 &   7.47 $\pm$ 0.01 &  0.01\\
Na  &  6.37 & 6.36 & 6.34 &   6.36 $\pm$ 0.02 & 0.02 \\
Mg  &  7.64 & 7.61 & 7.62 &   7.62 $\pm$ 0.02 & 0.06 \\
Al  &  6.44 & 6.47 & 6.48 &  6.47 $\pm$ 0.02 & 0.02 \\
Si  &  7.55 & 7.58 & 7.59 &  7.57 $\pm$ 0.02 & 0.06 \\
Ca  &  6.34 & 6.35 & 6.33 &  6.34 $\pm$ 0.02 & 0.08\\
Sc  &  3.19 & 3.14 & 3.15 &  3.16 $\pm$ 0.03 & 0.06\\
Ti  &  4.99 & 5.01 & 5.02 &  5.01 $\pm$ 0.02 & 0.05\\
V   &  4.00 & 4.03 & 4.07 &  4.03 $\pm$ 0.04 & 0.07\\
Cr  &  5.66 & 5.66 & 5.68 &  5.67 $\pm$ 0.01 & 0.07\\
Mn  &  5.41 & 5.41 & 5.51 &  5.44 $\pm$ 0.06 & 0.05\\
Co  &  4.91 & 4.92 & 4.91 &  4.91 $\pm$ 0.01 & 0.03\\
Ni  &  6.26 & 6.26 & 6.26 &  6.26 $\pm$ 0.00 & 0.05\\
Cu  &  4.03 & 4.02 & 4.12 &  4.06 $\pm$ 0.06 & 0.08\\
Zn  &  4.54 & 4.57 & 4.57 &  4.56 $\pm$ 0.02 & 0.10\\
 Y  &  2.17 & 2.15 & 2.16 &  2.15 $\pm$ 0.01 & 0.07\\
Zr  &  2.61 & 2.78 & 2.81 &  2.73 $\pm$ 0.10 & 0.14\\
Ba  &  2.35 & 2.44 & 2.47 &  2.42 $\pm$ 0.06 & 0.25\\
Ce  &  1.61 & 1.60 & 1.61 &  1.61 $\pm$ 0.01 & 0.08 \\
Nd  &  1.47 & 1.53 & 1.51 &  1.50 $\pm$ 0.03 & 0.06 \\
\hline
\end{tabular}
\tablefoot{
	\tablefoottext{a}{ Average value and standard deviation of the stellar parameters and $\log{\epsilon(X)}$.}\\
	\tablefoottext{b}{Average internal uncertainties.}
}
\end{table}
\begin{figure*}
	\centerline{
		\includegraphics[scale=0.50]{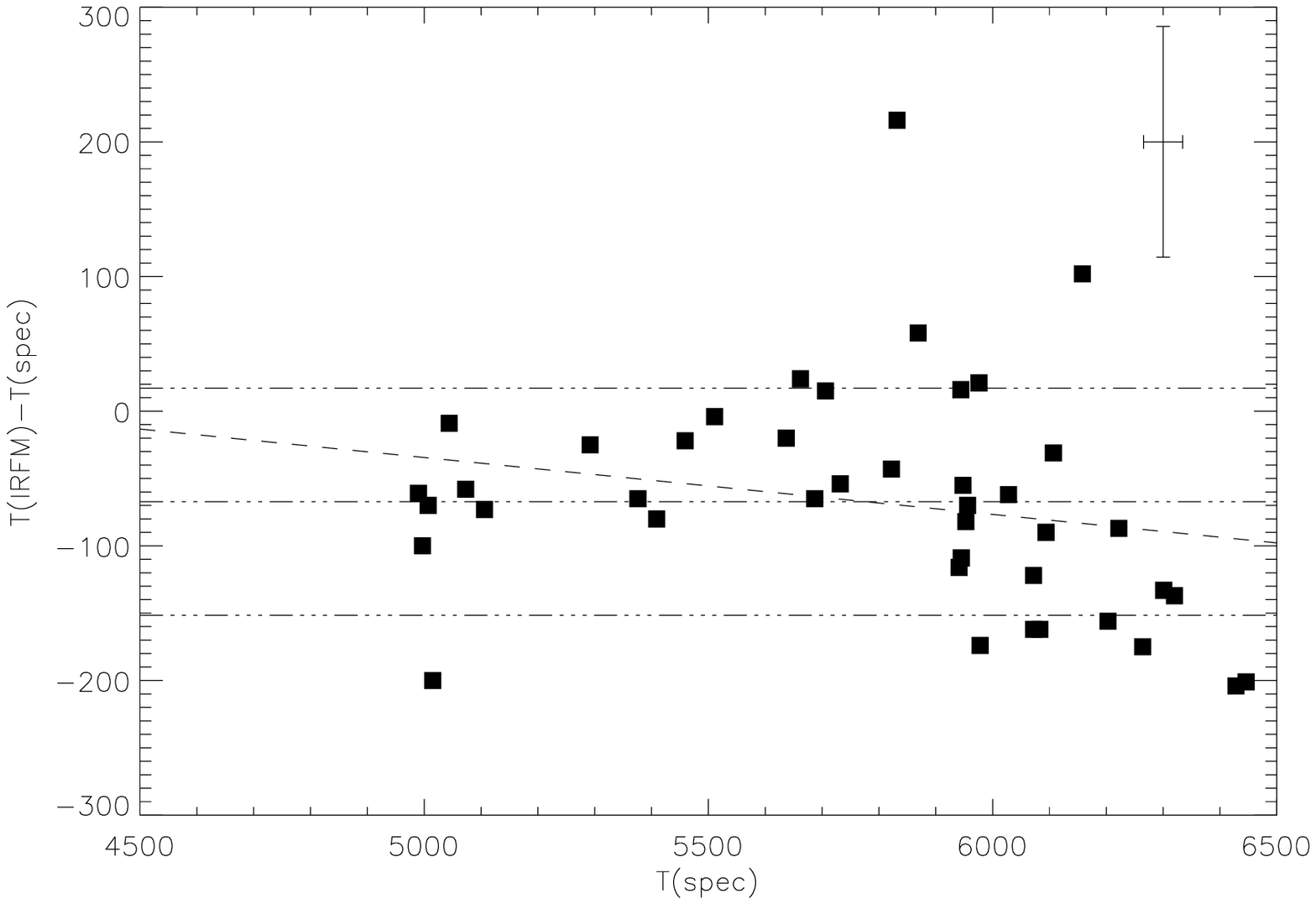}
	\includegraphics[scale=0.50]{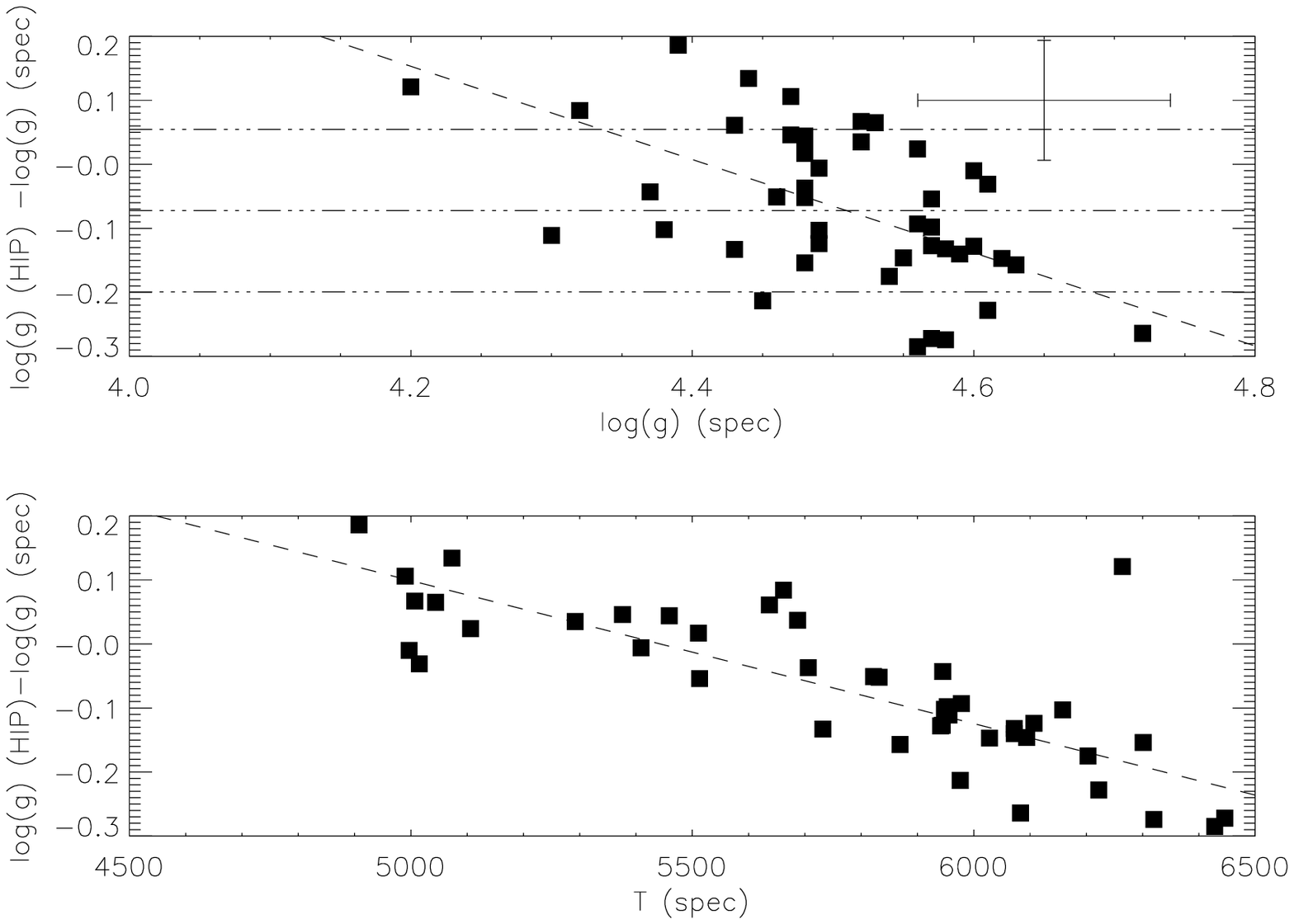}}
	\caption{Left panel: Comparison plot of $T_{\rm IRFM}$--$T_{\rm eff}$ vs $T_{\rm eff}$.  Right panel: Comparison plots of  $\log{g}_{\rm Hip}$--$\log{g}_{\rm spec}$ vs $\log{g}_{\rm spec}$ (top right), and  $\log{g}_{\rm Hip}$--$\log{g}_{\rm spec}$ vs $T_{\rm eff}$ (bottom right). In all panels, dashed-dotted lines represent the mean difference value and the standard deviation. Dashed-lines represent the ordinary least squares fit through the points.}
	\label{figirfm}
\end{figure*}

\subsection{IRFM based effective temperatures}

We have applied the infrared flux method~\citep[hereafter IRFM;][and references therein]{bla90} to the stellar sample presented in this work to also determine IRFM based effective temperatures, $T_{\rm IRFM}$, as in~\citet{gon09}.
We collected Johnson $V$ photometric data from the General Catalogue of Photometric Data~\citep[GCPD][]{mer97}. We also use the Johnson $V$ photometric data of {\sc Hipparcos}-selected nearby stars from \citet{koe10}. The $T_{\rm IRFM}$ and its uncertainty ($\Delta$~$T_{\rm IRFM}$) were derived as the weighted average of the three individual temperatures and uncertainties derived from J, H, and K.
For some stars of the sample we use the Tycho $V$ magnitudes from the Tycho-2 catalogue~\citep{hog00}, transformed into Johnson $V$ using the expression given in~\citet{mam02}.

We also collected 2MASS infrared $JHK_S$ photometry~\citep{skr06} for the stars of the stellar sample. The extinction in each photometric band, $A_i$, is derived using the relation $A_i = R_i E(B-V)$, where $R_i$ is given by the coefficients provided in~\citet{mcc04}.
Reddening corrections, $E(B-V)$, were estimated from the dust maps
of~\citet{sch98}, and corrected using the expressions {given
in~\citet{bon00a,bon00b}}. Parallaxes are the same as those used in Section~2.

The photometric data and $T_{\rm IRFM}$ values are given in Table~\ref{tblirfm}. The stars $\gamma$ Lep A/B and  $\xi$ Boo are perhaps too bright for 2MASS and thus the error on $JHK_S$ magnitudes is too large to provide an accurate determination of $T_{\rm IRFM}$. In addition, we estimated $T_{\rm IRFM}$ values assuming $E(B-V)=0$ for comparison, $T_{\rm IRFM,0}$, although the effect is typically well within the uncertainties of $T_{\rm IRFM}$. 
In Fig.~\ref{figirfm} we compare the $T_{\rm IRFM}$ values with the spectroscopic $T_{\rm eff}$ values. We found a mean  difference, alongside its standard deviation, of $T_{\rm IRFM}$--$T_{\rm eff}$~$= -67$ $\pm$ $84$~K. The average error bar on $T_{\rm IRFM}$ is $\sim 76$~K. These differences may be slightly correlated with our $T_{\rm eff}$, especially for the hottest stars of $T_{\rm eff}$ above 6000~K, this problem has been addressed in several studies \citep[see ][]{sou08,amm09a,amm09,mor13,tsa13}, but it seems to be inherent to the differences between the IRFM method and the iron EW approach.  
We verify that the correlation between 
$T_{\rm IRFM}$--$T_{\rm eff}$ and $T_{\rm eff}$ is not significant. We find a correlation coefficient of r~$=-$0.21~$\pm$~0.15 (illustrated by an ordinary least squares fit in Fig.~\ref{figirfm}). The goodness of fit is represented by the determination  coefficient (given by r$^2$~$=$~0.04~$\pm$~0.06). We have also performed a t-test to  assess the significance of the correlation coefficient (40 degrees of freedom). Thus, with a significance of 95\% we find that the correlation is not significant for our spectroscopic $T_{\rm eff}$. Also, the overall offset and scatter is small enough, possibly indicating that the $T_{\rm IRFM}$ is really similar to our spectroscopic $T_{\rm eff}$. Thus, we decided 
to adopt the spectroscopic $T_{\rm eff}$ for our abundance analysis.

\subsection{{\sc Hipparcos} based gravities}

We derived {\sc Hipparcos} gravities based on the obtained spectroscopic $T_{\rm eff}$ for the stellar sample in this study. This 
alternative gravity derivation requires the Johnson $V$ photometric data discussed in Section~4.2, as well as the parallaxes 
discussed in Section~2. We use the web interface\footnote{ http://stev.oapd.inaf.it/cgi-bin/param} of the PARSEC5
isochrones \citep[see][]{bre12} to derive the {\sc Hipparcos} surface gravity \citep[as in ][]{sou08}. The web interface 
only requires $T_{\rm eff}$, $V_{\rm Johnson}$, [Fe/H], and the parallax, as well as their uncertainties to compute 
the  desired {\sc Hipparcos} surface gravities. 

The photometric data and {\sc Hipparcos} gravity values are given in Table~\ref{tblirfm}. In Fig.~\ref{figirfm} we compare 
the $\log{g}_{\rm spec}$ values with the spectroscopic $\log{g}_{\rm Hip}$ values and we find a mean difference of
$\log{g}_{\rm Hip}$--$\log{g}_{\rm spec}$~$=$ $-0.07$ $\pm$ 0.13~dex.  We find a correlation coefficient, for $\log{g}_{\rm Hip}$ vs $\log{g}_{\rm spec}$,
$r$~$=-$0.49~$\pm$~0.15 ($r^2$~$=$~0.24~$\pm$~0.21). In order to assess the significance of this correlation value 
we performed  a t-test. As in Section~4.2 we used a confidence level of 95 \% and 43 degrees of freedom. On the other 
hand for $\log{g}_{\rm Hip}$ vs $T_{\rm spec}$ the correlation coefficient results in $r$~$=-$0.74~$\pm$~0.07
($r^2$~$=$~0.55~$\pm$~0.10). In both cases the correlation seems to be significant, with a 95 \% confidence 
level, as it is shown in the  right-hand panels of Fig.~\ref{figirfm}. Since the correlation is significant in the case 
of surface gravities, the effect on the chemical abundances must be considered (we refer the reader to the 
next section for further details).

On average, the stars tend to show lower $\log{g}_{\rm Hip}$ than $\log{g}_{\rm spec}$. The spectroscopic methodology 
we employ to derive surface gravities gives reliable $T_{\rm eff}$ estimates, but it is rather inefficient~in estimating
the surface gravity \citep[see][]{sou08,tsa13,mor13}. For the cooler stars the $EW$s of the $\ion{Fe}{ii}$ get weaker
as $T_{\rm eff}$ drops. The hottest stars may also pose a problem in this sense, possibly due to arising difficulty 
in measuring the $EW$ of the $\ion{Fe}{ii}$ lines, thus losing sensitivity to surface gravity as these lines get 
weaker and less reliable with increasing temperatures (see bottom panel in Fig.~\ref{figirfm}).

\begin{figure}
	\centerline{\includegraphics[scale=0.55]{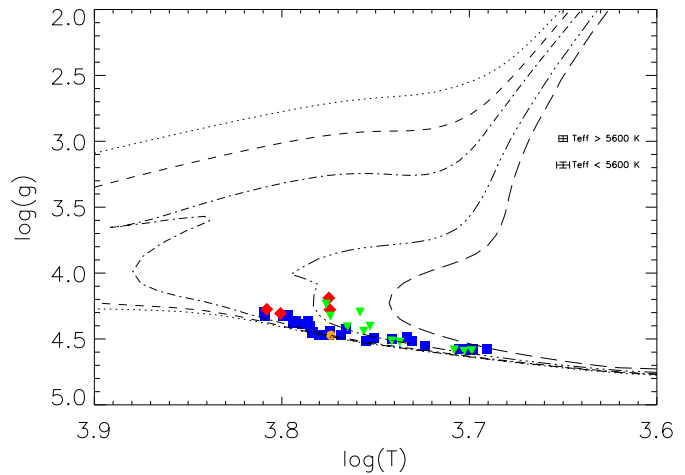}}
	\caption{Spectroscopic $\log{T_{\rm eff}}$ vs.  $\log{g}_{\rm Hip}$ for the candidate stars. We have 
		employed the Yonsei-Yale isochrones \citep{dem04} for Z~=~0.0, and age~=~0.1, 0.3, 0.5, 1, 4, and 13 Gyr (from left to right). Mean
error bars are represented at the middle right. Blue squares represent stars selected as members by the {\it chemical tagging} 
approach, red diamonds represent those stars that have similar Fe abundances, but different values of other elements. The orange
circle represents the reference star HD~115043. Inverted green triangles represent those stars that do 
not have similar Fe abundances (as well as dissimilar values of other elements). For the membership criterion  we refer 
the reader to Section~5.}
	\label{umaiso}
\end{figure}

As a complementary stellar parameter test, we create a $\log{g}_{\rm Hip} -\log{T_{\rm eff}}$ diagram show 
in Fig.~\ref{umaiso}. The selected stars fit within 
the depicted isochrones, being consistent with the isochrone for 0.3 Gyr \citep[for a Ursa Major age reference, see][]{kin03}. 

\subsection{Chemical abundances}

The selection of the chemical elements in this study is the same as in \citet{tab12} (see Table~\ref{lltab}) whose line 
list comes from a combination of atomic line data from \citet{gon10}, \citet{pom11}, and \citet{sou08}. A total 
of 20 elements were analyzed: Fe, the $\alpha$-elements (Mg, Si, Ca, and Ti), the Fe-peak elements (Cr, Mn, Co, and Ni), the odd-Z elements (Na, Al, Sc, and V), 
Cu, Zn, and the s-process elements (Y, Zr, Ba, Ce, and Nd), see Tables~\ref{galtab2} and \ref{galtab3}. Chemical 
abundances were calculated using the $EW$ method. The $EW$s were determined using the ARES code \citep{sou07}, following
the approach described in Sect.~4.1.

Once the $EW$s are measured, the analysis is carried out with the LTE MOOG code \citep[2002 version, see ][]{sne73} using 
the ATLAS model corresponding to the derived atmospheric parameters. We determine chemical abundances (see Tables~\ref{galtab2} 
and \ref{galtab3}) relative to solar values using the spectrum of the asteroid Vesta, and two lunar spectra acting as solar
reference for each instrument. 
We compute the mean of the line-by-line differences of each chemical element and candidate star with respect to our solar 
references (one for each spectrograph, see Table~\ref{tablaparsol} for the solar reference elemental abundances). However, to 
avoid incorrect $EW$ measurements (e.g. caused by a wrong continuum placement), we rejected those lines separated by more 
than a factor of two of the standard deviation ($\sigma$) from the median differential abundance derived for each line. Finally, in 
case of stars observed with two or three spectrographs, we simply take the average value of the available results.
We have compared the solar abundances obtained with different instruments and the differences seem to be very 
small (0.10 dex or better) for the majority of the elements treated in this study (see Table~\ref{tablaparsol}). 

The differential abundances were also determined to establish the membership of each stellar candidate using the star 
HD~115043 as reference (see Tables~\ref{diftable2} and \ref{diftable3}). The internal 
uncertainties of the derived stellar parameters (using {\scshape StePar}, see Sect.~4.1) are 28~K for 
$T_{\rm eff}$, 0.07~dex for $\log{g}$, 0.05~km~s$^{-1}$ for $\xi$, and 0.03~dex for [Fe/H]. These average
errors are in fact quite small, reflecting the relative internal precision of the obtained parameters. However, using 
these average values to assess the error bar on element abundance would be too optimistic. 
We found that systematic errors for $T_{\rm eff}$  and $\log{g}$ are~67~K and 0.07~dex respectively (see
Sects.~4.2 and 4.3). Combining these systematic errors  with our internal uncertainties, we obtained 
a total uncertainty of 72~K for $T_{\rm eff}$ and 0.10~dex for $\log{g}$. 

However, for stars at $T_{\rm eff}$~$>$~6000 K, these combined uncertainties can raise up to 115~K and 
0.27~dex. Therefore, in order to work with more conservative and reliable uncertainties, we used the values 
given by \citet{nev09}, i.e.: $\Delta T_{\rm eff}$~$=$~$\pm$~100~K, $\Delta\log{g}$~$=$~$\pm$~0.30~dex, 
$\Delta\xi$~$=$~0.50~km~s$^{-1}$, and $\Delta$[Fe/H]~$=$~$\pm$~0.30~dex. Using these uncertainties we derived the 
abundance sensitivities to changes in the stellar atmospheric parameters (see Table~\ref{tablasense} and 
\ref{tablasense2}). Then, we combined the sensitivities to give an estimation of the error 
bar for  [X/H] and [X/Fe]. The final errors are usually driven by $T_{\rm eff}$. However, some are 
dominated by $\xi$ (e.g., for Ba) or by $\log{g}$ (e.g., for Ce and Nd).

We performed a careful evaluation of the impact due to systematic errors on stellar atmospheric parameters
derived with {\scshape StePar}. From subsections~4.2 and 4.3  the parameter most severly affected appears 
to be surface gravity. Thus, one might wonder whether its spectroscopic derivation may have an effect on 
the derived abundances. Therefore, we re-computed the differential abundances (with respect to HD~115043) with 
the {\sc Hipparcos} surface gravities to verify possible differences. We find small differences at about 
hundredths of dex. For example, Ba and Ni remain nearly unaltered (with mean differences of 0.00 $\pm$ 0.02 
and 0.01 $\pm$ 0.01dex), whereas for Ca and Ce we find variations of -0.01 $\pm$ 0.03 dex, and 0.02 $\pm$ 0.05 dex 
respectively. As an additional check for systematic deviations, we compared two stars
in common with \citet{bia12} and \citet{dor12} ($\gamma$ Lep A/B). Their abundances differ from ours 
by up to 0.11 dex in the worst case (i.e. for [Al/Fe]). In the best case, i.e. for [Ni/Fe], they differ 
only by 0.01 dex. The above differences are similar to those due to the internal scatter (typically 0.01-0.10 dex). Finally, in order to be consistent with the previous 
study from \citet{tab12}, we decided to use the stellar atmospheric parameters coming from {\scshape StePar}.

\section{Discussion}

We will compare our derived element abundances with those of thin disc stars \citep{gon10,gon13} to determine
whether our values follow Galactic trends. We will also verify the chemical homogeneity of the Ursa Major MG 
and whether some of the stars indeed have  homogeneous abundances of all the considered elements.

\subsection{Element abundances}

The element abundances were determined in a fully differential way by comparing them with those derived 
for a solar spectrum (as stated in Section~4.1). The choice of elements is taken from \citet{tab12} (see also Table~\ref{lltab}) as explained  in Section~4.4.\\

\begin{figure*}
\centerline{
\includegraphics[scale=0.45]{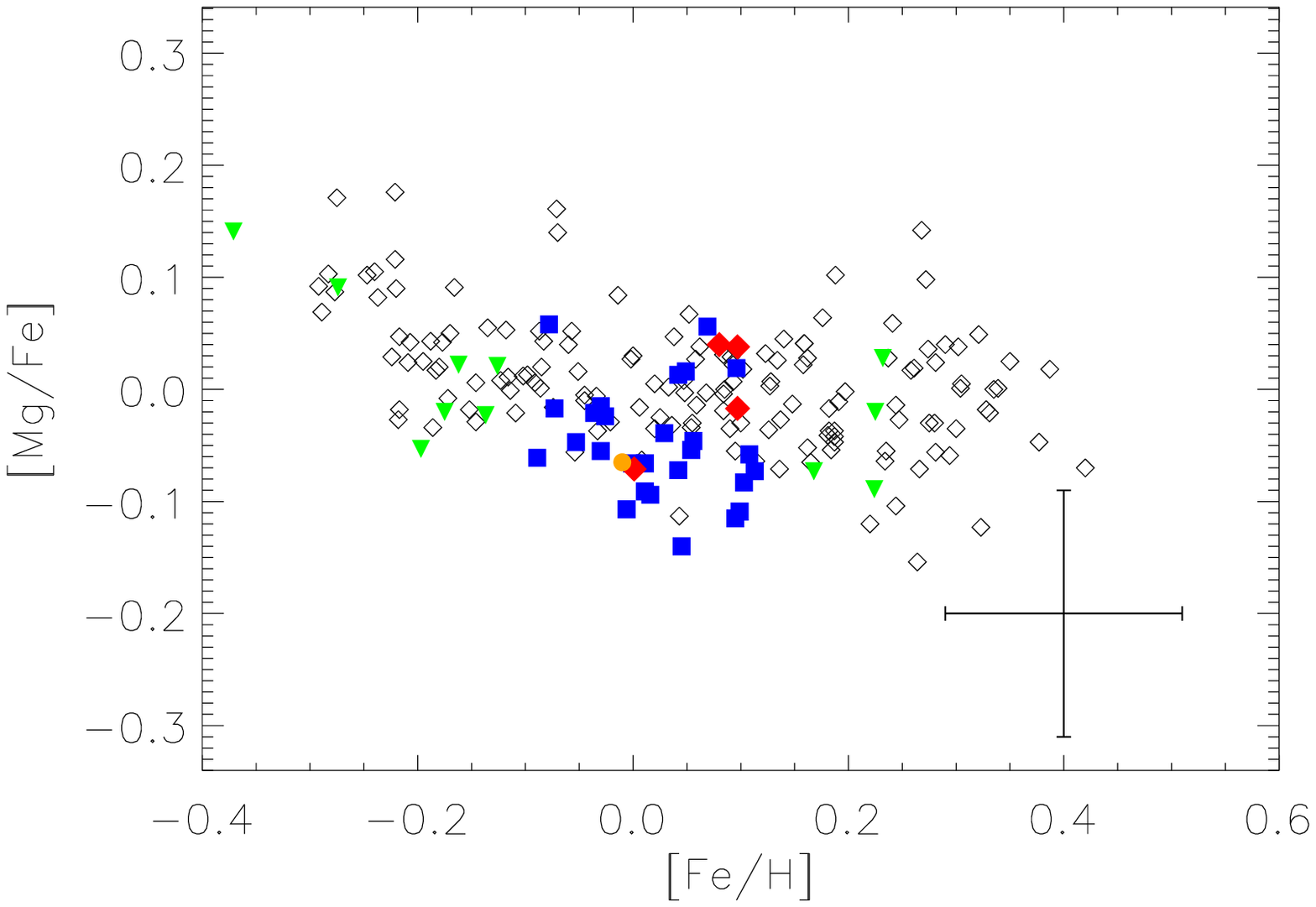}
\includegraphics[scale=0.45]{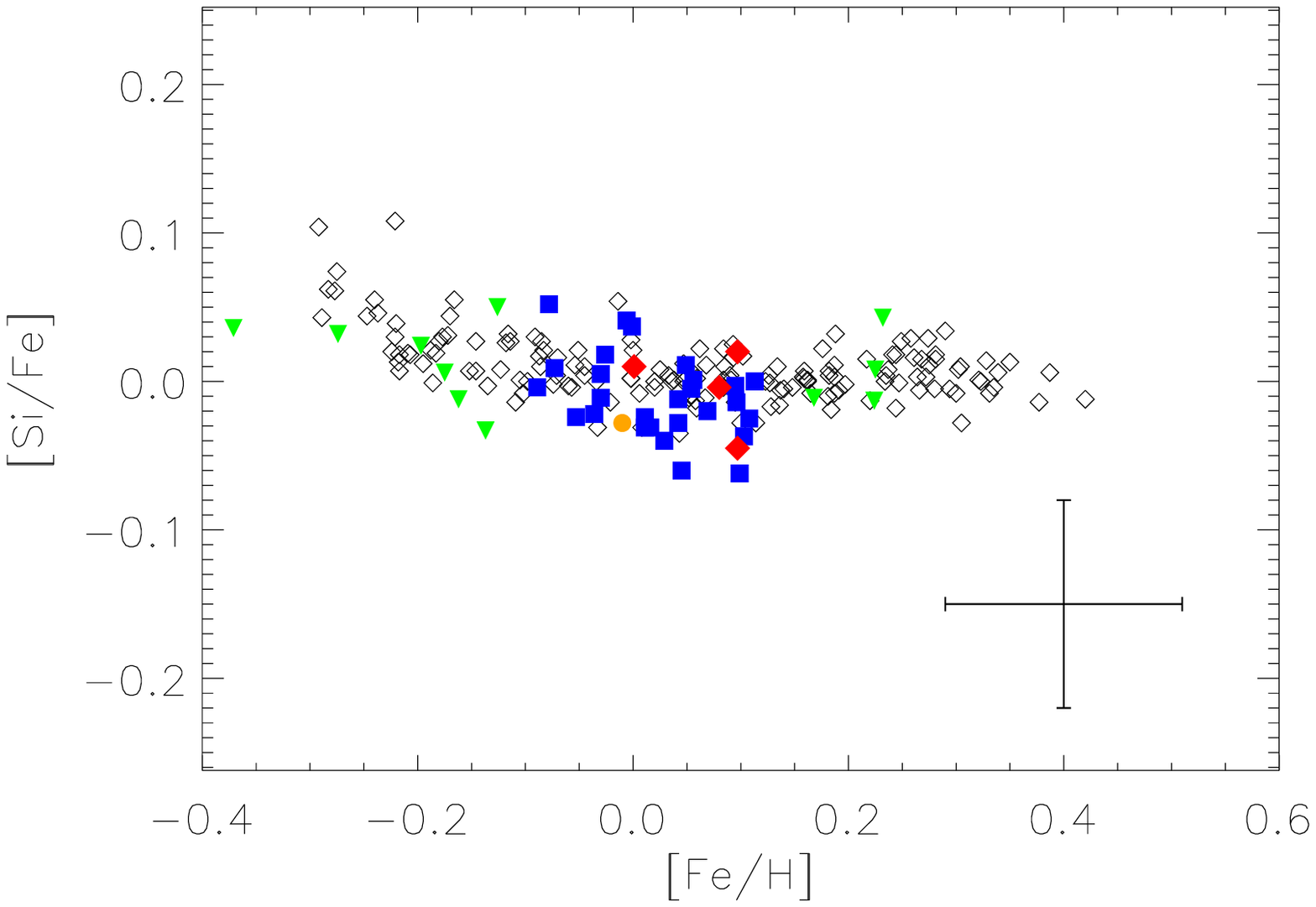}}
\centerline{
\includegraphics[scale=0.45]{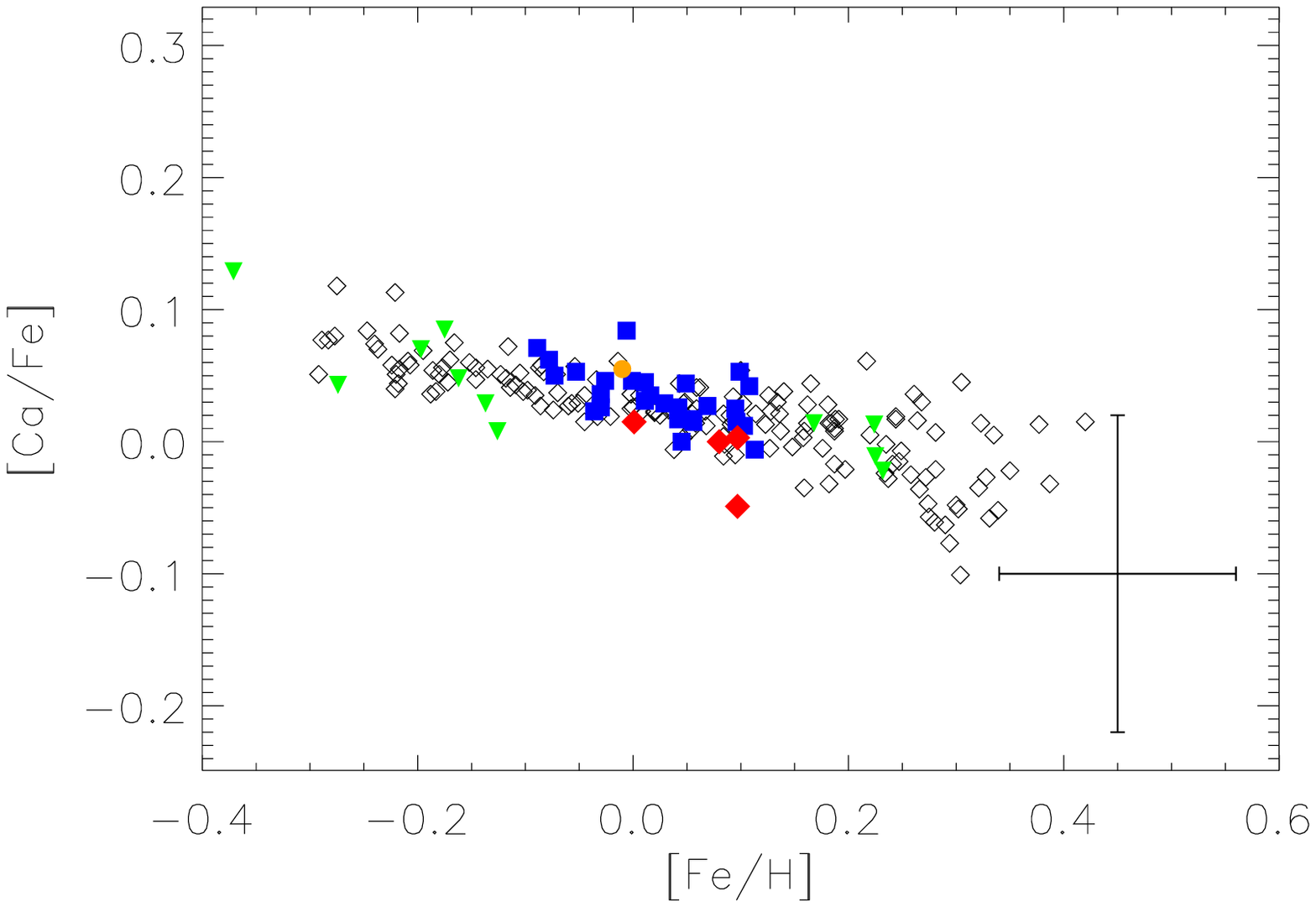}
\includegraphics[scale=0.45]{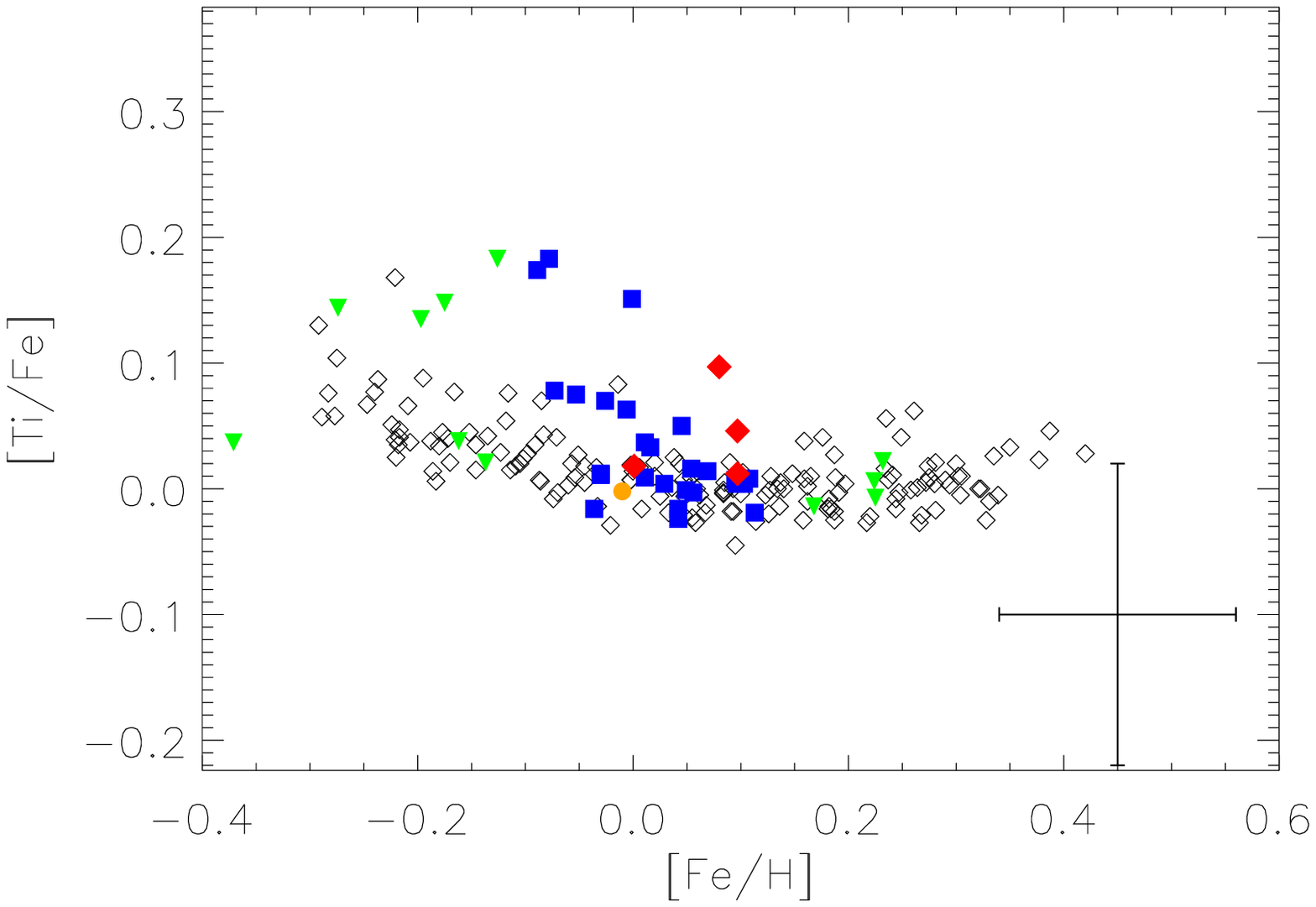}}
\centerline{
\includegraphics[scale=0.45]{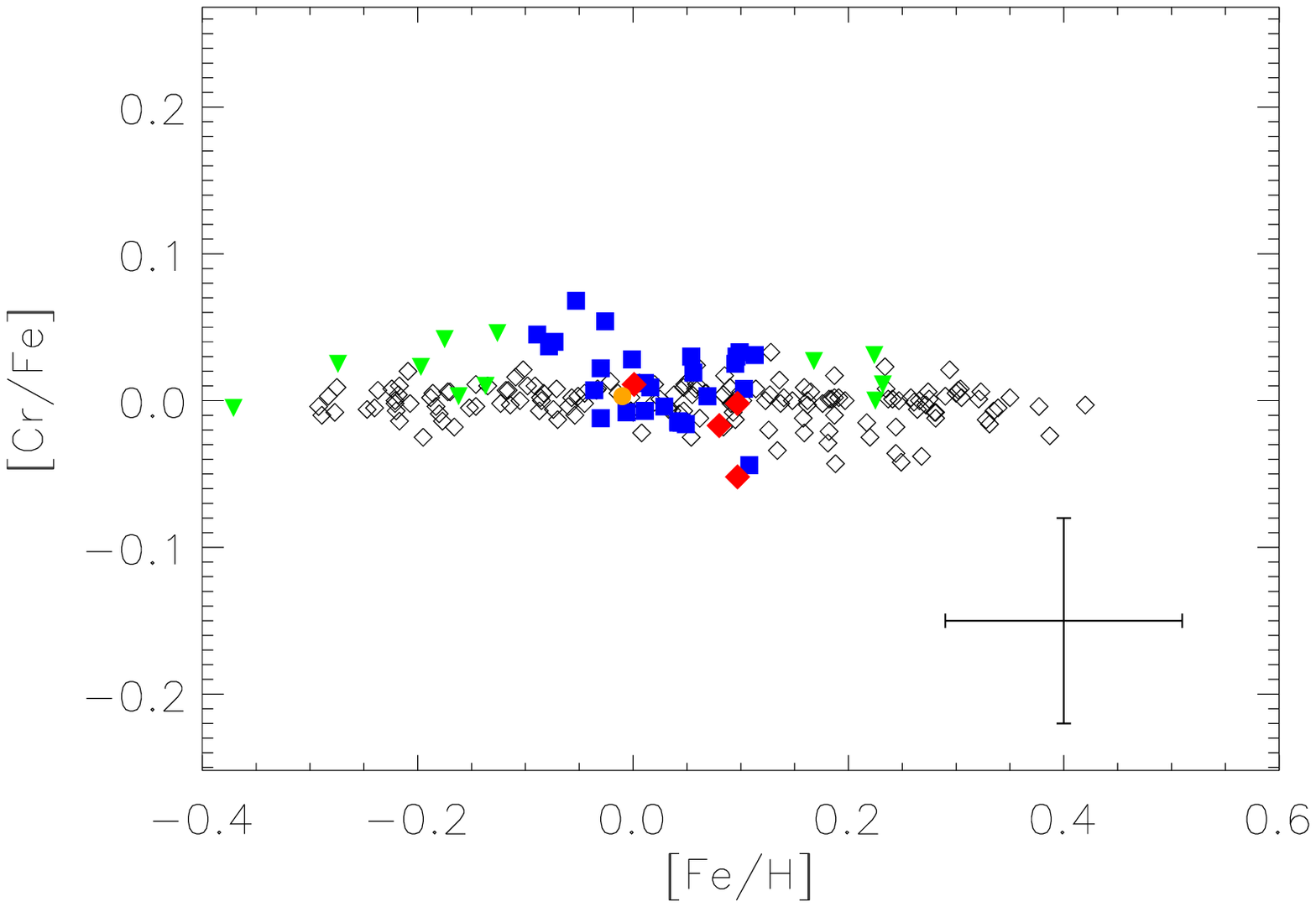}
\includegraphics[scale=0.45]{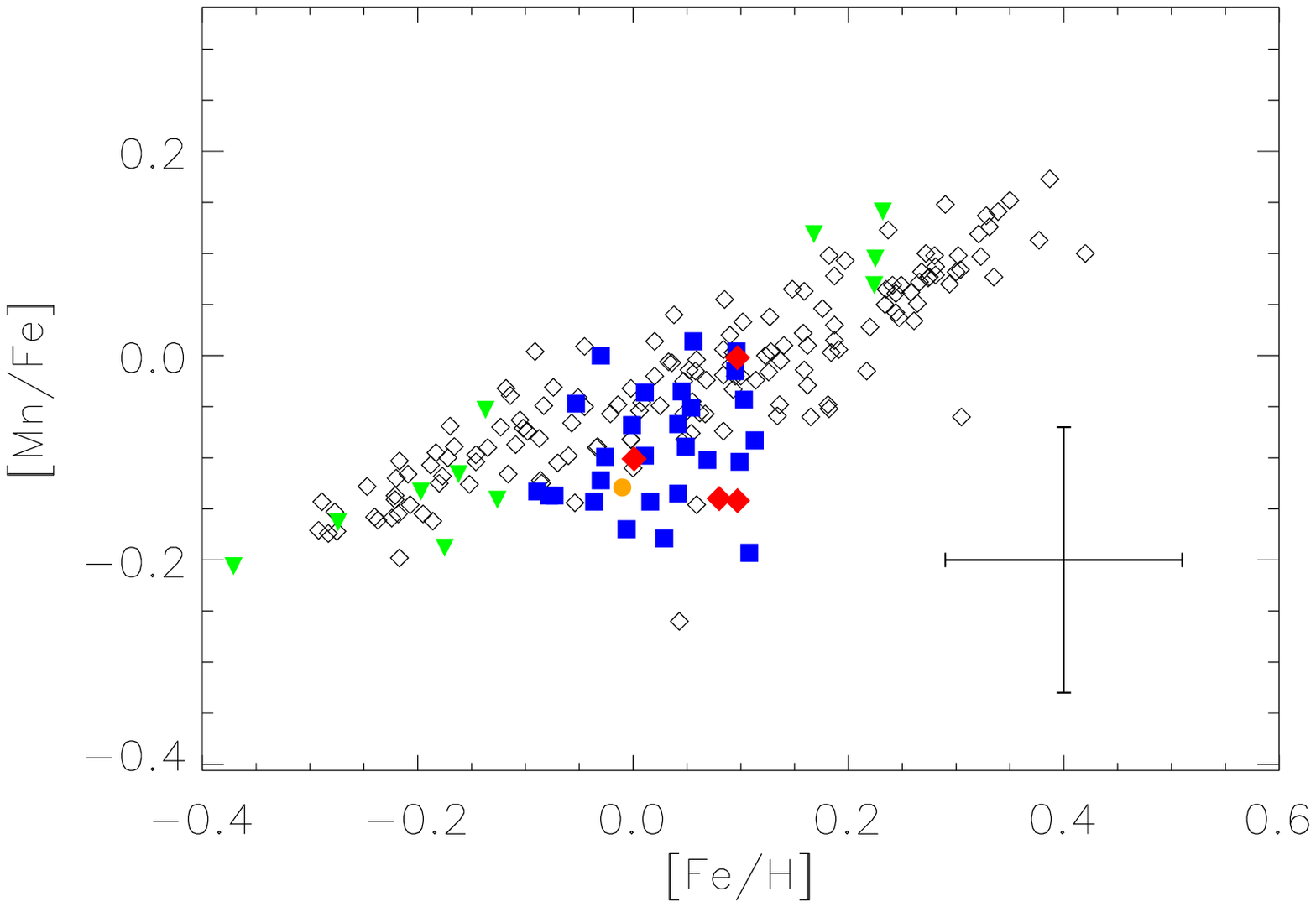}}
\centerline{
\includegraphics[scale=0.45]{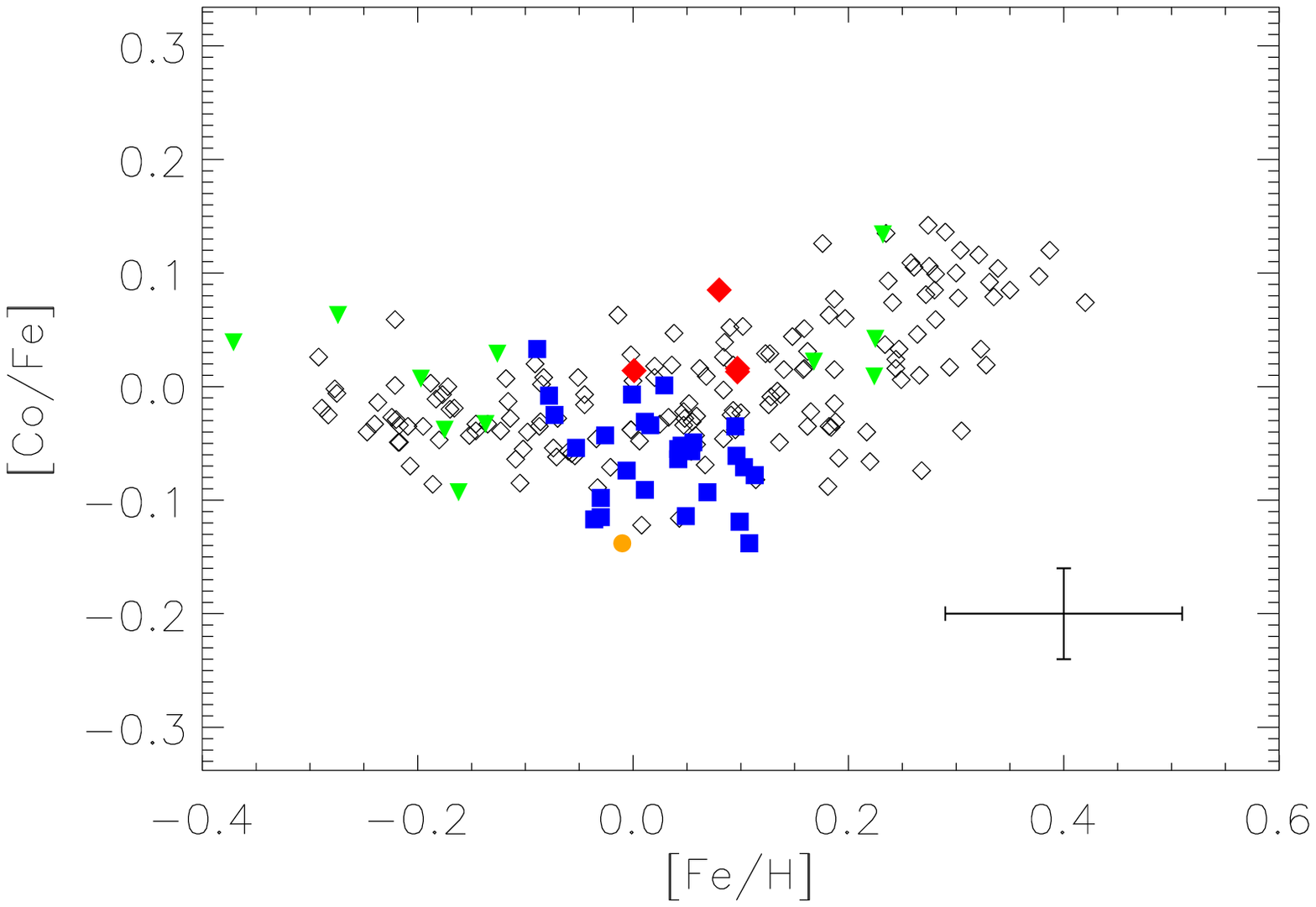}
\includegraphics[scale=0.45]{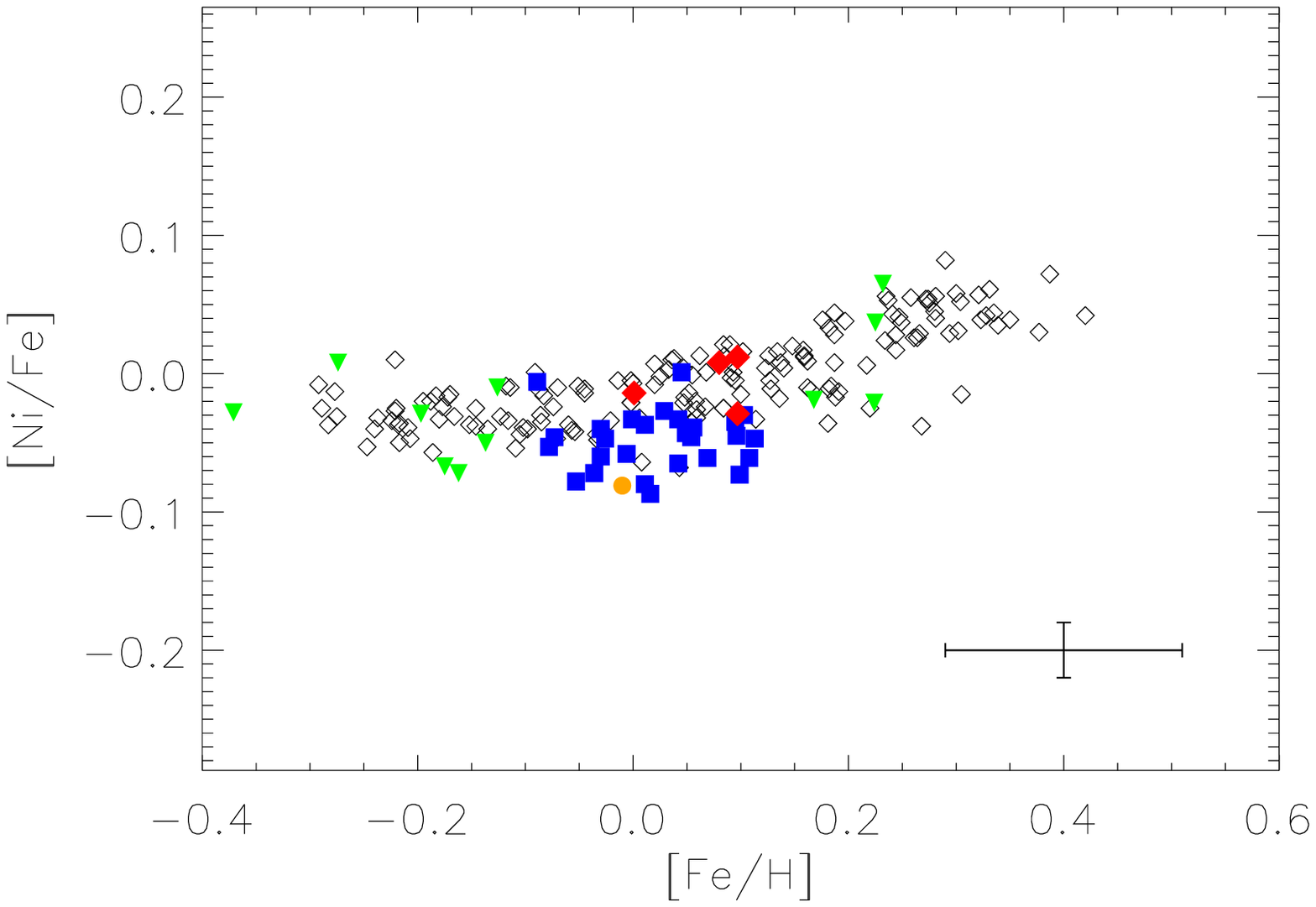}}

\caption{[X/Fe] vs. [Fe/H] for the $\alpha$-elements (Mg, Si, Ca, and Ti), and the Fe-peak elements (Cr,Mn,Co, and Ni): open 
diamonds represent the thin disc data \citep{gon10,gon13}, red diamonds are our stars compatible 
to within 1-rms with  the Fe abundance but not for all elements, blue squares are the candidates 
selected to become members of the Ursa Major MG. Inverted green 
triangles show incompatible stars. The reference star, HD~115043, known to be a member of the Ursa Major
nucleus is marked with an orange circle. Also, representative error bars are displayed in each graph.}
\label{umagal1}
\end{figure*}

In the case of the $\alpha$-elements (see Fig.~\ref{umagal1}) Si and Ca seem  to follow the Galactic 
trends \citep[see][]{ben05,red06,gon10,gon13}. Mg is slightly sub-solar for stars around solar metallicity. Ti seems 
to follow the trends, but the scatter tends to increase as [Fe/H] decreases for this narrow metallicity range. It has 
been suggested that Ti may suffer from NLTE effects, especially for cool stars. Therefore, to further check this issue, we 
have derived the difference $\log{\ion{Ti}{ii}}$--$\log{\ion{Ti}{i}}$. For the coolest stars ($T_{\rm eff}$~$\leq$~5500~K), we 
obtain $\log{\ion{Ti}{ii}}$--$\log{\ion{Ti}{i}}$~$=$~0.14~$\pm$~0.09. For the hottest stars ($T_{\rm eff}$~$\ge$~5500~K), we 
obtain 0.06~$\pm$~0.06 dex.  At 1--$\sigma$ level the difference is significant for the coolest stars ($T_{\rm eff}$~$\leq$~5500~K). Other studies 
have attributed that difference to Ti over-ionization \citep{lai08,dra09,bia12,adi12,sil13}. However, the total error bar for $\log{\ion{Ti}{i}}$ is 0.21 dex (see Table~\ref{tablasense}),  maybe implying that the Ti abundance difference is not significant for the coolest stars, even if an observable offset is present. 
This difference may be connected to deviations either from excitation or ionization equilibrium 
\citep{adi12}. Another possible explanation for the observed over-ionization could be an incorrect T--$\tau$ 
relationship in the adopted model atmospheres \citep{lai08}. Whereas this effect can be compensated 
for [Fe/H] by changing $\xi$, it does not necessarily apply to other elements \citep{adi12}. 

For the iron peak elements (Cr, Mn, Co, and Ni, see Fig.~\ref{umagal1}), we find a small scatter 
in Ni and Cr. We note that most of the stars lie below the Galactic trend, and that Mn has a larger scatter. Ni, Mn, and Co
show on average sub-solar values.

For the odd-Z elements (Na, Al, Sc, and V, see Fig.~\ref{umagal2}), Na and Al seem to be sub-solar in 
composition as it happens for some Fe-peak elements.  A high dispersion is observed for Sc, however it seems 
compatible with the Galactic trend. We confirm a large dispersion for V, which some authors 
interpret as a NLTE effect \citep[e.g.][]{bod03,gil06,nev09} affecting mostly the coolest stars. Vanadium 
lines are indeed difficult to measure and may require very high signal-to-noise data.

Cu, Zn, and the s-process elements (Y, Zr, Ba, Ce, and Nd, see Figs.~\ref{umagal2}) follow similar 
trends to those seen in solar analogues \citep{gon10}. We find some enhancement for Ba above the solar level 
as observed in open clusters and moving groups with ages below 1 Gyr. \citet{dor09} and \citet{dor12} showed 
that for 0.3 Gyr \citep[the Ursa Major attributed age, see][]{kin03,amm09} one might expect to find 
0.2-0.3 dex for [Ba/Fe]. In our sample, we find similar values for a majority of Ursa Major MG 
stars (see Fig.~\ref{umagal2}). The Ba over-abundance is not reflected for Y, Zr, and Ce. Although the scatter 
is relatively large the  average abundance values are not enhanced. Cu and Zn seem to be solar in 
spite of the high scatter found for these elements. Nd seems to be really high compared to the Galactic 
abundance pattern, although there are some stars following the Galactic trend. At solar metallicities, however, it 
raises up to 0.2 dex. The higher than solar Nd abundance of
many Ursa Major candidate stars does not endanger the subsequent differential analysis, since the reference
star HD~115043 is also enhanced ([Nd/Fe]~$=$~0.15~$\pm$~0.03, see Table~\ref{galtab3}).

\subsection{Differential abundances with respect to HD~115043}

\begin{figure*}
\centering
\centerline{
\includegraphics[scale=0.35]{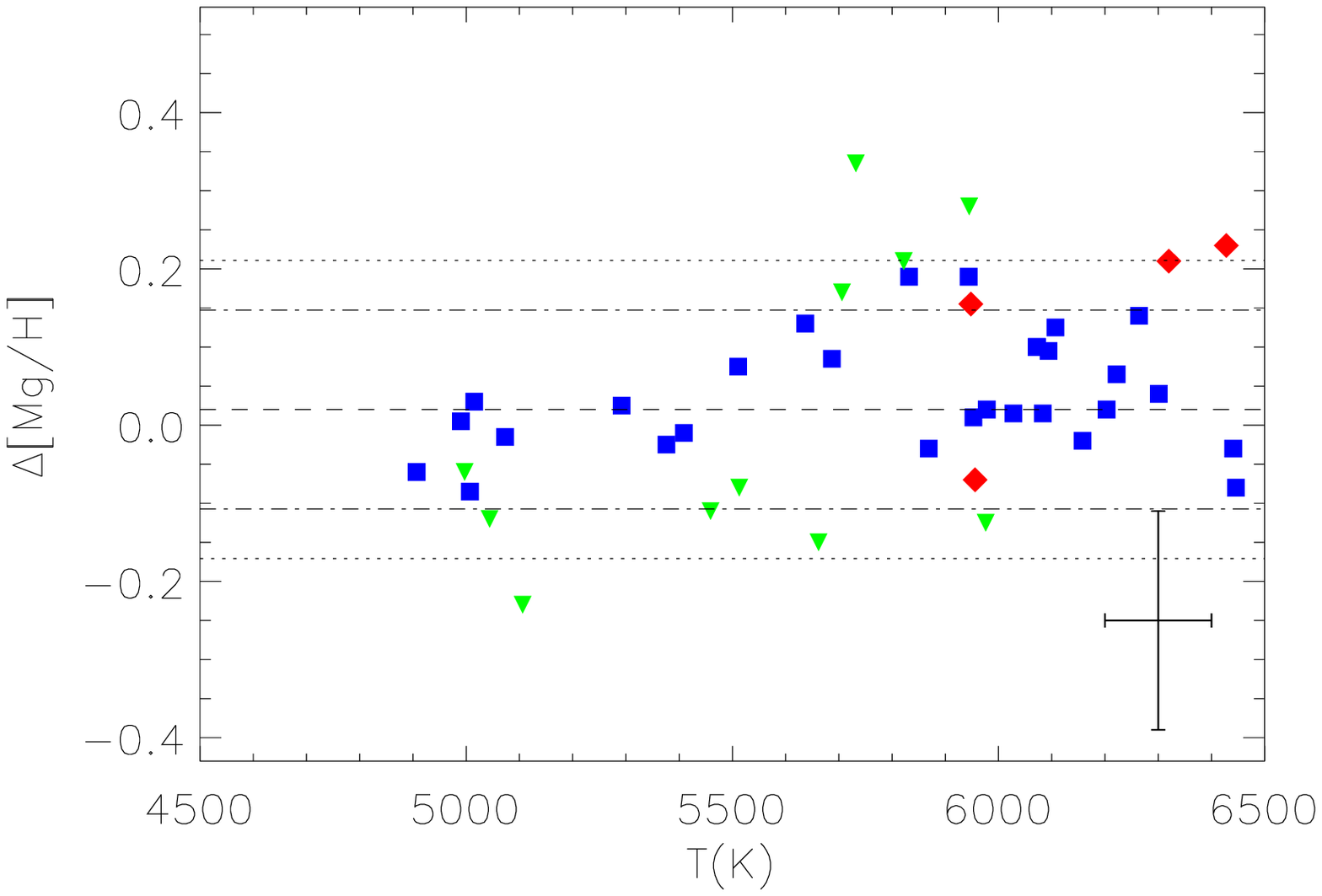}
\includegraphics[scale=0.35]{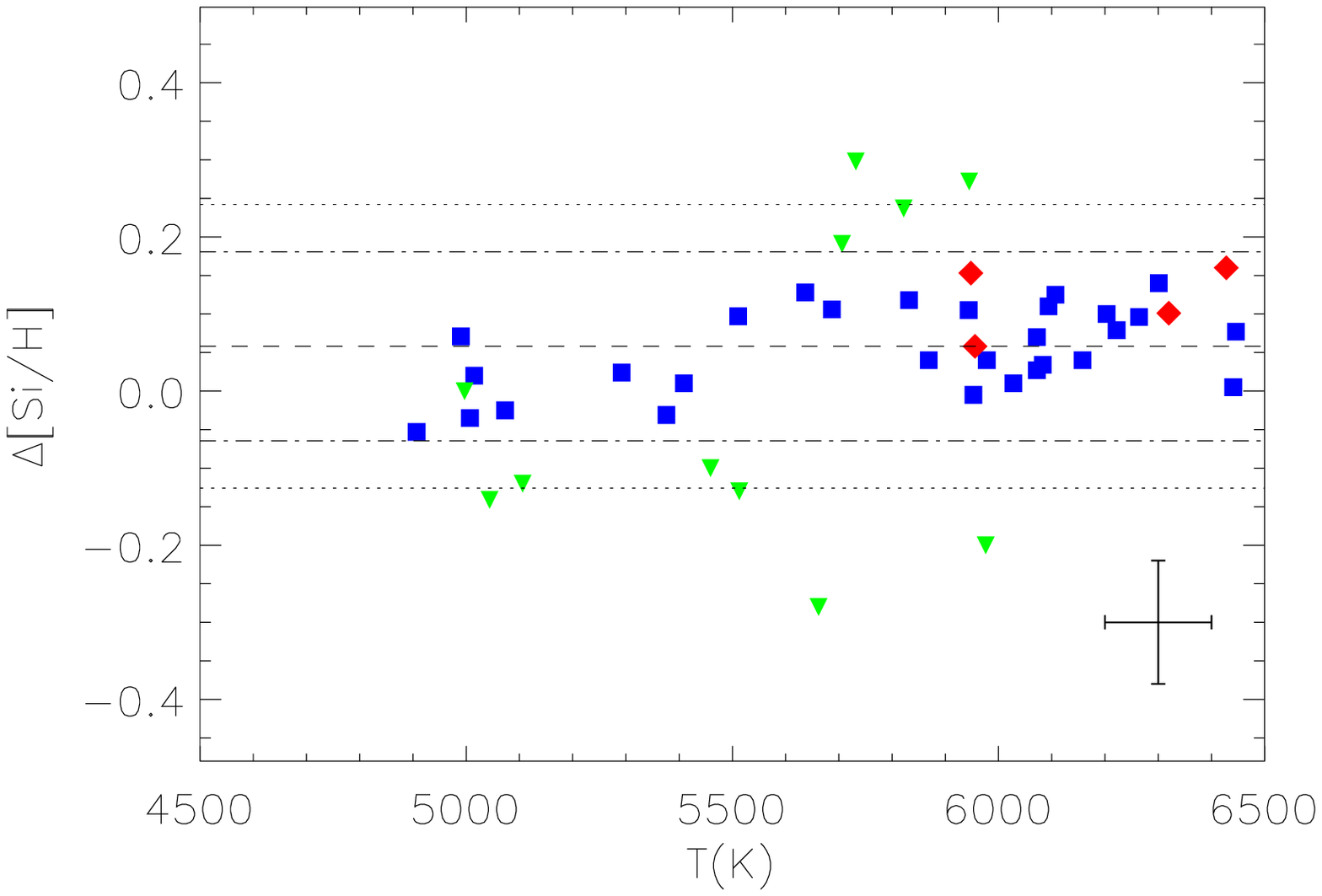}
\includegraphics[scale=0.35]{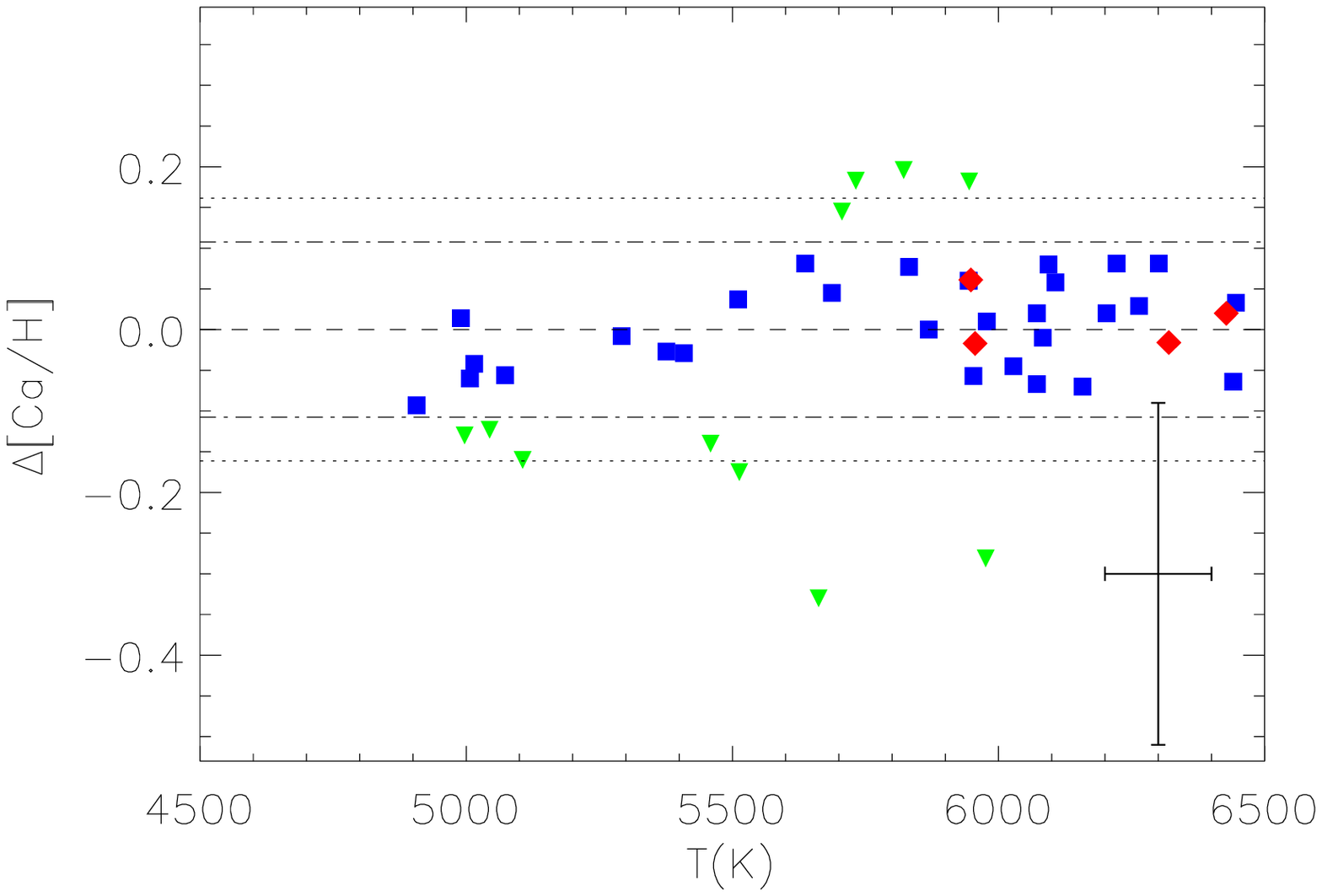}}
\centerline{
\includegraphics[scale=0.35]{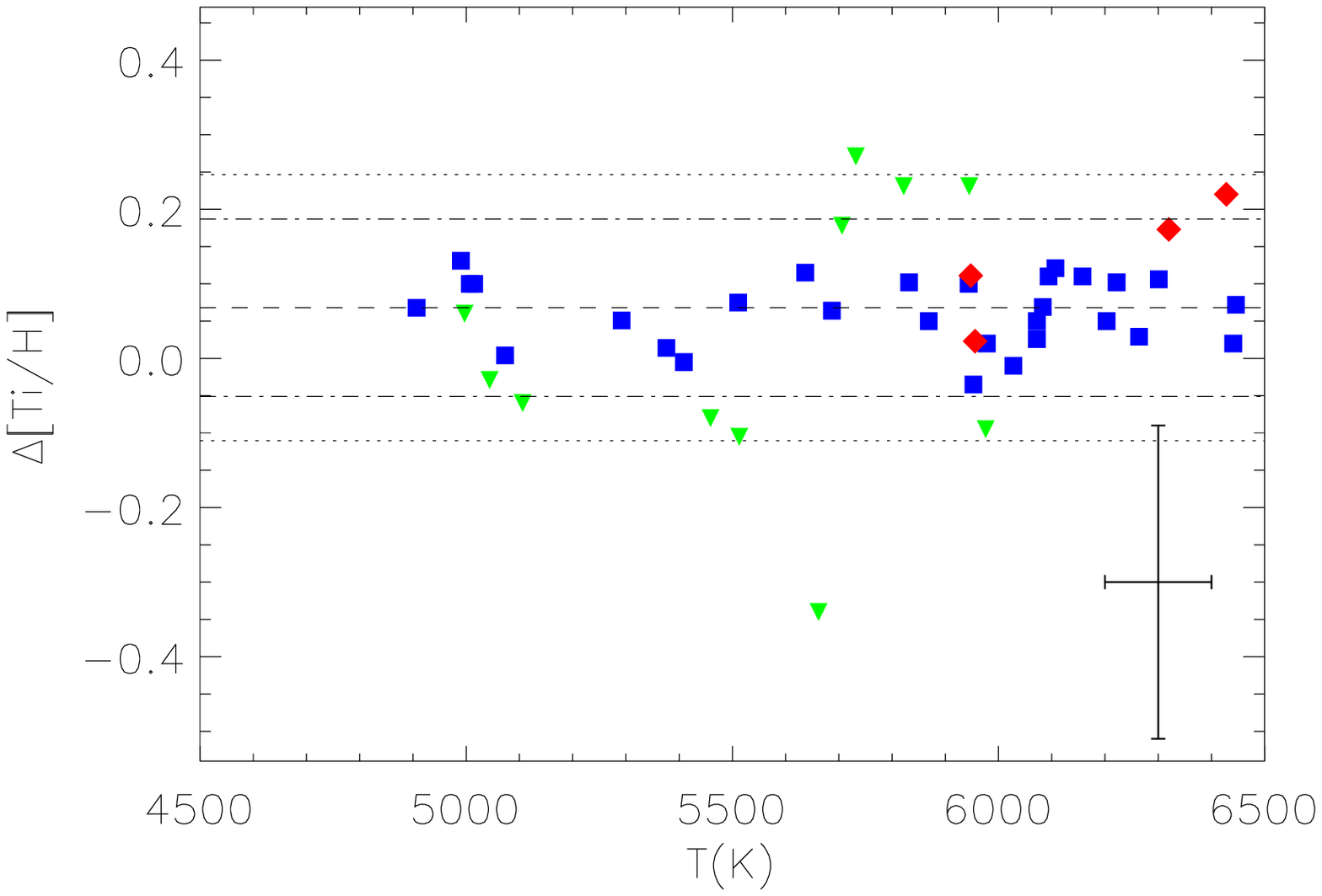}
\includegraphics[scale=0.35]{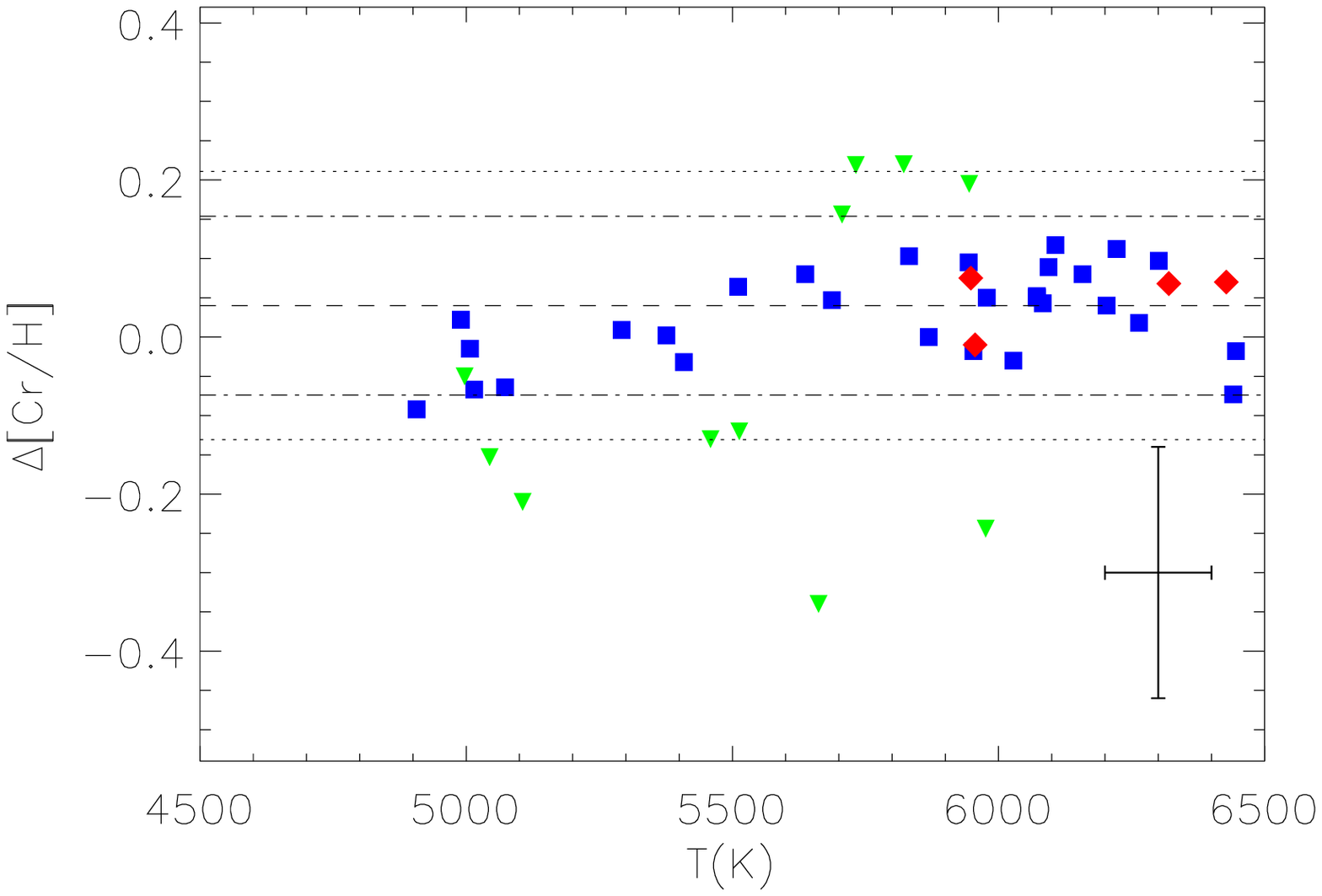}
\includegraphics[scale=0.35]{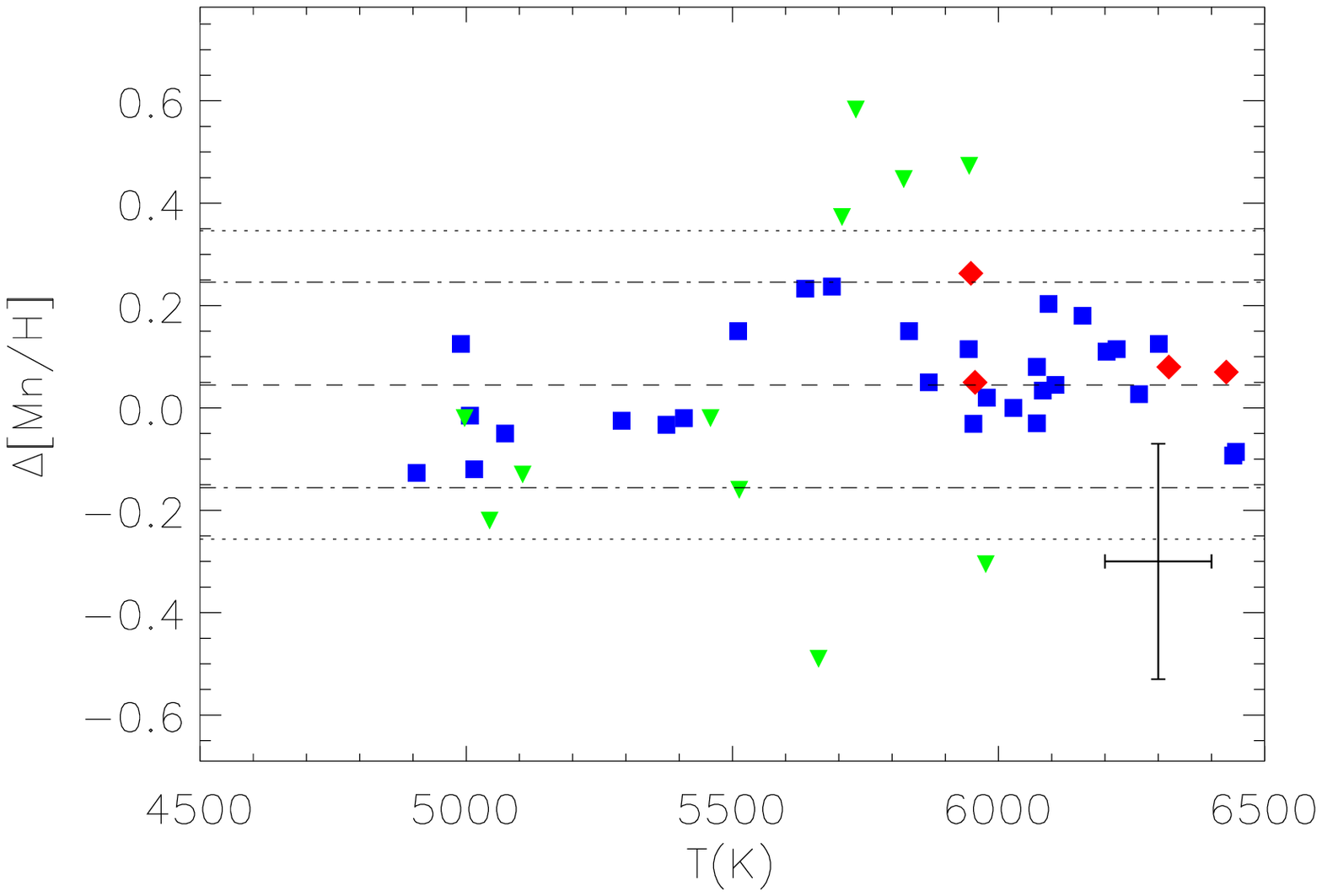}}
\centerline{
\includegraphics[scale=0.35]{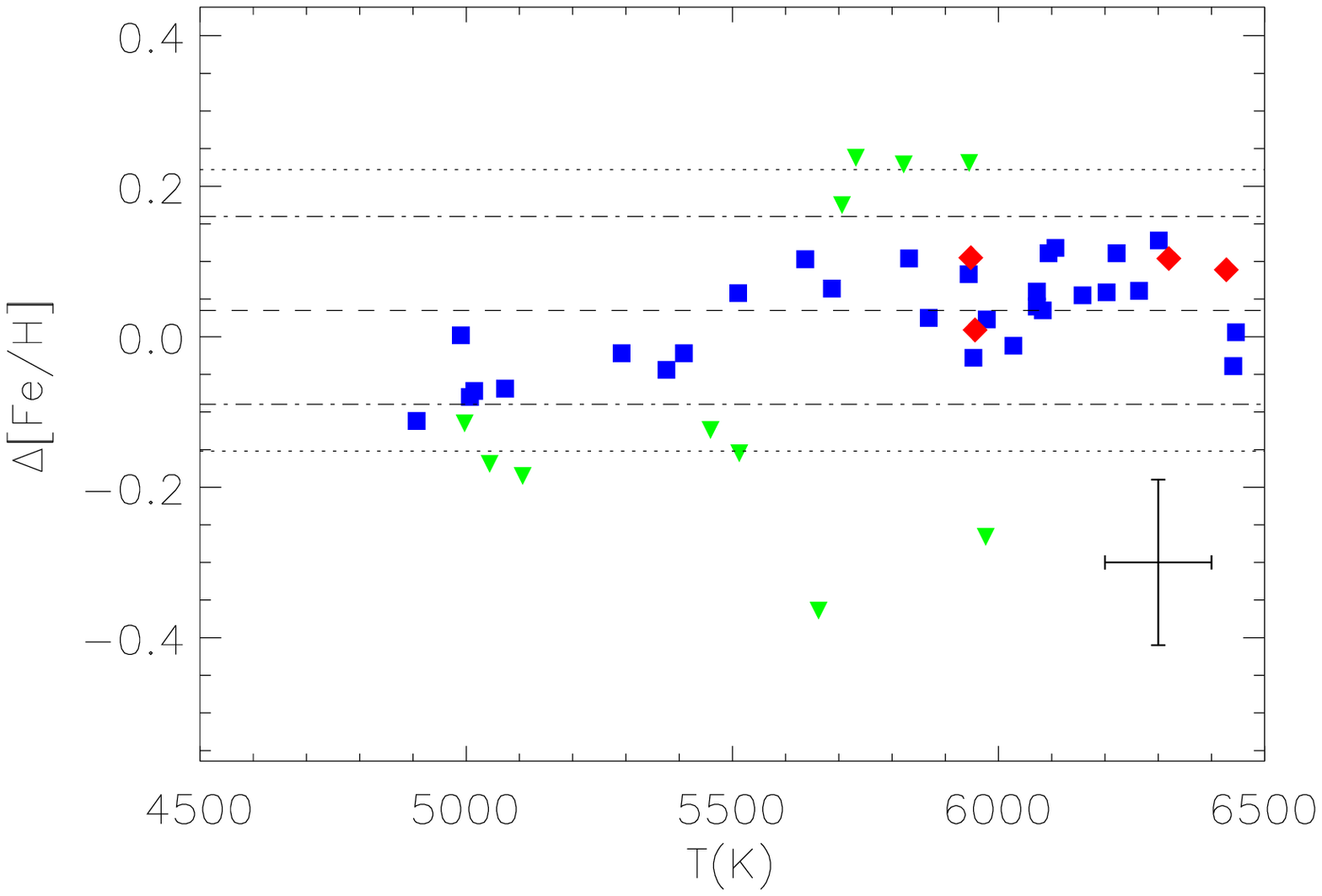}
\includegraphics[scale=0.35]{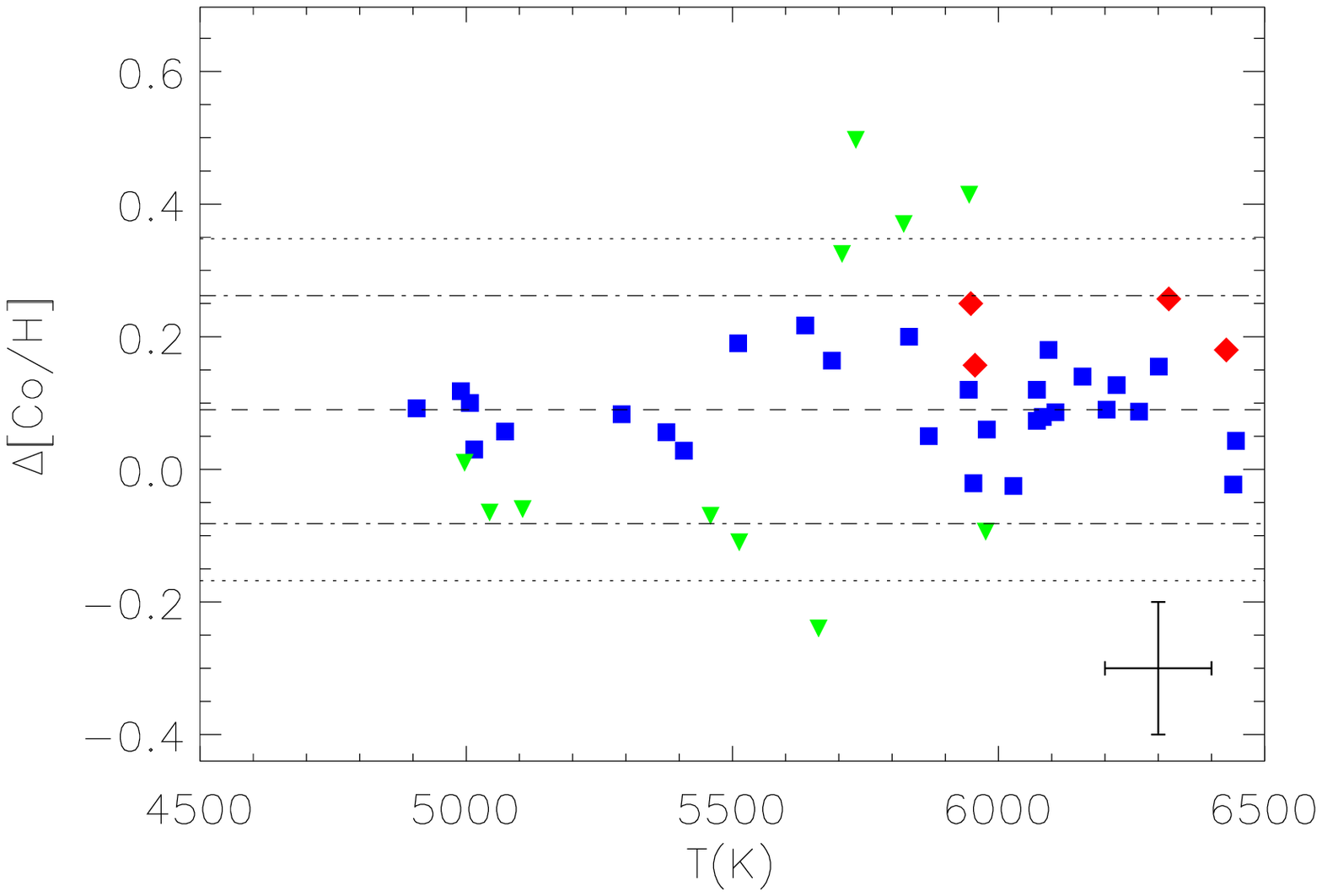}
\includegraphics[scale=0.35]{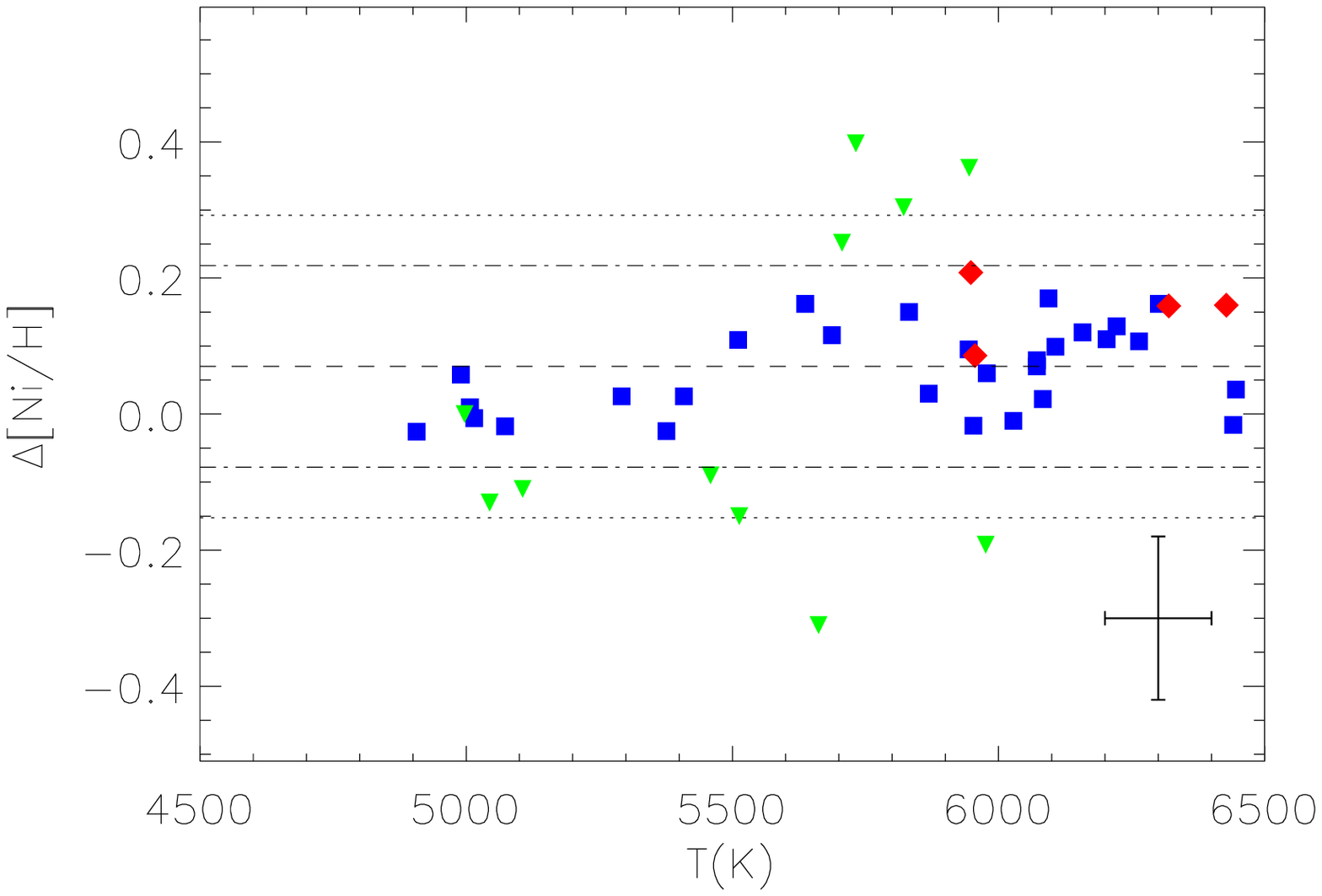}}
\caption{$\Delta$[X/H] differential abundances (with respect to HD~115043) for the 
$\alpha$-elements (Mg, Si, Ca, and Ti), Fe, and the Fe-peak elements (Cr, Mn, Co, and Ni) vs. $T_{\rm eff}$. Dashed-dotted 
lines represent 1-rms over and below the median for our sample,  whereas dotted lines represent 
the 1.5-rms level. Dashed lines represent the mean differential abundance. The meaning of the
symbols is the same as
in Fig.~\ref{umagal1} }
\label{umadif1}
\end{figure*}

We determine differential abundances $\Delta$[X/H] by comparing our measured abundances with those 
of a reference star known to be a member of the Ursa Major nucleus \citep[HD~115043, see ][]{kin03} on a 
line-by-line basis. The candidate selection within the sample was determined by applying a one 
root-mean-squared (rms, thereafter) rejection over the median for almost every chemical element 
studied. The rejection process considers the rms in the abundances of the sample for each element. At first, we discarded 
every star that deviates by more than 1-rms from the median abundance denoted by the dashed-dotted lines in 
Figs.~\ref{umadif1}, and \ref{umadif2}. The initial rms values considered during the candidate 
selection are given in Table~\ref{tablarms}. 

The initial 1-rms rejections lead to the identification of 15 candidate members. We subsequently 
apply a more flexible criterion allowing stars to become members when their abundances
were within the 1-rms interval for 90 \% of the elements considered and the remaining 
10 \% within the 1.5-rms interval (i.e. 18 elements and 2 elements respectively). 
The final rms is referred to the selected candidates of the Ursa Major moving group. The error
analysis considers only the standard deviation in the line-by-line differences. Using this flexible
approach allows us to find 29 members that may share similar abundances among the whole sample
containing 44 stars (i.e., a 66\%).  

\begin{figure*}
\centering
\centerline{\includegraphics[scale=0.50]{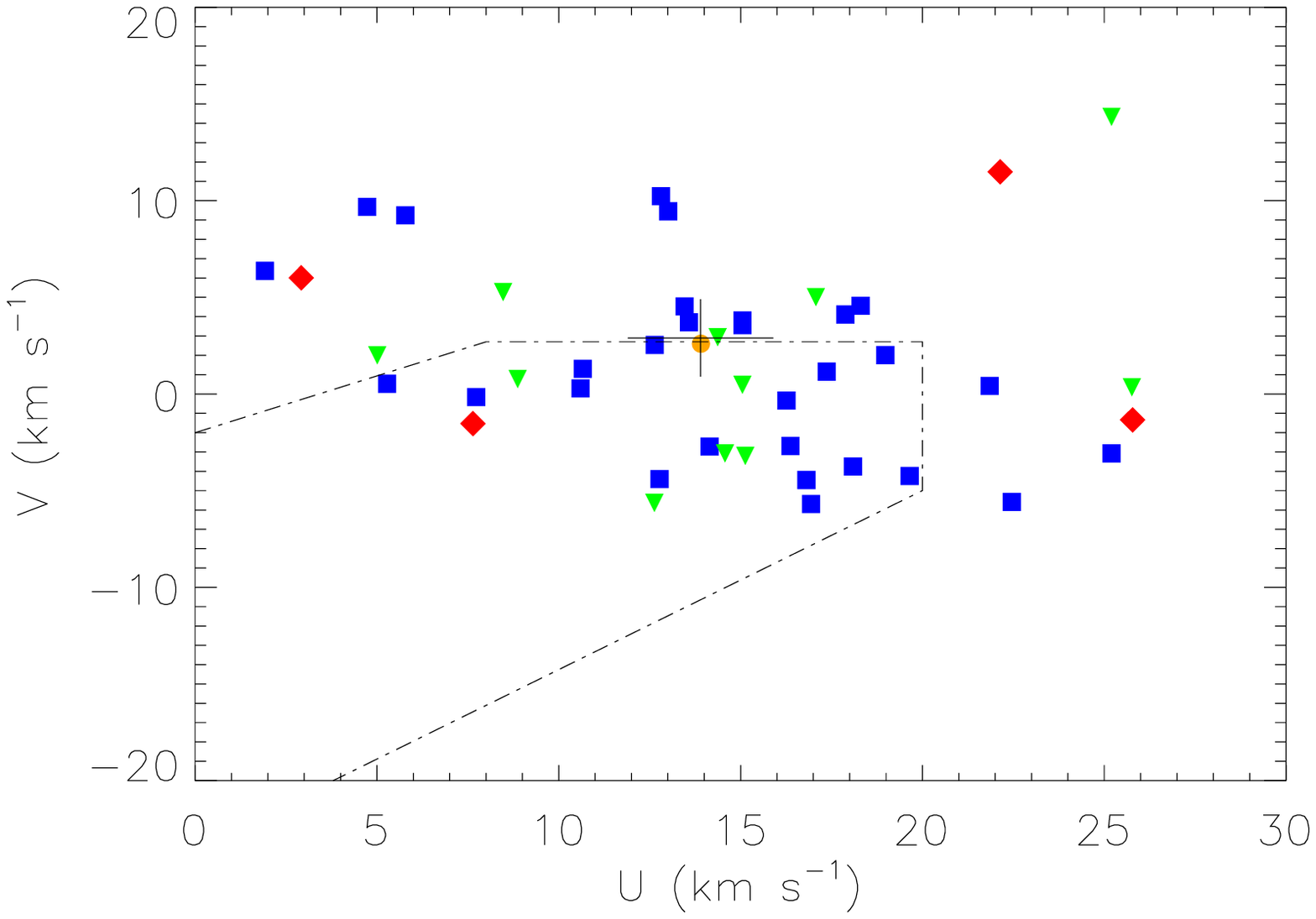}
\includegraphics[scale=0.50]{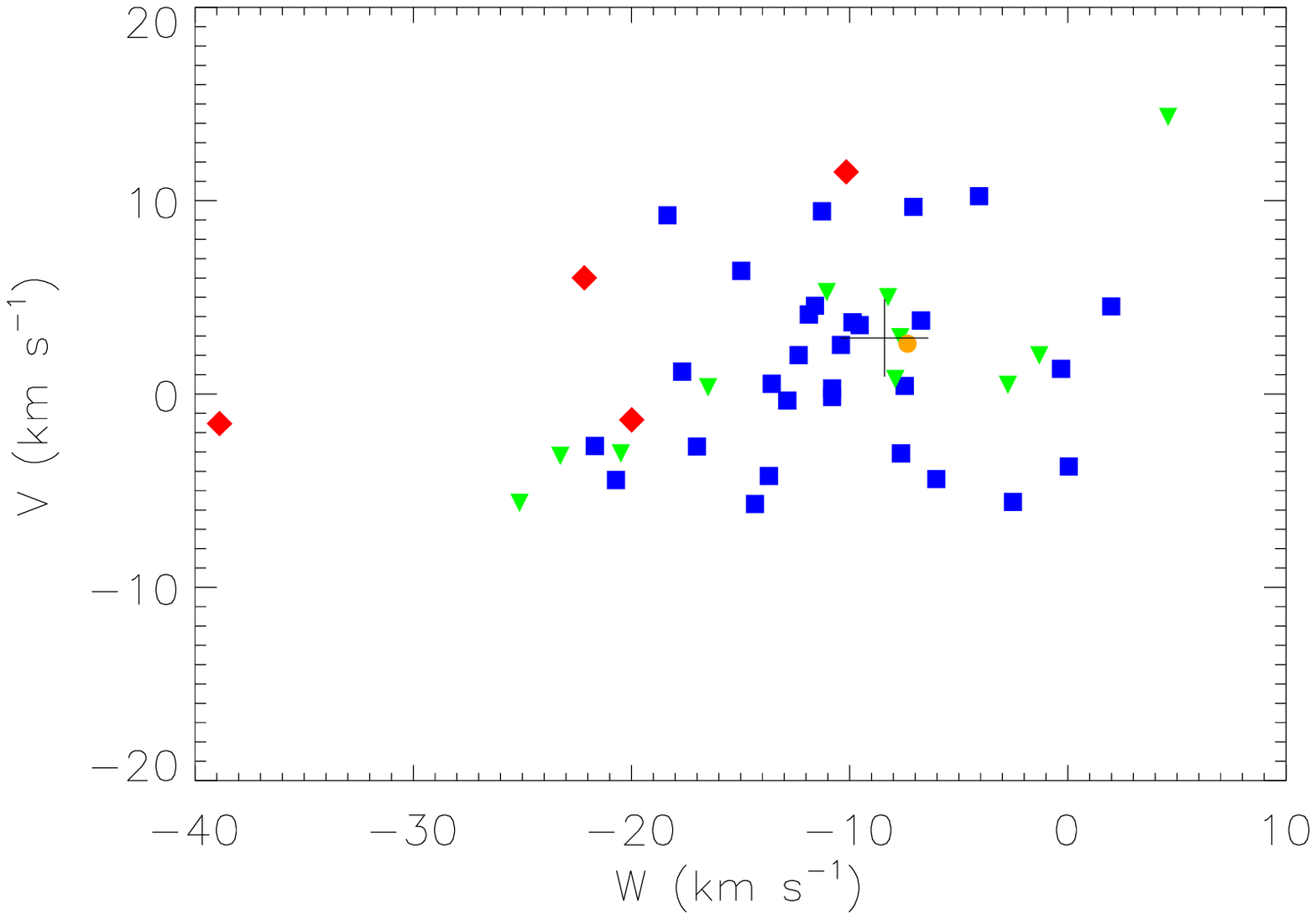}}
\caption{\scriptsize $U$, $V$, and $W$ recalculated velocities for the possible members of the 
Ursa Major MG. For a full explanation on the final selection criterion, we refer the reader to Section~5. The big black cross 
indicates the $U$, $V$, and $W$ central location of the Ursa Major MG \citep[see ][]{kin03}. Dashed lines
show the region where the majority of the young disc stars  tends to be according to \citet{egg84,egg89}. 
The meaning of the symbols is the same as in Fig.~\ref{umagal1}}
\label{uvwuma}
\end{figure*}

This more flexible rms-based analysis was made in order to identify the degree at which 
the sample is homogeneous, and to account for the likely contamination of the sample by field stars. Therefore, to assess 
this degree of homogeneity one must take into account the number of stars that lie within 1-rms, 1.5-rms, 2-rms, and 3-rms  
intervals (see Table~\ref{tablatotal}). The last three columns of Table~\ref{tablapar} give information about membership 
based on the differential abundances (with respect to HD~115043) of Fe and the other elements following these criteria. Combining
the pure 1-rms rejection and the more flexible criteria we  obtain that from 34\% to 66\% of the candidates 
are members of UMa, for a pure 1-rms rejection and the flexible criterion, respectively. The rms of the final 
selection for different elements ranges about 0.1 dex to 0.05 dex.  We find that Si, Ca, Cr, Fe, and Ce exhibit  an internal dispersion equal or better than 0.08 dex. On the other hand, Na, Mg, Ti, Ni, Zn, Zr, and Nd display a disperssion of less than 0.1 dex. The remaining elements have a rms scatter around 0.1 dex (see Table~\ref{tablarms}). Interestingly, the
present chemical analysis can eliminate some outliers in the space velocity diagram (see Fig~\ref{uvwuma}). In addition, our 
final set of selected candidates tends to concentrate nearby the mean velocity of the Ursa Major MG.

\begin{table}
\small
\caption{Median differential abundances (with respect to HD~115043), and both initial and 
final rms values for all considered elements.}
\label{tablarms}
\centering
\begin{tabular}{lrrr}
\hline\hline
Element  & $\Delta$ [X/H] & rms$_o$ & rms$_f$ \\
\hline
Na &  0.03 & 0.15 & 0.09 \\
Mg  & 0.02 & 0.13 & 0.08 \\
Al  & 0.10 & 0.15 & 0.11 \\
Si  & 0.06 & 0.12 & 0.08 \\
Ca  &  0.00 & 0.11 & 0.05 \\
Sc &  0.05 & 0.17 & 0.12 \\
Ti &   0.07 & 0.12 & 0.08 \\
V &  0.10 & 0.18 & 0.14 \\
Cr  & 0.04 & 0.11 & 0.06 \\
Mn  & 0.05 & 0.20 & 0.11 \\
Co &  0.09 & 0.17 & 0.11 \\
Ni &  0.07 & 0.15 & 0.09 \\
Fe &  0.04 & 0.12 & 0.07 \\
Cu &  0.13 & 0.22 & 0.17 \\
Zn &  0.03 & 0.15 & 0.09 \\
Y  &  0.09 & 0.15 & 0.10 \\
Zr  &  0.01 & 0.16 & 0.09 \\
Ba  & $-0.14$ & 0.22 & 0.15 \\
Ce  & $ -0.01$ & 0.10 & 0.06 \\
Nd  & $-0.01$ & 0.15 & 0.08 \\
\hline 
\end{tabular}
\end{table}

As a final test, we compared the Hyades SC abundances \citep{tab12} and thin disc \citep{gon10} with those of UMa (see Fig.~\ref{abhyum}). For the thin disc, each value is 
derived from the average of those stars within one $\sigma$ around the [Fe/H] of each MG. It is interesting 
to check whether the two moving groups have different abundance patterns. Some of the 20 individual 
abundances seem to be marginally distinguishable, with a few noticeable exceptions, namely Ca, V, 
Y, Ba, and Zr. The other elements  might behave differently for the two groups, the
Ursa Major MG being less metallic (nearly solar) than the Hyades SC (super-solar composition). 

Different abundance patterns  can indicate a different formation site \citep{fre02}. Therefore, it
would be possible to distinguish different stars from different moving groups 
when using the {\it chemical tagging} approach. In spite of the fact that the abundances seem 
to be different from those of field stars, the internal dispersion does not give a clear
hint on any palpable difference. We also note that a detailed treatment of several different 
elements is important to have a good picture of the composition of moving groups. In conclusion, the two
MGs might behave differently in the abundance space.

\begin{table}
\small
\caption{Percentage analysis based on the rejection level of the differential abundances (with respect to HD~115043).}
\label{tablatotal}
\centering
\begin{tabular}{c c c}
		\hline\hline
		rms & \#stars & \%stars \\
		\hline
		1.0 & 15 &  34 \\
		1.5 & 31 &  71 \\
		2.0 & 36 &  82 \\
		2.5 & 38 &  86 \\
		3.0 & 41 &  93 \\
		\hline
	\end{tabular}
\end{table}

\section{Conclusions}

We have computed the stellar parameters and their uncertainties for 45 Ursa Major MG candidate stars, and 
obtained their chemical abundances for 20 elements (Fe, Na, Mg, Al, Si, Ca, Ti, V, Cr, Mn, Co, Fe, 
Ni, Cu, Zn, Y, Zr, Ba, Ce, and Nd), using a fully differential abundance approach with solar spectra 
of Vesta and the Moon as solar references.

We derive the Galactic space velocity components for each star and use them to check the 
original selection based on Galactic velocities \citep{mon01a,lop10}, which was then improved using 
the radial velocities derived from our data. We employ the new {\sc Hipparcos} proper motions and 
parallaxes \citep{hog00,lee07} using the procedures described in \citet{mon01a}. To perform a 
preliminary consistency check, we analysed the $U$, $V$, and $W$ Galactic velocities (see Fig.~\ref{uvwuma}) of
the final selected stars to not include any outliers in $V$.

\begin{figure}
	\centering
	\includegraphics[scale=0.50]{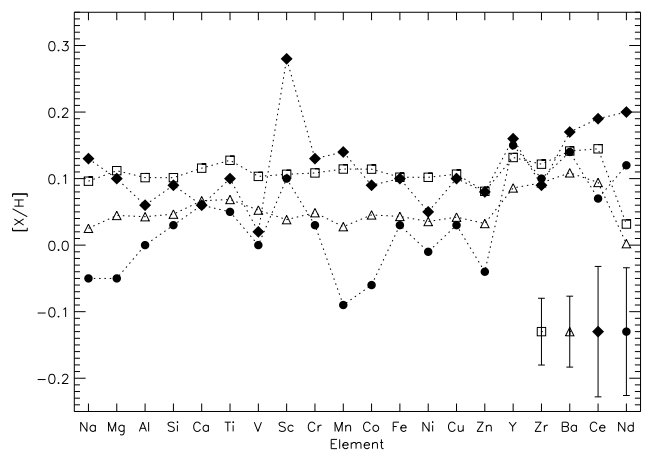}
	\caption{Elemental abundances for the Hyades SC \citep[filled diamonds, see ][]{tab12} and the Ursa Major
		MG (filled circles). Thin disc 
		        abundance values  for field solar analogues \citep{gon10} are represented by open 
			triangles (at Ursa Major MG [Fe/H]) and 
			        open squares (at the Hyades SC [Fe/H]). Right bottom error bars represent 
				the standard deviation. Dotted lines 
				        simply join the points for each moving group.  }
					\label{abhyum}
				\end{figure}

As a complementary test of the stellar parameters, we compile a $\log{g}$ vs. $\log{T_{\rm eff}}$ diagram to
verify the consistency of the method  employed to determine the stellar parameters. This diagram 
shows that most of the stars fall on the isochrone for the Ursa Major attributed age 
\citep[0.3 Gyr, see ][]{kin03,amm09}. This is an important but insufficient condition to ascertain 
that they have a common origin. The differential abundance analysis ({\it chemical tagging}) shows that
the finally 29 selected stars are compatible with the accepted age isochrone, as expected if they have 
evaporated from a single star forming event. The membership percentage that we find in this
work (66\%) may indicate that the Ursa Major MG is likely to originate from a dispersing cluster. This
result was also pointed out by 
other studies \citep[such as][and references therein]{kin03,kin05,amm09}. Furthermore, we 
also verify that different moving groups (Hyades SC and Ursa Major) might be distinguished by 
the individual element abundances (see Fig.~\ref{abhyum}).

A yet more detailed analysis of different age indicators and chemical homogeneity 
is in progress and will be presented in future publications. This analysis will lead to a more 
consistent means of confirming a list of candidate members from the abundance analysis.

\begin{acknowledgements}

H.M.T and D.M acknowledge financial support from the Universidad Complutense de Madrid (UCM), the Spanish Ministry  of Economy 
and Competitiveness (MINECO) from pojects AYA2011-30147-C03-02, and The Comunidad de Madrid 
under PRICIT project S2009/ESP-1496  (AstroMadrid).  H.M.T also acknowledges the financial support of 
the Spanish Ministry of Economy and Competitiveness (MINECO) under grants BES-2009-012182 and 
EEBB-I-12-04038. J.I.G.H. acknowledges financial support from the  Spanish Ministry project 
MINECO AYA2011-29060, and also from the Spanish Ministry of Economy and Competitiveness (MINECO) under 
the 2011 Severo Ochoa Program MINECO SEV-2011-0187. M.A. acknowledges help by Klaus Fuhrmann who adapted 
the FOCES reduction pipeline for use with the Tautenburg Coud\'e-Echelle spectrograph. In addition, M.A. thanks Eike W. Guenther 
for providing observing time with the 2m telescope in Tautenburg. Furthermore, M.A. would like to thank
the staff at the Calar Alto and Tautenburg observatories. M.A. is supported 
by DLR (Deutsches Zentrum f\"ur Luft-und Raumfahrt) under the project 50 OW 0204. 
We would like to thank the anonymous referee for helpful comments and corrections. This publication 
makes use of data products from the Two Micron All Sky Survey, which is a joint project of the 
University of Massachusetts and the Infrared Processing and Analysis Center/California 
Institute of Technology, funded by the National Aeronautics and Space Administration and the National 
Science Foundation. This research has made use of the SIMBAD database, operated at the CDS, Strasbourg, France.
\end{acknowledgements}

\begin{appendix}
\section{On-line material}
\begin{figure*}
\centering
\centerline{
\includegraphics[scale=0.45]{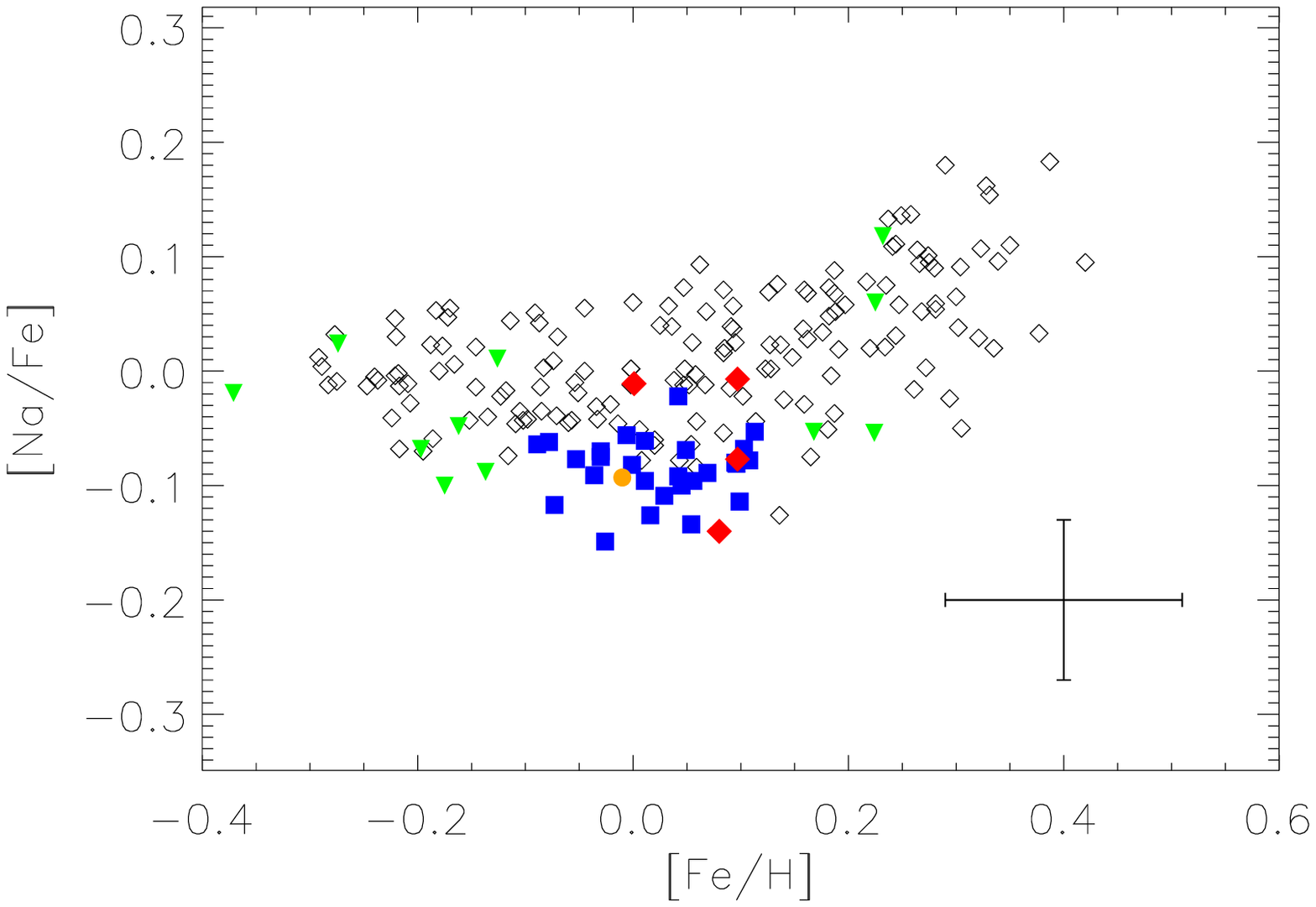}
\includegraphics[scale=0.45]{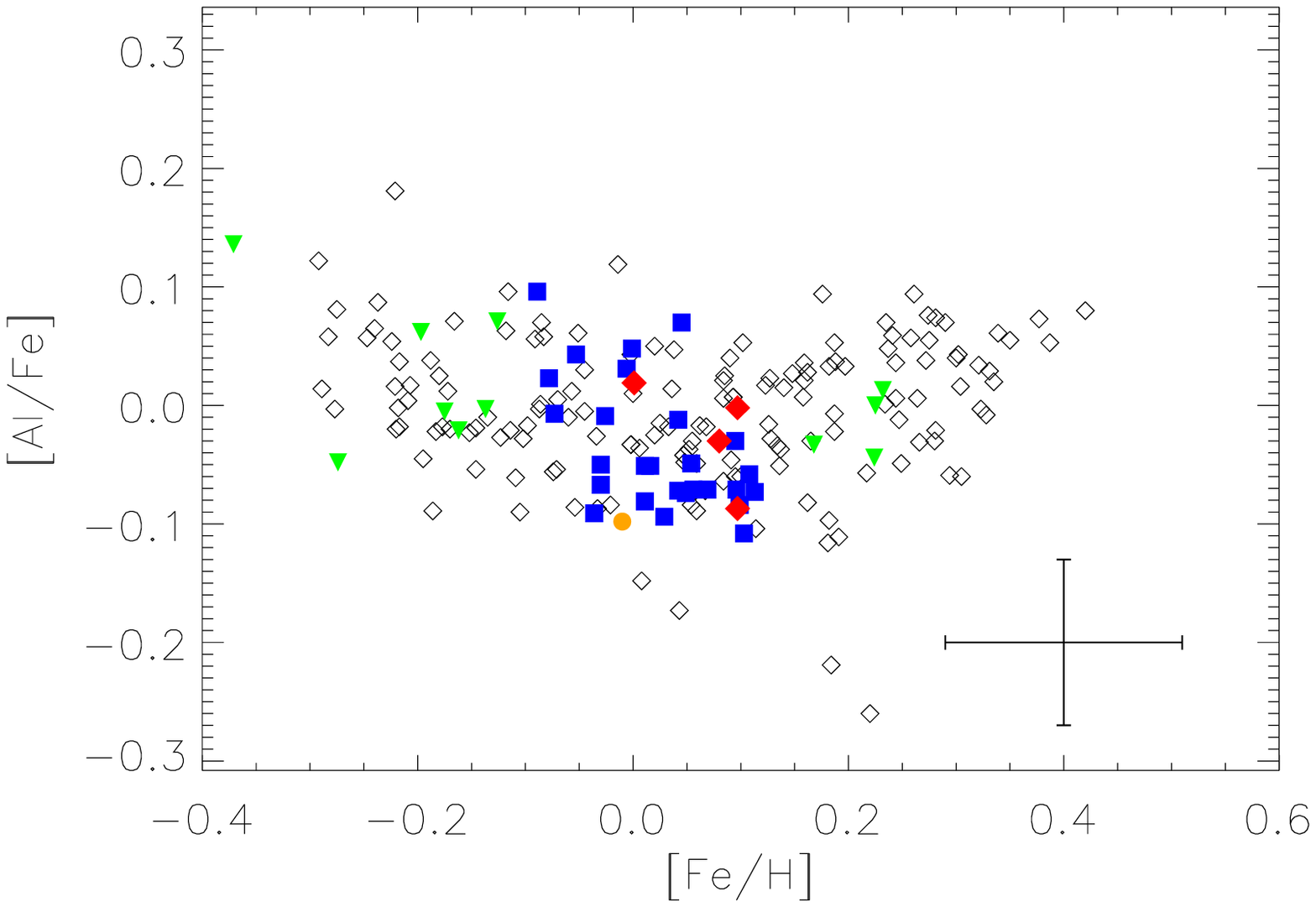}}
\centerline{
\includegraphics[scale=0.45]{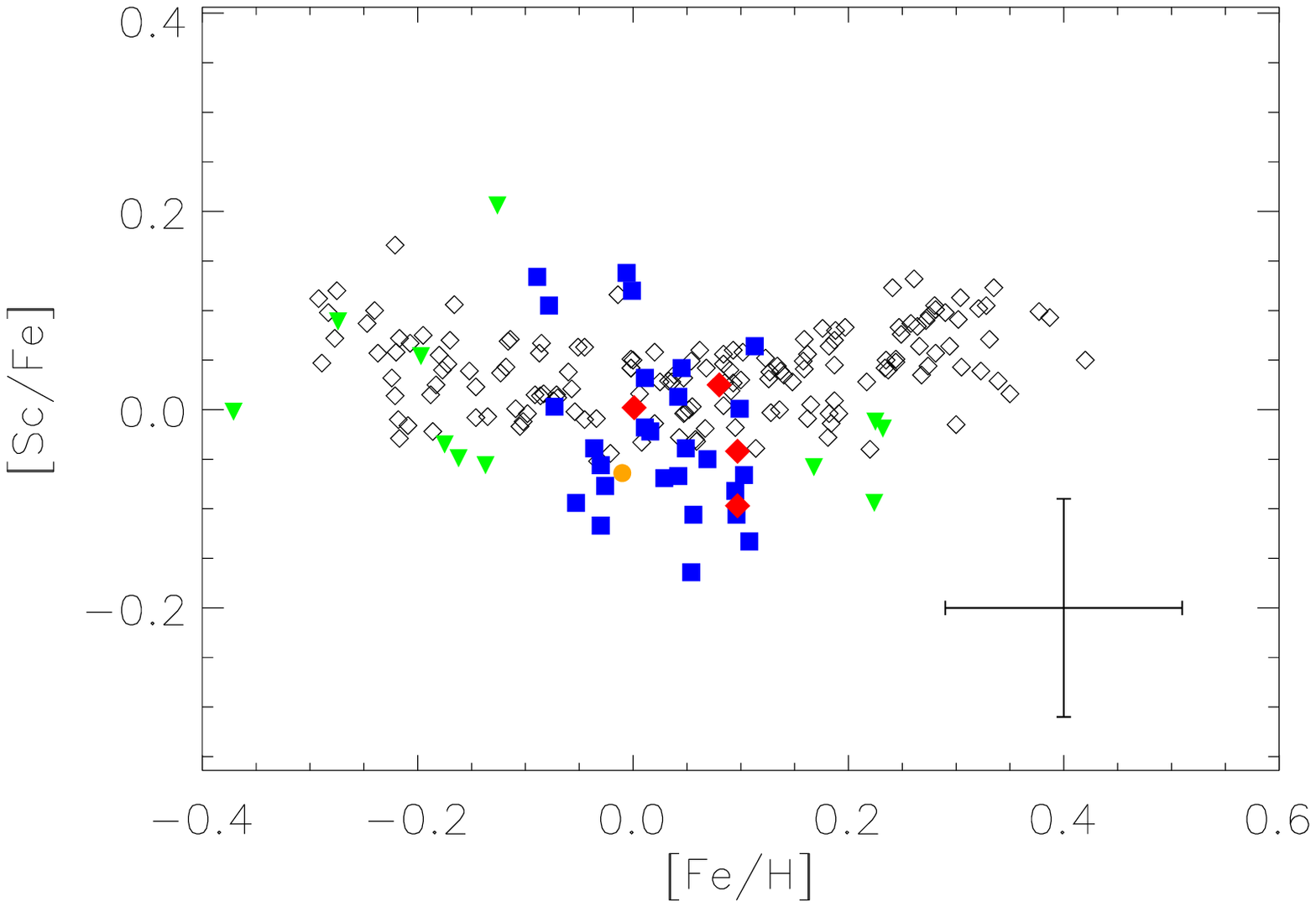}
\includegraphics[scale=0.45]{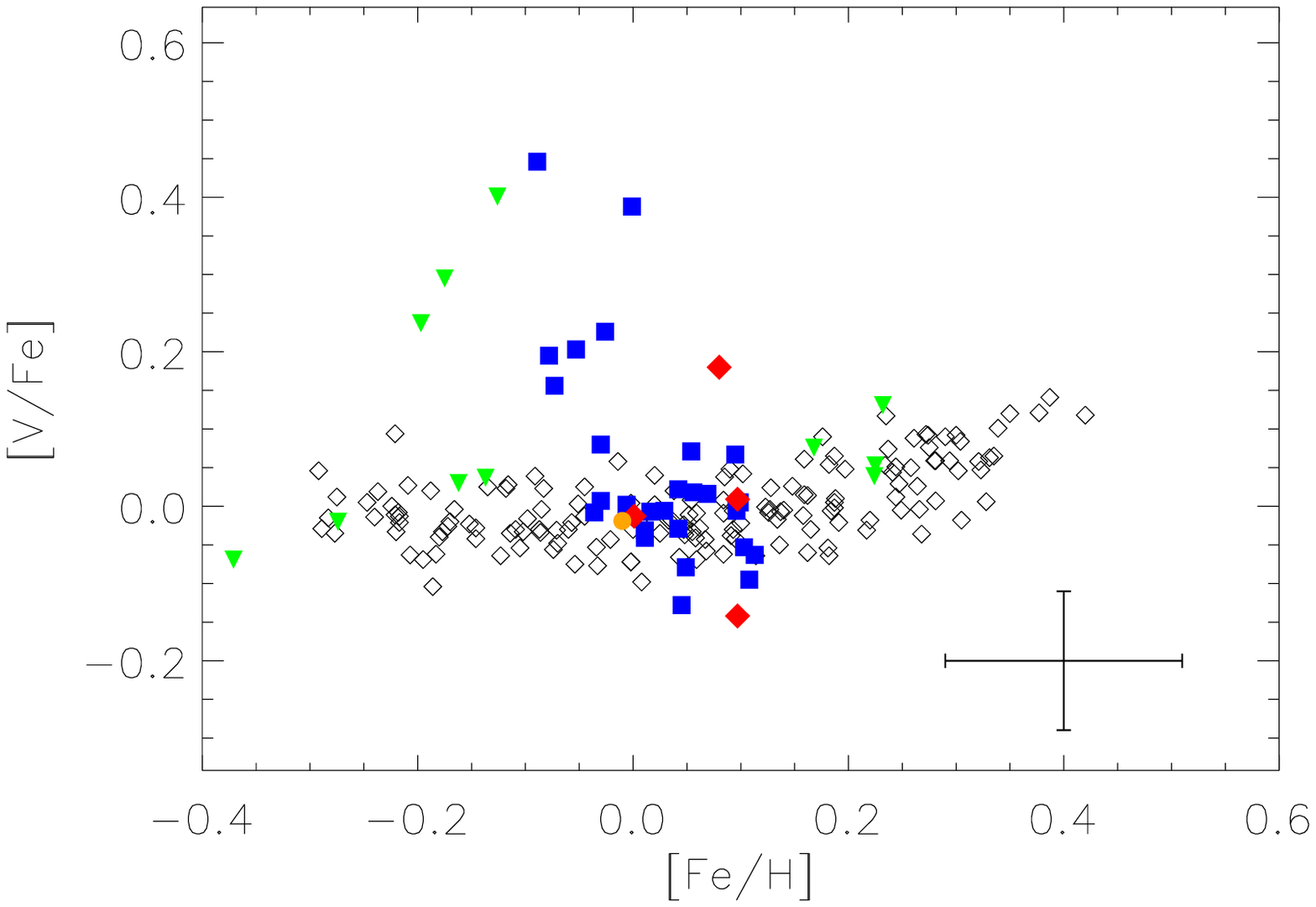}}
\centerline{
\includegraphics[scale=0.45]{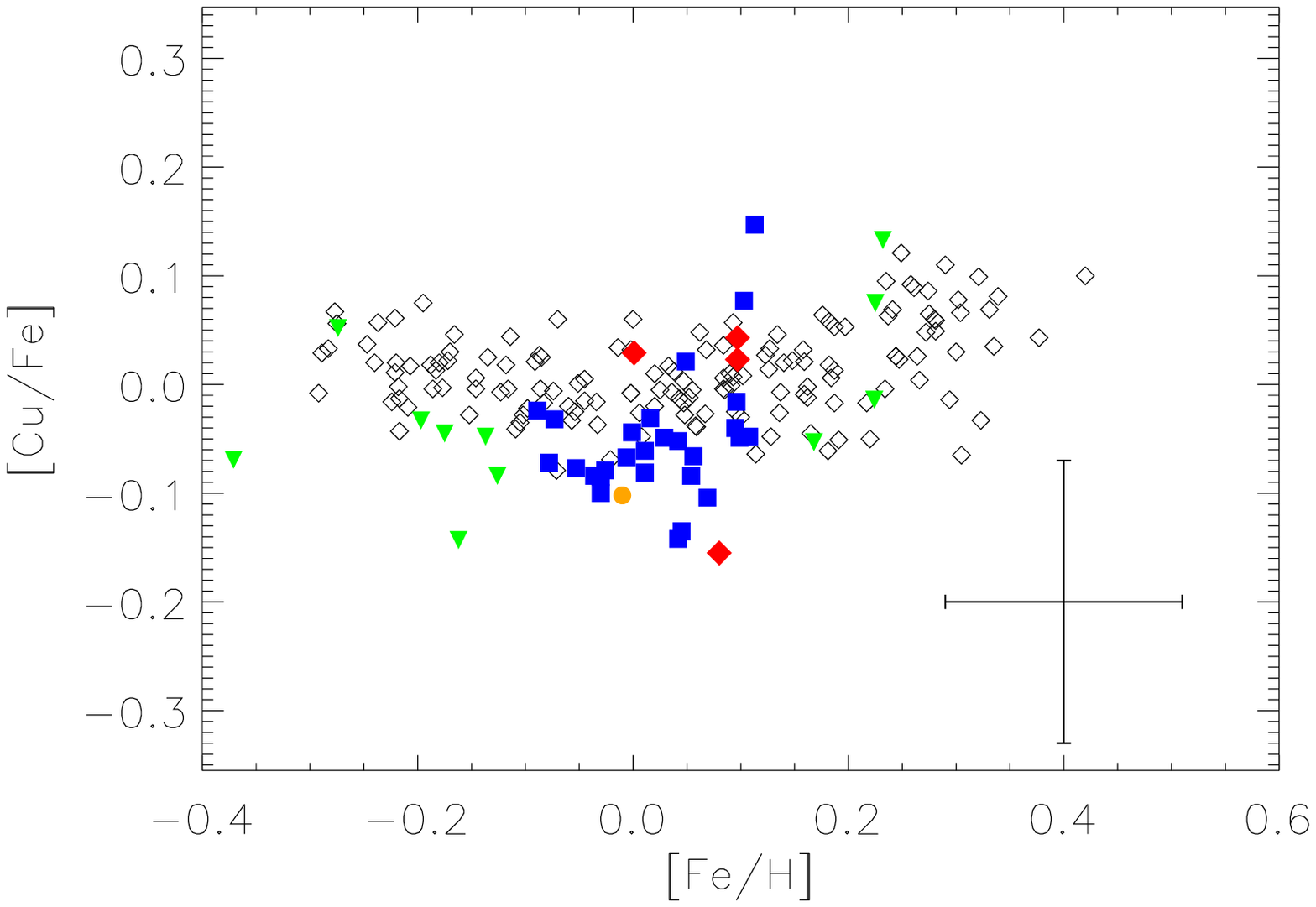}
\includegraphics[scale=0.45]{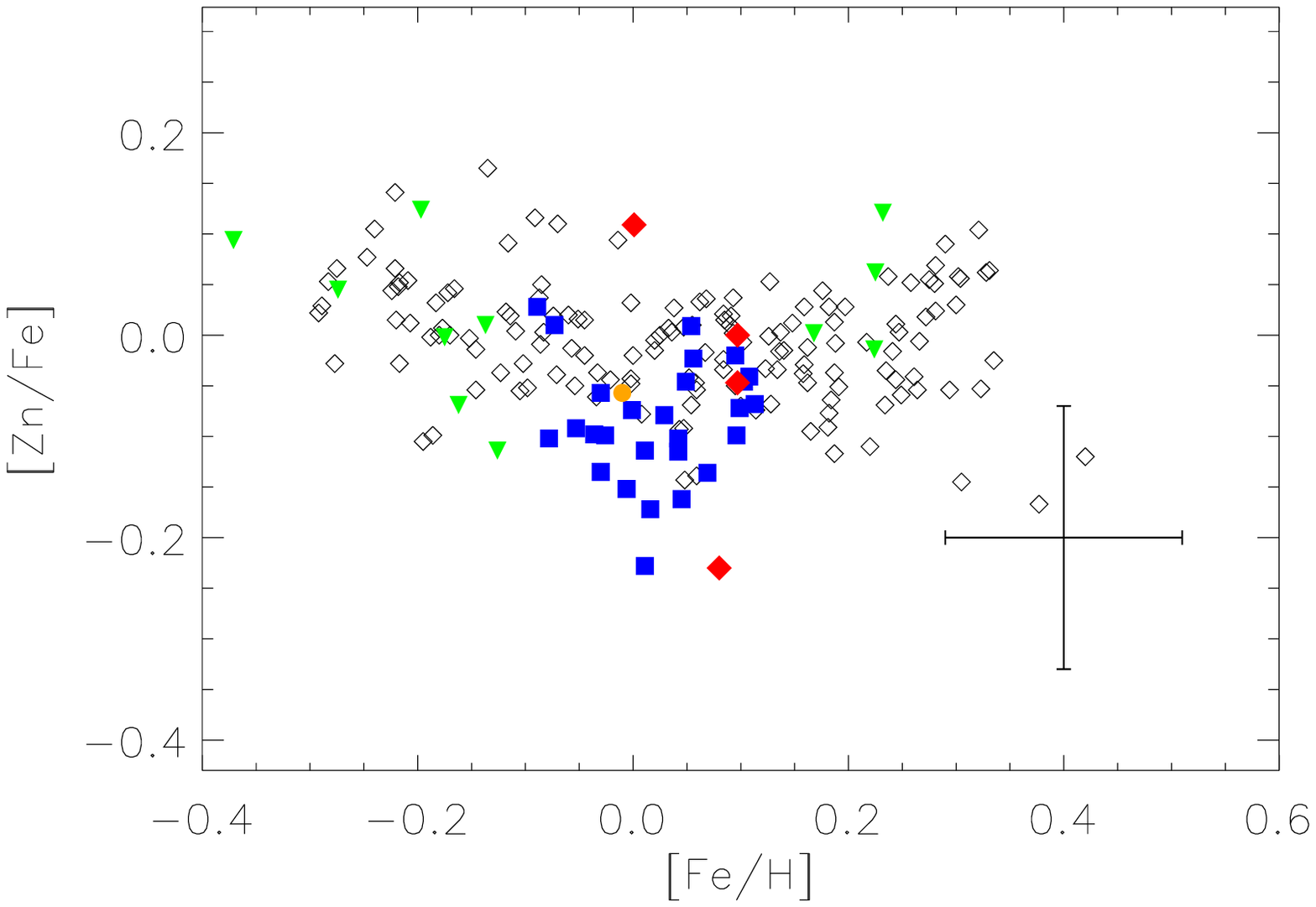}}
\centerline{
\includegraphics[scale=0.45]{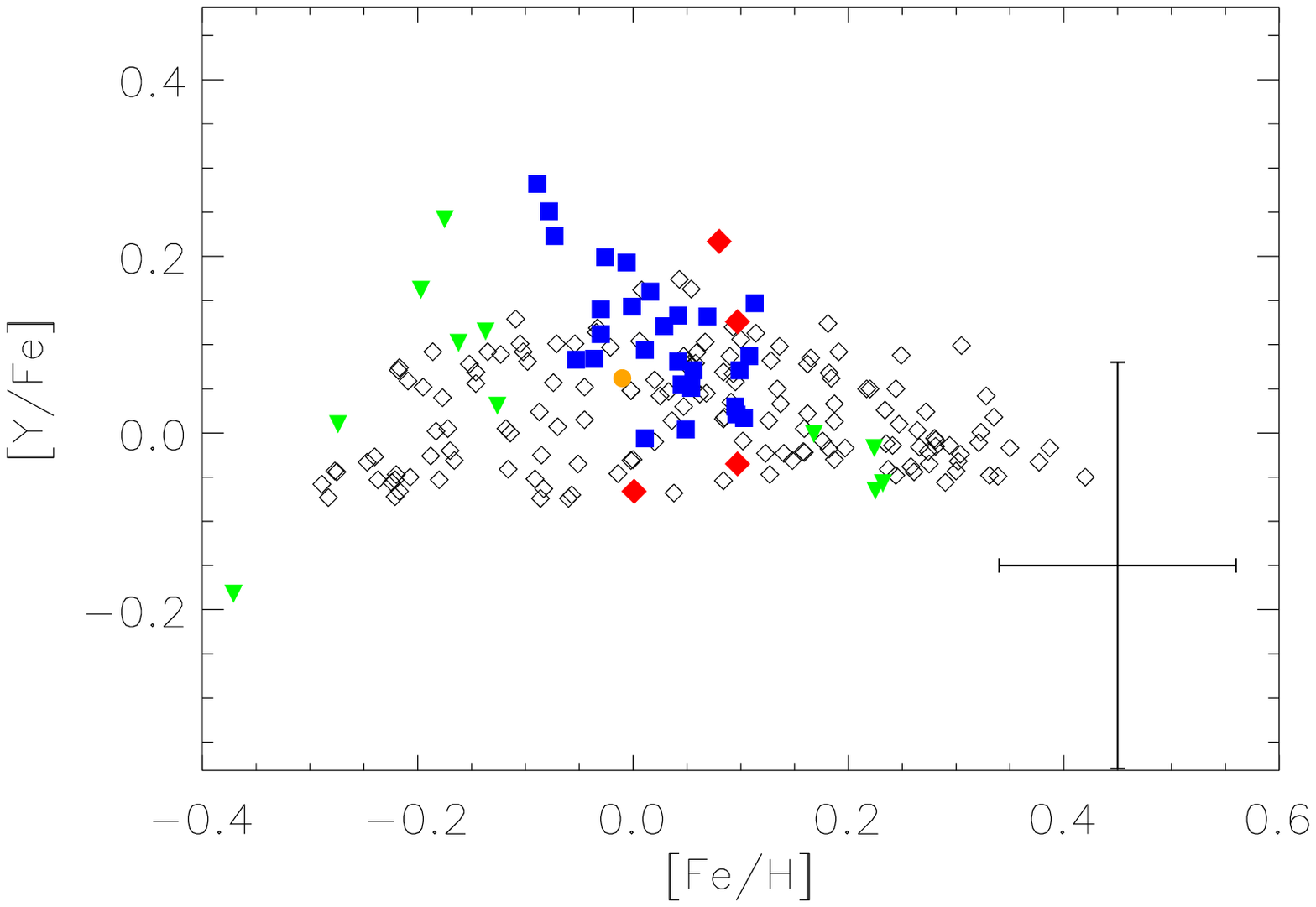}
\includegraphics[scale=0.45]{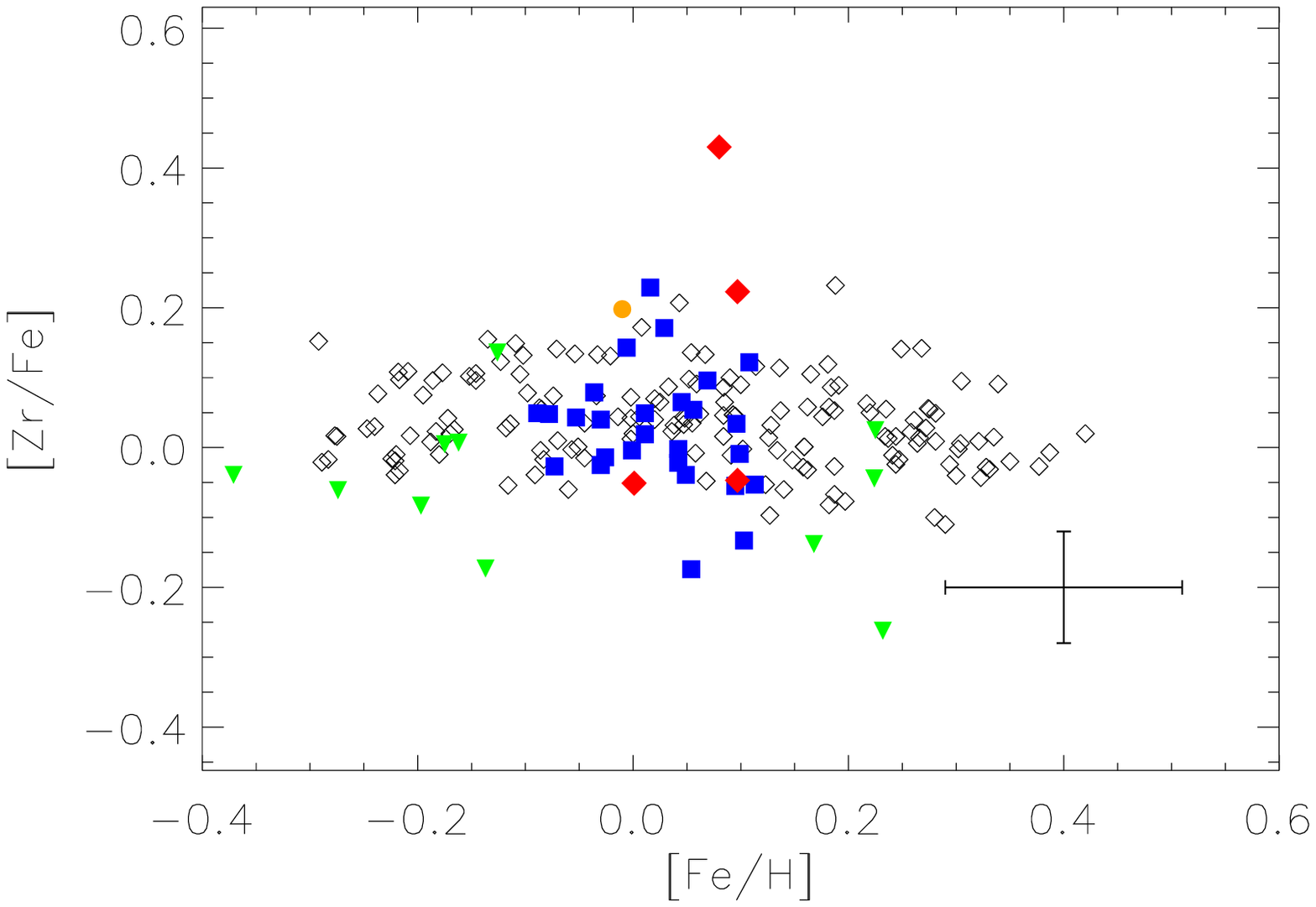}}
\caption{Same as Fig.~\ref{umagal1} but for the odd-Z elements (Na, Al, Sc, and V),  Cu, Zn, and 
the s-process elements (Y, Zr, Ba, Ce, and Nd).}
\label{umagal2}
\end{figure*}

\begin{figure*}
\centerline{\includegraphics[scale=0.45]{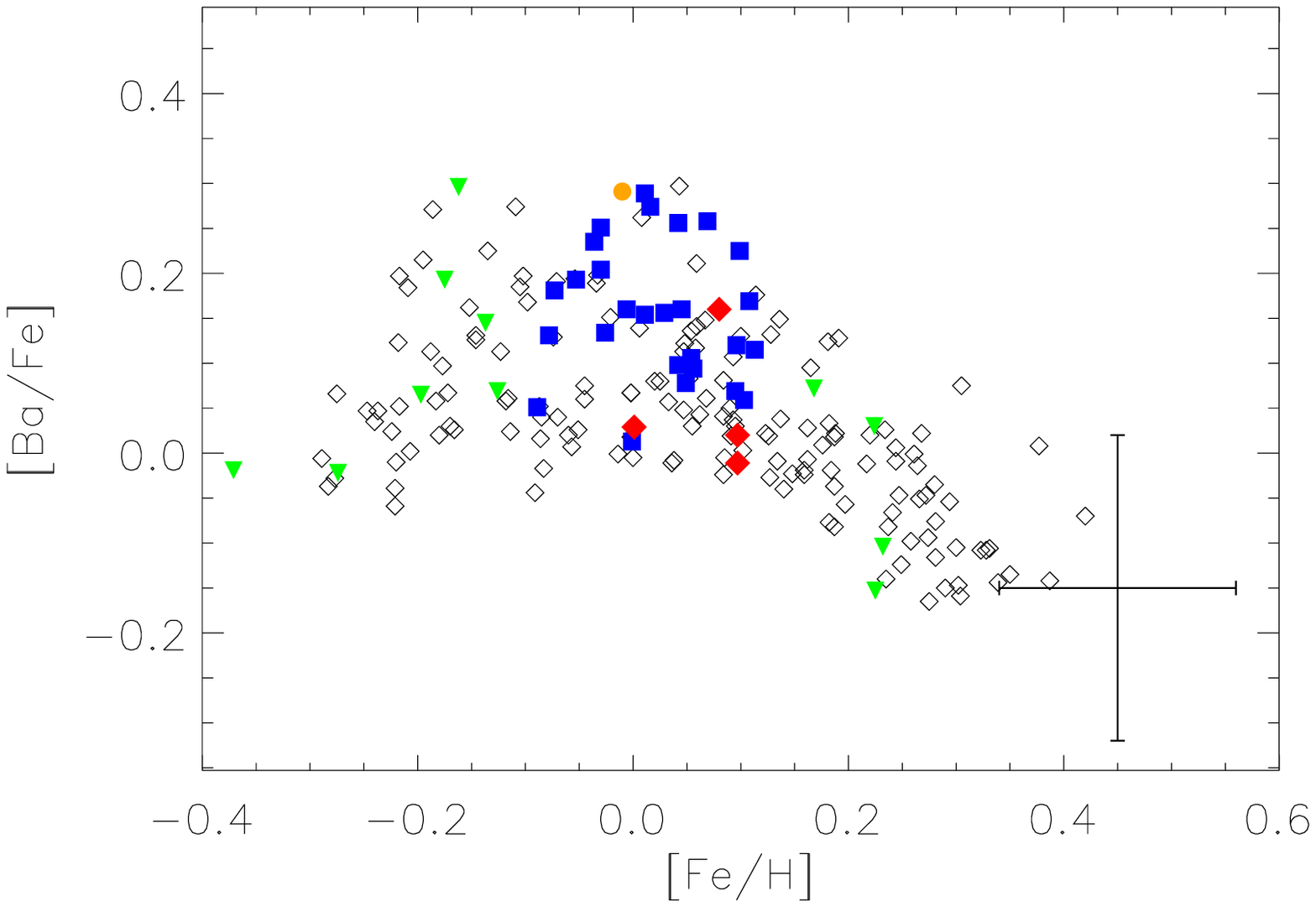}
\includegraphics[scale=0.45]{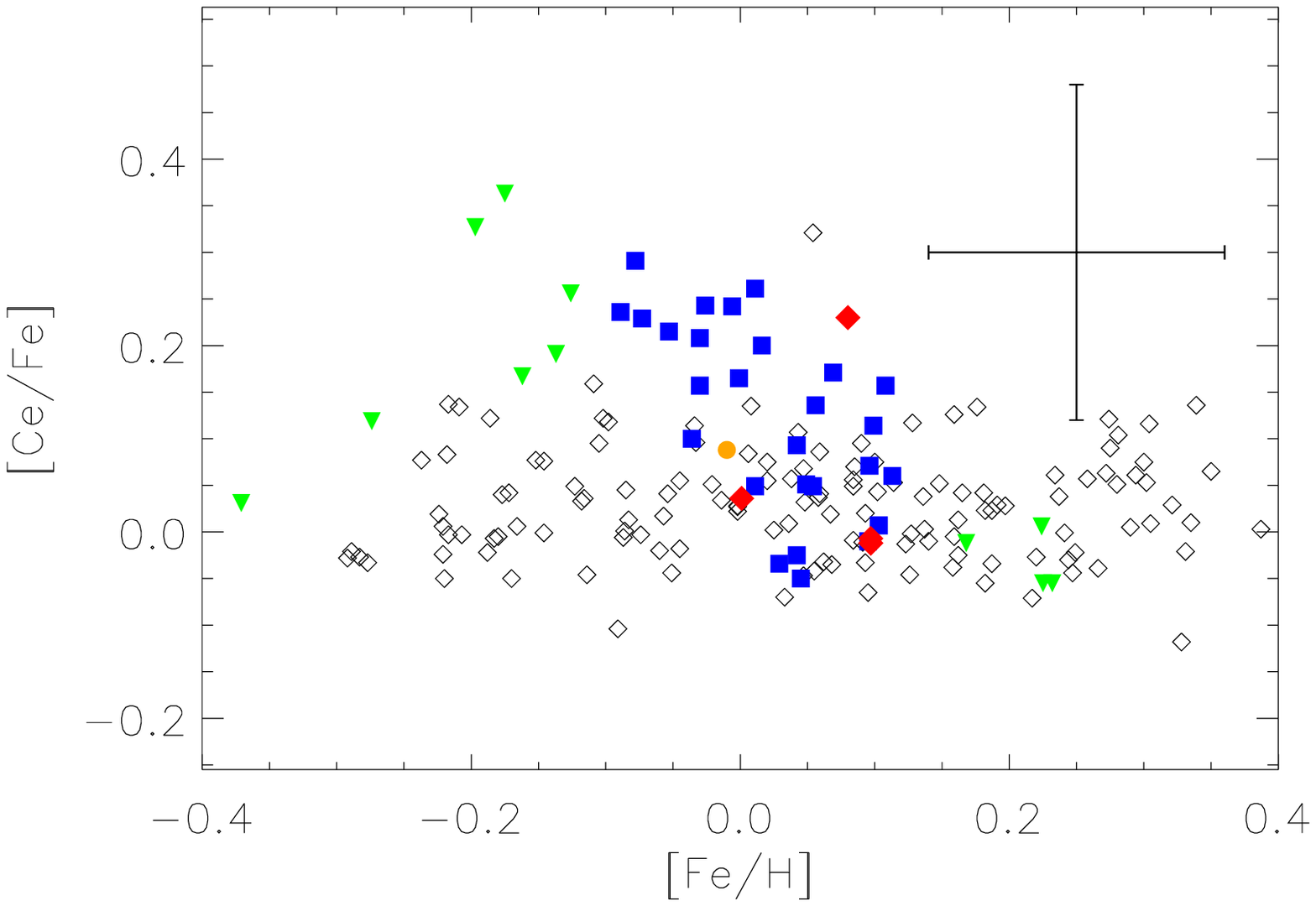}}
\centerline{\includegraphics[scale=0.45]{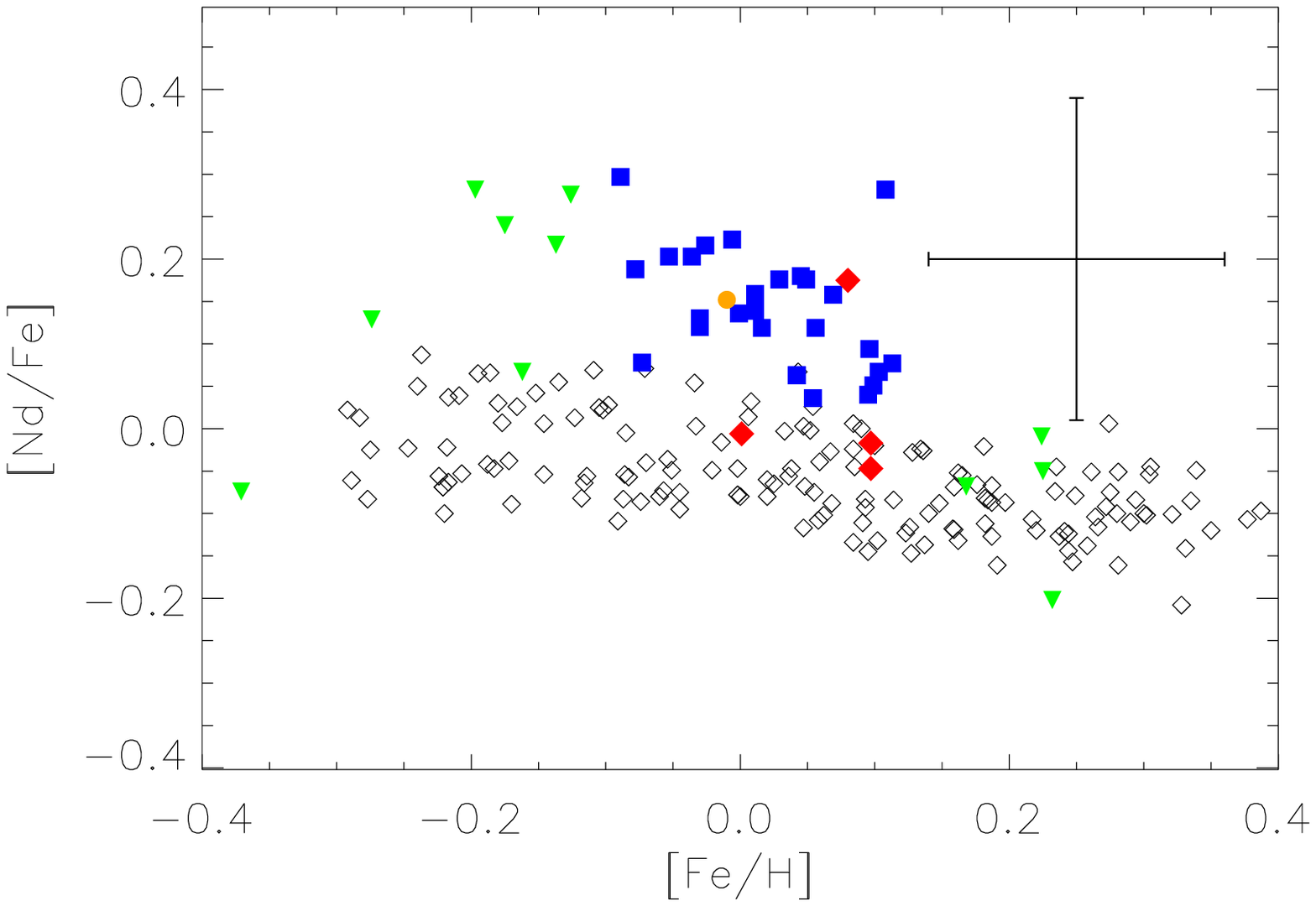}}
\caption{Fig.~\ref{umagal2} Continued.}
\end{figure*}

\begin{figure*}
\centerline{
\includegraphics[scale=0.35]{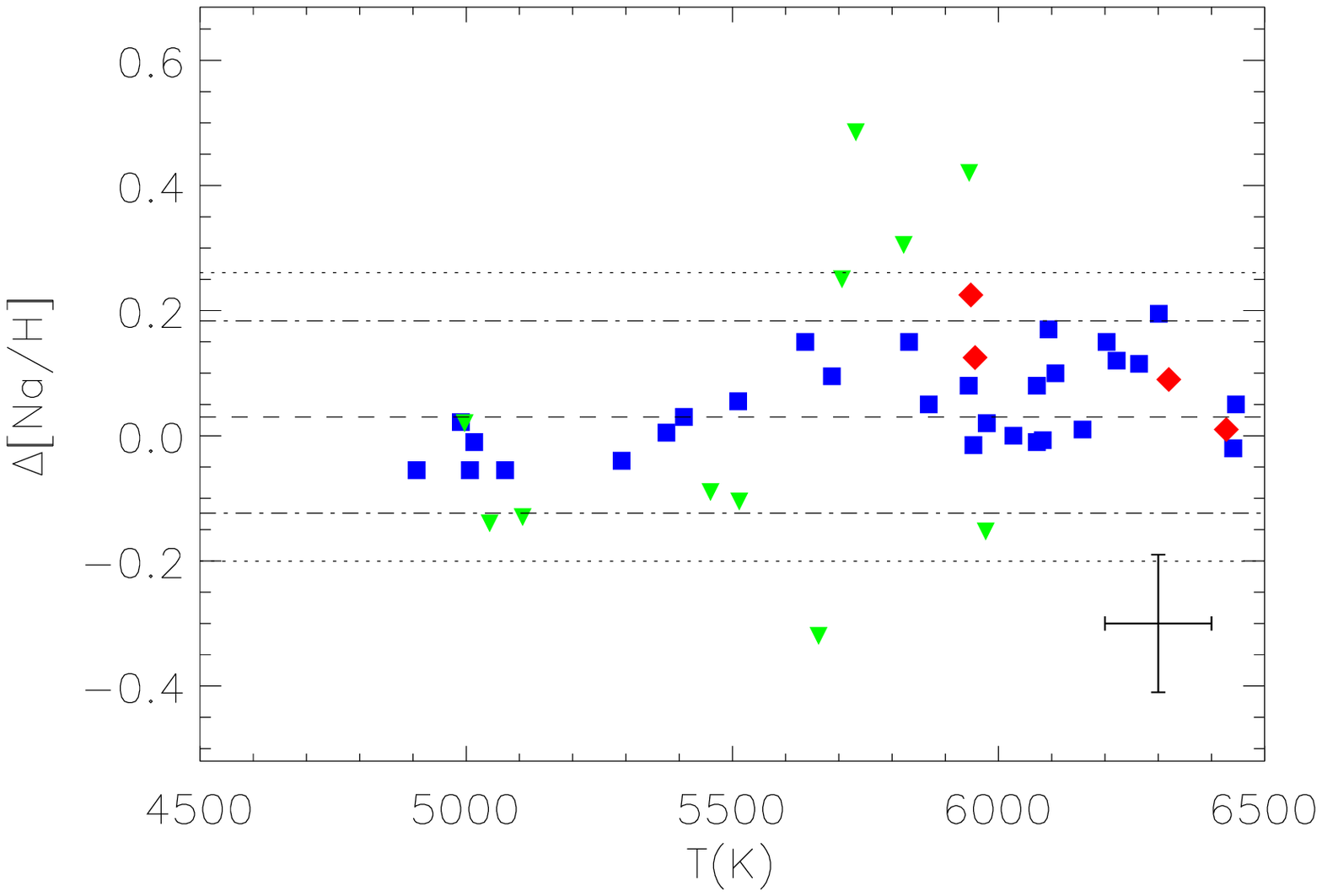}
\includegraphics[scale=0.35]{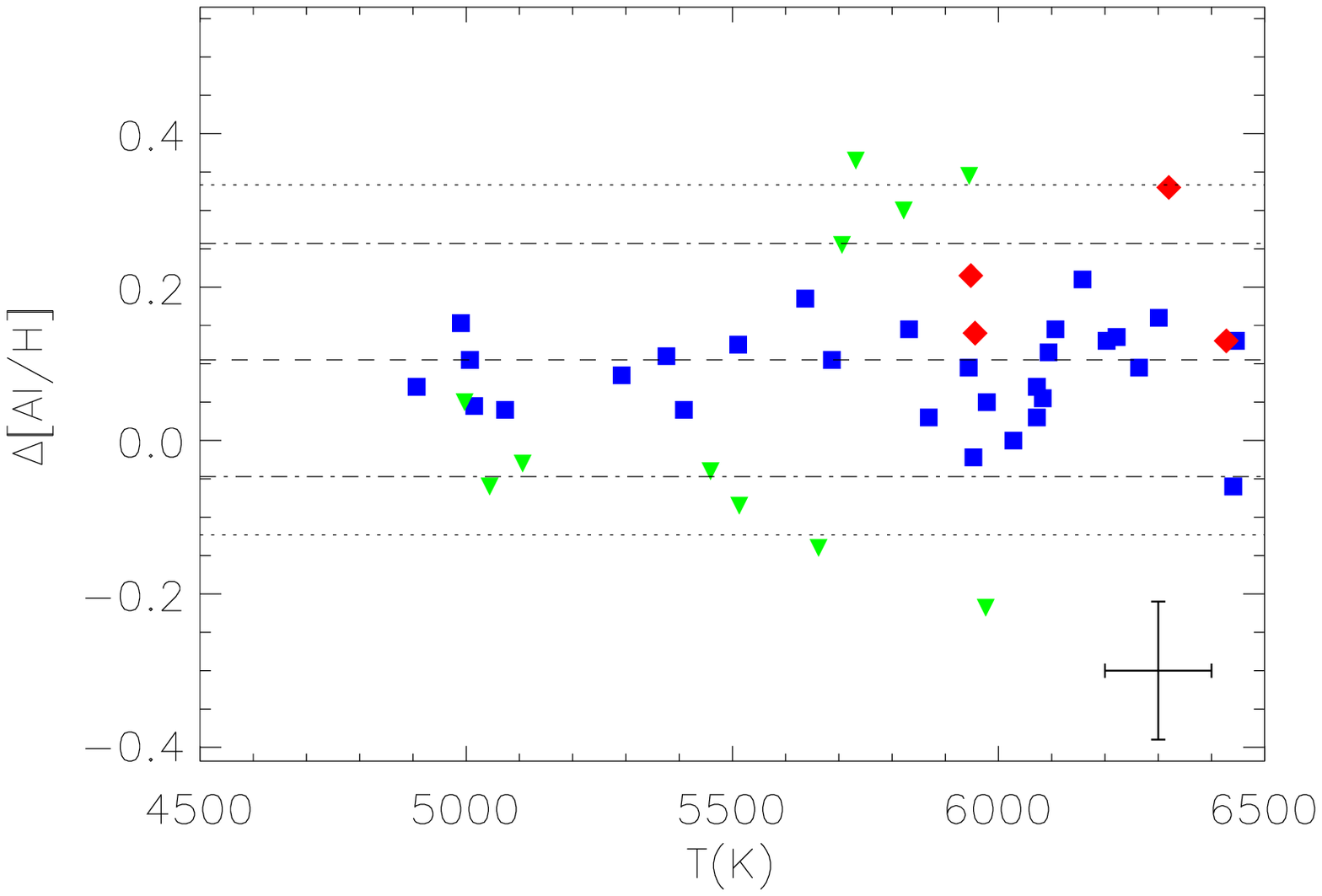}
\includegraphics[scale=0.35]{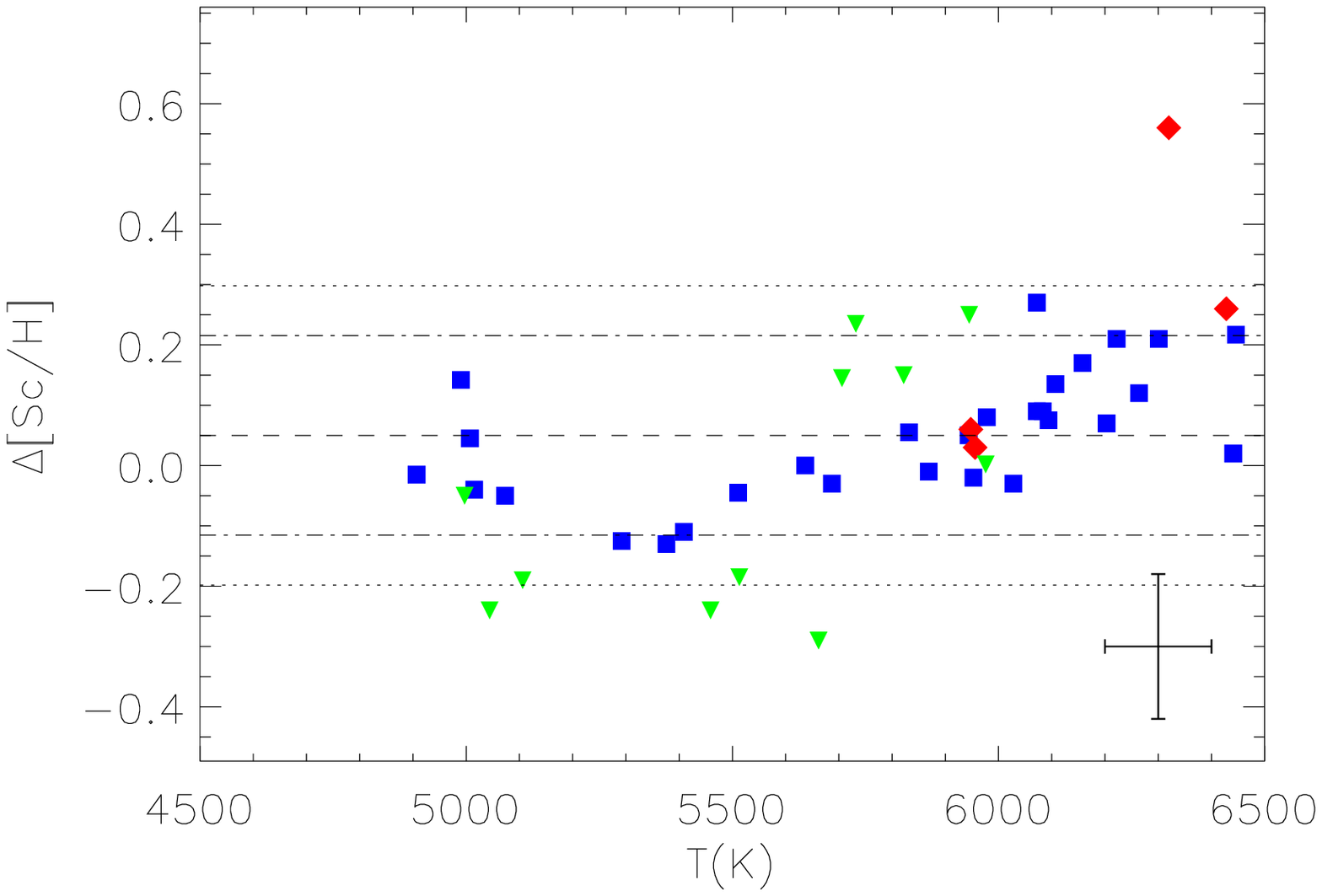}}
\centerline{
\includegraphics[scale=0.35]{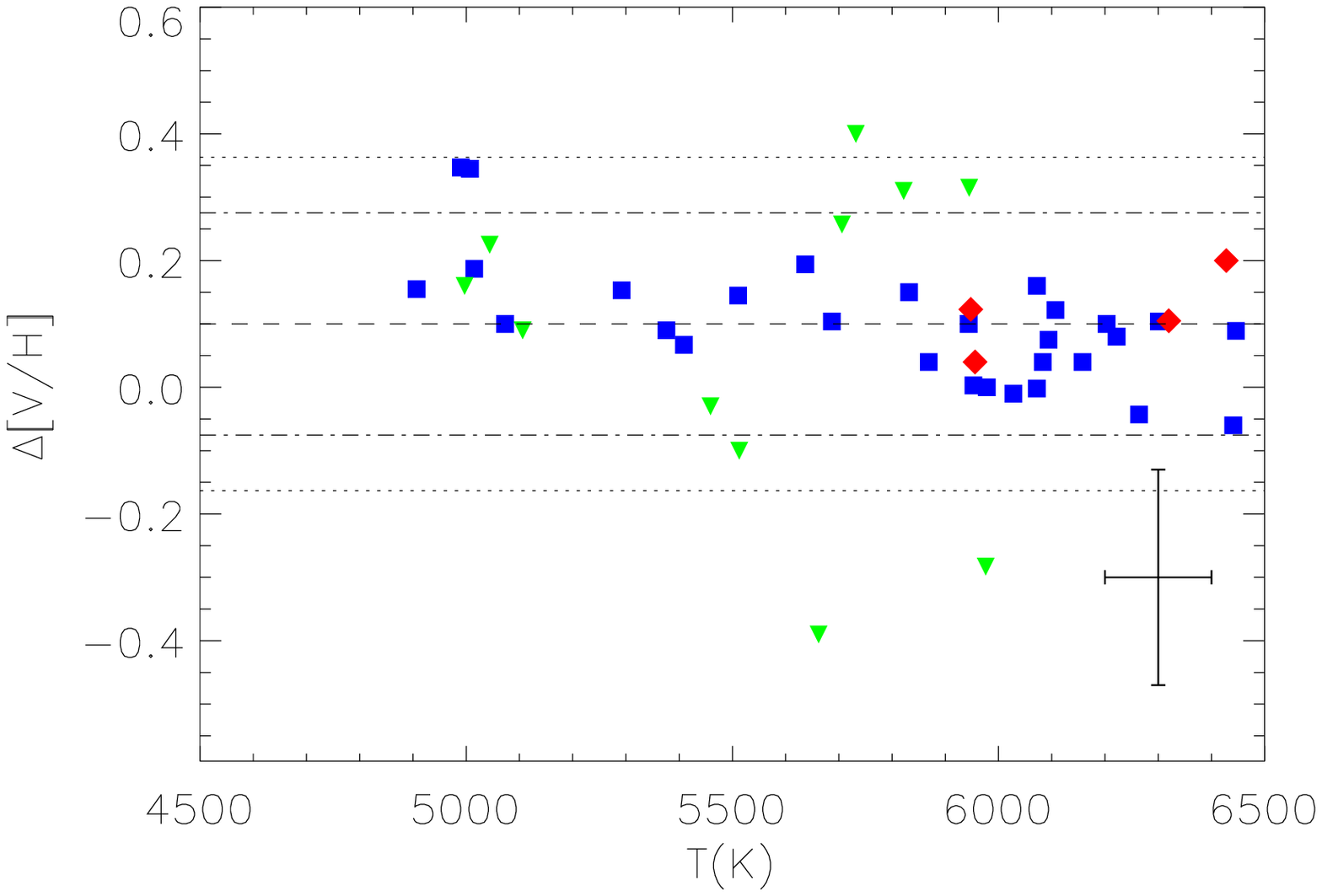}
\includegraphics[scale=0.35]{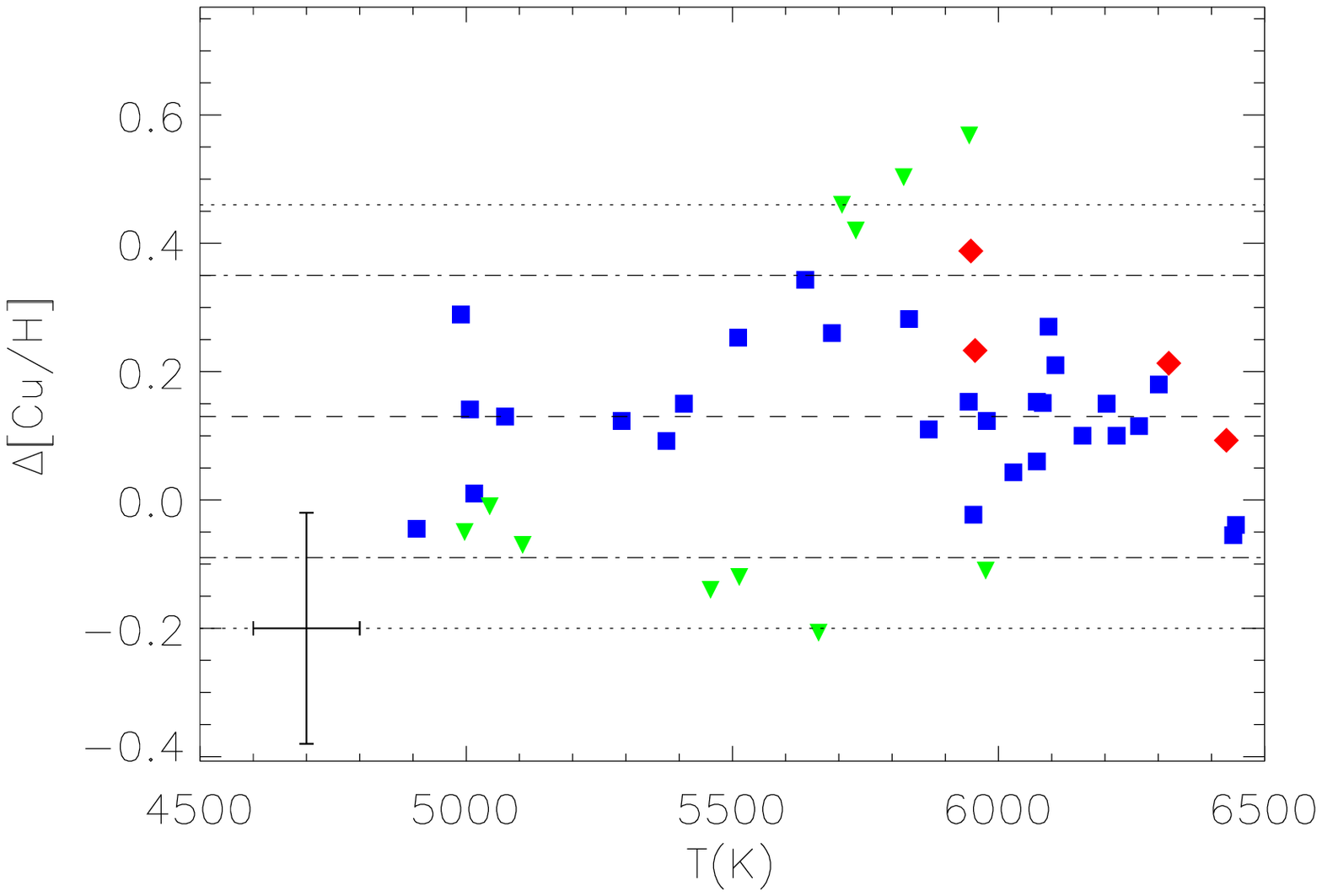}
\includegraphics[scale=0.35]{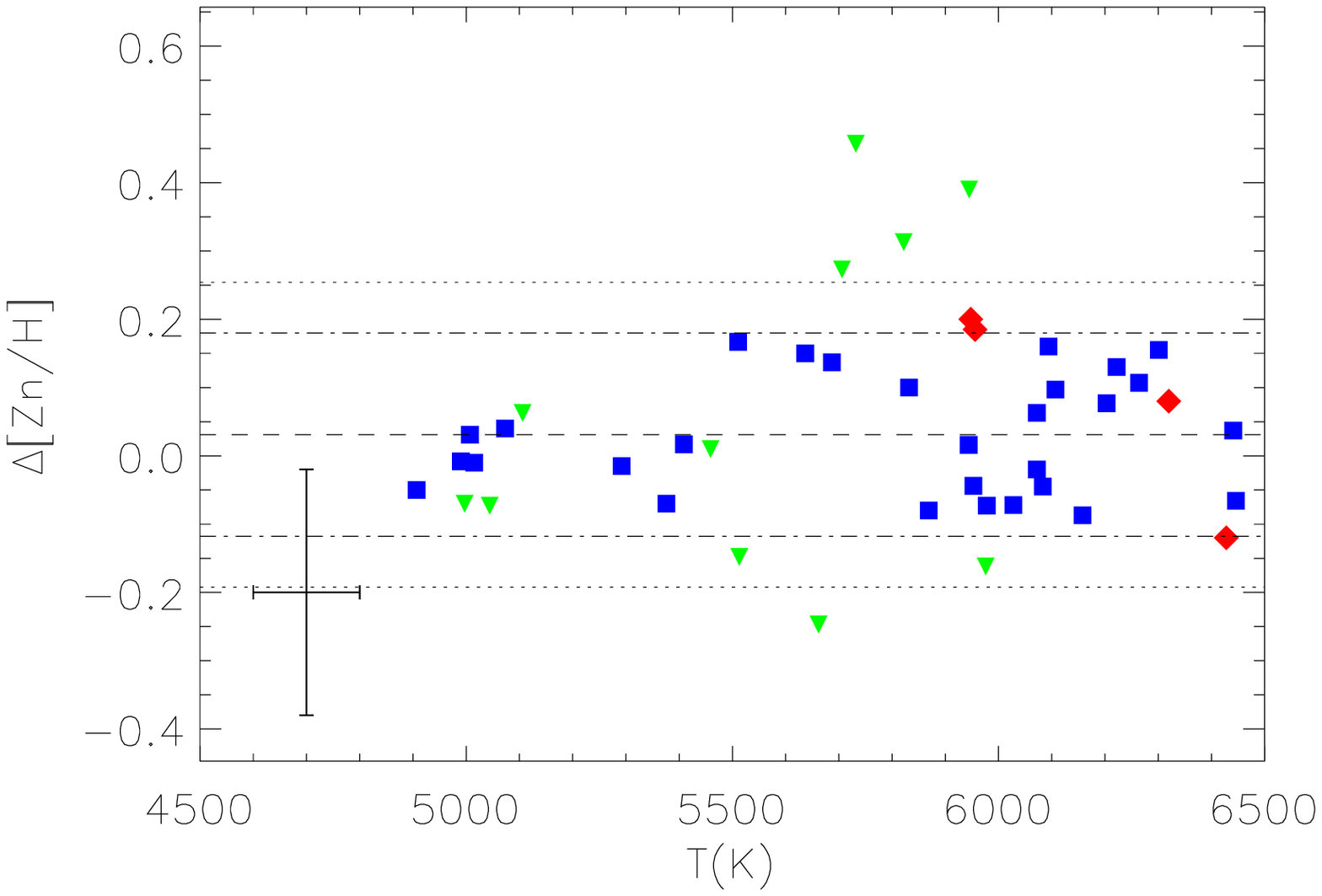}}
\centerline{
\includegraphics[scale=0.35]{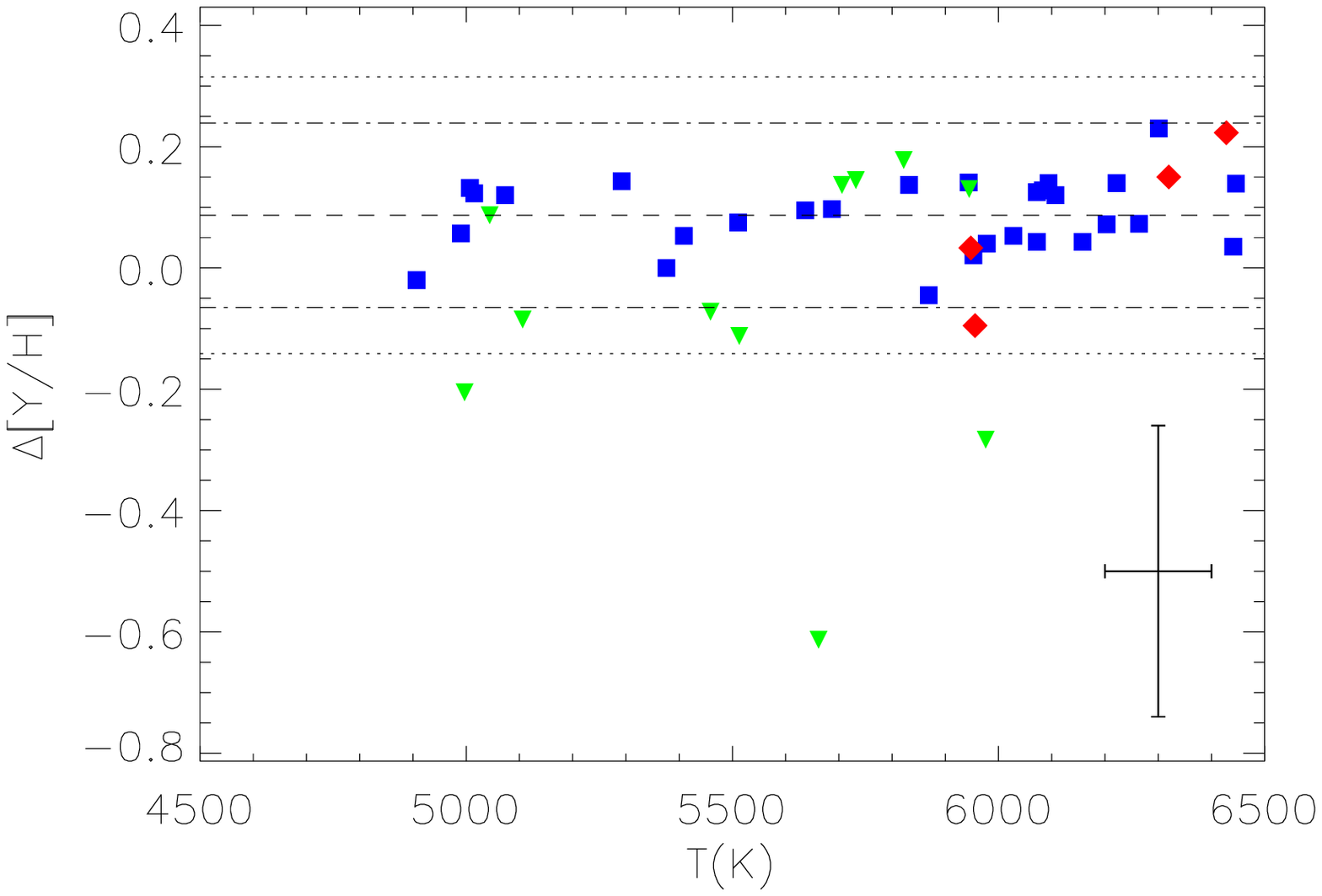}
\includegraphics[scale=0.35]{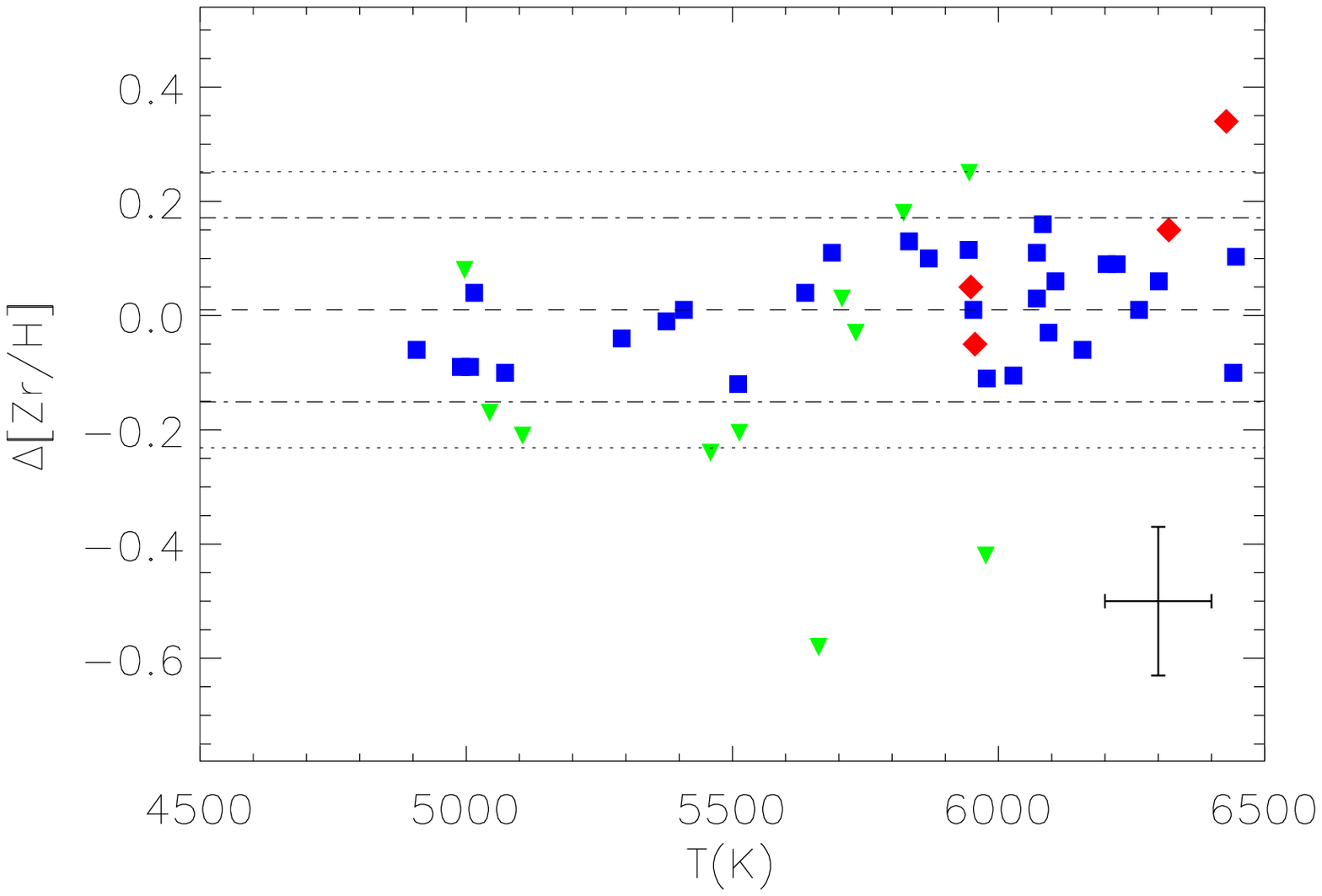}
\includegraphics[scale=0.35]{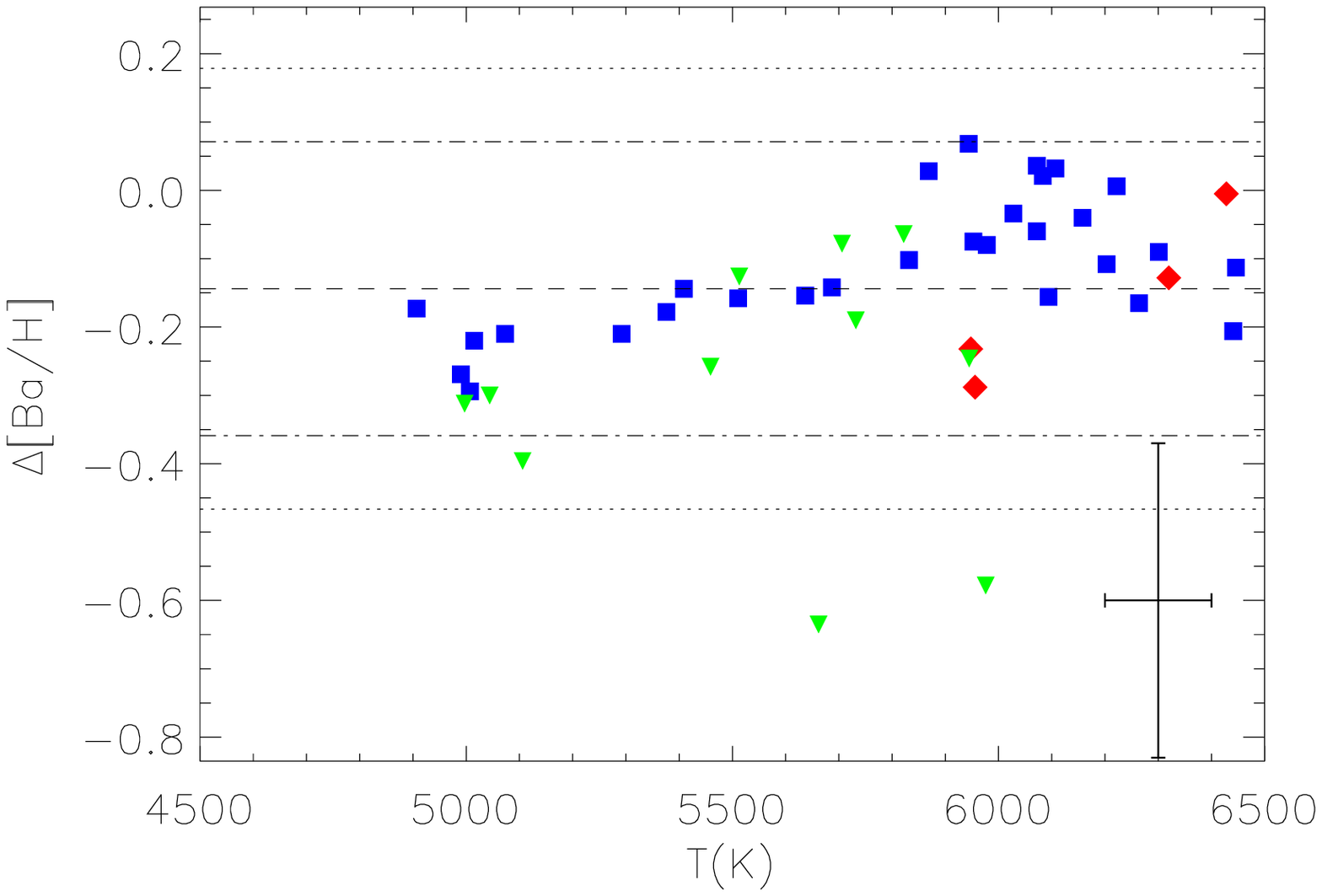}}
\centerline{
\includegraphics[scale=0.35]{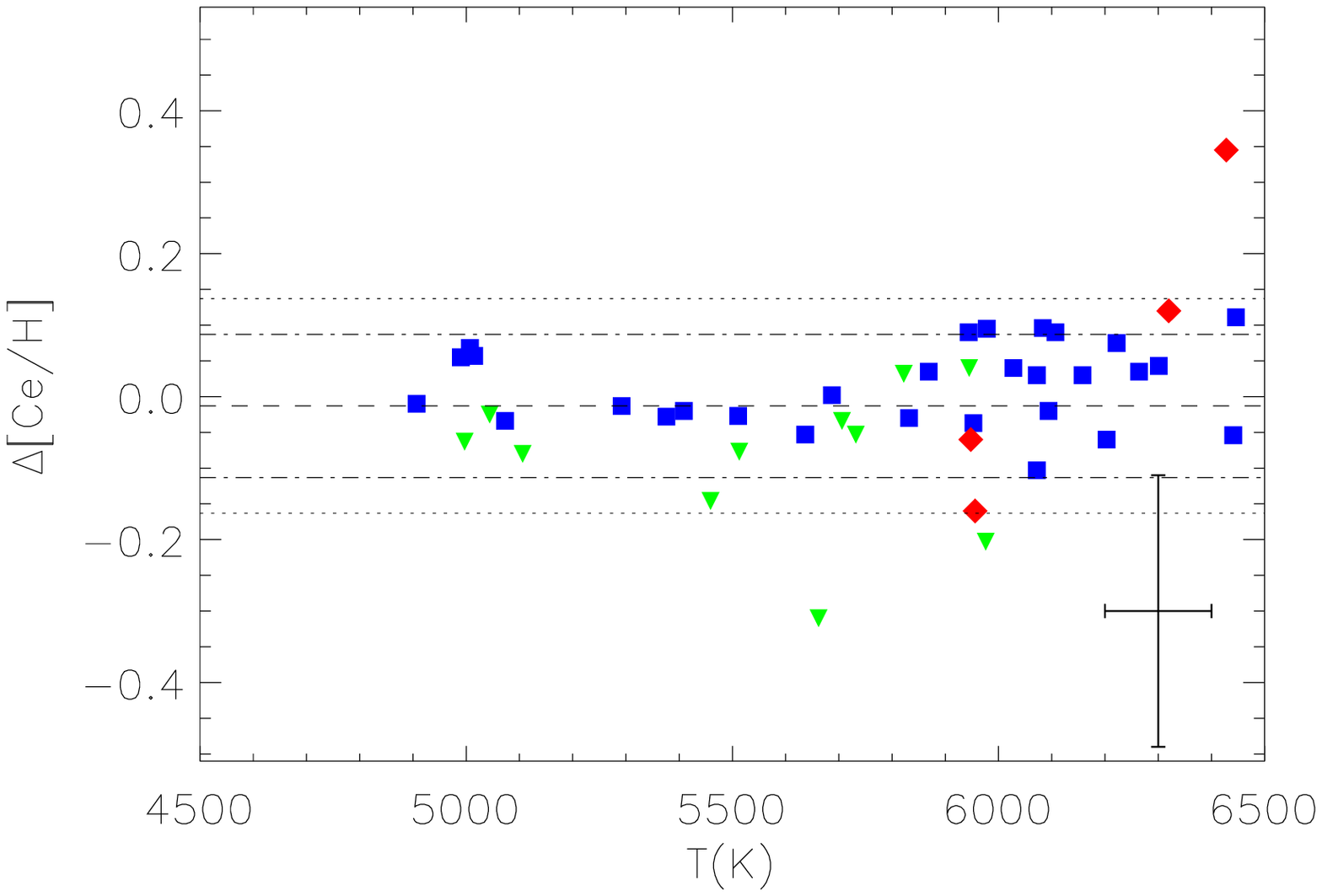}
\includegraphics[scale=0.35]{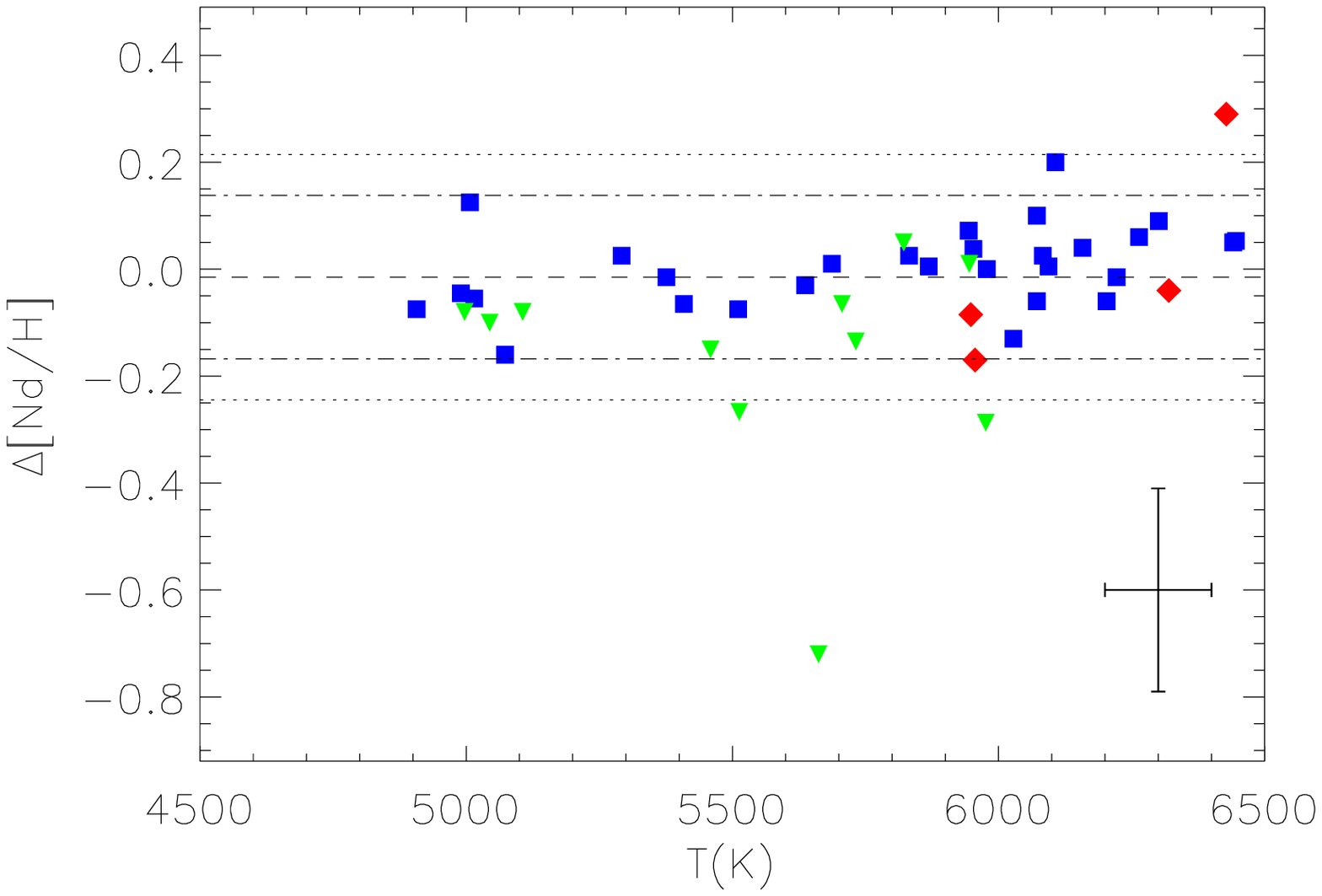}}
\caption{Same as Fig.~\ref{umadif1} but for the odd-Z elements (Na, Al, Sc, and V),  Cu, Zn, and 
the s-process elements (Y, Zr, Ba, Ce, and Nd).}
\label{umadif2}
\end{figure*}

\begin{landscape}
\begin{table}
\scriptsize
\caption{$V$ magnitude, JHK 2MASS photometry, color excess, spectroscopic $T_{\rm eff}$, reddening-uncorrected $T_{\rm IRFM,0}$, reddening-corrected  $T_{\rm IRFM}$, angular diameter $\theta$, spectroscopic and {\sc Hipparcos} $\log{\rm g}$,
iron abundance [Fe/H], and bolometric flux for the sample stars. }
\label{tblirfm}
\centering
\begin{tabular}{lccccccccccccc}
\hline\hline
\noalign{\smallskip}
Name & $V_{\rm Johnson}$ & $J_{\rm 2MASS}$ &  $H_{\rm 2MASS}$ &  $K_{\rm 2MASS}$  & E(B-V) & $T_{\rm eff}$  & $T_{\rm IRFM,0}$ &  $T_{\rm IRFM}$       &   $\theta$   &  $\log{\rm g}$  &  $\log{\rm g}_{Hip}$ &  [Fe/H]    &  F$_{bol}$\\
     &  (mag)        &  (mag) & (mag) & (mag) & (mag)   &   (K)  &   (K)            &    (K)       &    (mas)     &  (dex) &  (dex) & (dex)  & erg s$^{-1}$ cm$^{-2}$\\
\noalign{\smallskip}
\hline
\noalign{\smallskip}
HD 4048        &  6.643 &  5.687 $\pm$ 0.024  &  5.422 $\pm$ 0.040 &  5.342 $\pm$ 0.024 &  0.00877  &  6428  &  6179  $\pm$  99 &   6224 $\pm$   98 &  0.351 $\pm$ 0.016 &  4.56 & 4.28 $\pm$  0.03 &  0.08 &  2.9529E-08\\ 
V445 And       &  6.600 &  5.515 $\pm$ 0.018  &  5.258 $\pm$ 0.029 &  5.177 $\pm$ 0.016 &  0.00440  &  6028  &  5946  $\pm$  71 &   5966 $\pm$   71 &  0.351 $\pm$ 0.016 &  4.62 & 4.47 $\pm$  0.01 & -0.03 &  2.9244E-08\\
HD 8004        &  7.216 &  6.088 $\pm$ 0.024  &  5.868 $\pm$ 0.033 &  5.789 $\pm$ 0.024 &  0.01060  &  6072  &  5902  $\pm$  90 &   5950 $\pm$   91 &  0.408 $\pm$ 0.015 &  4.58 & 4.45 $\pm$  0.01 &  0.03 &  8.3572E-08\\
HD 13289       &  7.617 &  6.565 $\pm$ 0.023  &  6.328 $\pm$ 0.040 &  6.287 $\pm$ 0.015 &  0.01481  &  6158  &  6035  $\pm$  78 &   6105 $\pm$   79 &  0.230 $\pm$ 0.007 &  4.49 & 4.39 $\pm$  0.03 &  0.05 &  1.2986E-08\\
HD 20367       &  6.412 &  5.325 $\pm$ 0.023  &  5.117 $\pm$ 0.026 &  5.039 $\pm$ 0.020 &  0.01885  &  6107  &  5987  $\pm$  79 &   6076 $\pm$   81 &  0.305 $\pm$ 0.010 &  4.49 & 4.37 $\pm$  0.03 &  0.11 &  3.4137E-08\\ 
$\gamma$ Lep A &  3.587 &  2.804 $\pm$ 0.276  &  2.606 $\pm$ 0.236 &  2.508 $\pm$ 0.228 &  0.00099  &  6441  &    --  $\pm$  -- &    --  $\pm$   -- &    --  $\pm$   --  &  4.72 & 4.33 $\pm$  0.02 & -0.04 &       --   \\
\end{tabular}
 \tablefoot{
 \tablefoottext{*}{ $V_{\rm Tycho}$ converted to $V_{\rm Johnson}$ from~\citet{mam02}  expression.}\\
 }
\end{table}
\end{landscape}

\begin{center}
\onecolumn
\begin{longtable}{ccccc}

\caption{Wavelength, elements, excitation potential, oscillator strengths for the spectral 
lines used in the present work. For the $\ion{Fe}{i,ii}$ linelist, please see \citet{sou08}
{\bf References.} (G10)~\citet{gon10}; (P11)~\citet{pom11}}
\label{lltab}
\\
		\hline
		\noalign{\smallskip}
		$\lambda$ ($\AA$) & Chemical species & $\chi_{\rm l}$ (eV) & $\log{gf}$  & Ref. \\*
		\hline
		\noalign{\smallskip}
		\endfirsthead

		\hline
		\noalign{\smallskip}
		$\lambda$ ($\AA$) & Chemical species & $\chi_{\rm l}$ (eV) & $\log{gf}$  & Ref. \\*
		\hline
		\noalign{\smallskip}
		\endhead
		
		\hline
		\endfoot
		\hline
		\endlastfoot

6154.23 & $\ion{Na}{i}$  & 2.10  & -1.622 & G10 \\
6160.75 & $\ion{Na}{i}$  & 2.10  & -1.363 & G10 \\
4730.04 & $\ion{Mg}{i}$  & 4.35  & -2.234 & G10 \\
5711.09 & $\ion{Mg}{i}$  & 4.35  & -1.777 & G10 \\
6319.24 & $\ion{Mg}{i}$  & 5.11  & -2.300 & G10 \\
6696.03 & $\ion{Al}{i}$  & 3.14  & -1.571 & G10 \\
6698.67 & $\ion{Al}{i}$  & 3.14  & -1.886 & G10 \\
5517.54 & $\ion{Si}{i}$  & 5.08  & -2.496 & G10 \\
\end{longtable}
\end{center}

\begin{landscape}
\begin{table}
\scriptsize
\caption{[X/Fe] ratios for the  $\alpha$-, Fe-peak, and odd-Z elements: Na, Mg, Al, Si, Ca, Sc, Ti, V, Cr, Mn, Co, and Ni.  Their
uncertainties are the line-to-line dispersion. }
\label{galtab2}
\centering
\begin{tabular}{llcccccccccccc}     
\hline
\hline
\noalign{\smallskip}
{\small Name} & {\small Instrument\tablefootmark{a}} & {\small [Na/Fe]} & {\small  [Mg/Fe]} & {\small  [Al/Fe]} &{\small  [Si/Fe]} & {\small  [Ca/Fe]} &{\small  [Sc/Fe]} &  {\small [Ti/Fe]} & {\small [V/Fe]} & {\small [Cr/Fe]} & {\small [Mn/Fe]} & {\small [Co/Fe]} &  {\small [Ni/Fe]} \\
\noalign{\smallskip}
\hline
\noalign{\smallskip}
  HD 115043 &      H, T & -0.09 $\pm$  0.01 & -0.07 $\pm$  0.05 & -0.10 $\pm$  0.02 & -0.03 $\pm$  0.01 &  0.05 $\pm$  0.01 & -0.06 $\pm$  0.08 & 0.00 $\pm$  0.01   &-0.02 $\pm$  0.04 &  0.00 $\pm$  0.01 & -0.13 $\pm$  0.02 & -0.14 $\pm$  0.01  & -0.08 $\pm$  0.01 \\
\noalign{\smallskip}
\hline
HD  4048       &       T & -0.14 $\pm$  0.07 &  0.04 $\pm$  ---  & -0.03 $\pm$  ---  & -0.00 $\pm$  0.01 &  0.00 $\pm$  0.02 &  0.02 $\pm$  0.22 &  0.10 $\pm$  0.02  & 0.18 $\pm$  0.12 & -0.02 $\pm$  0.12 & -0.14 $\pm$  0.08 &  0.09 $\pm$  0.03 &  0.01 $\pm$  0.01 \\
V445   And     &     F, T & -0.07 $\pm$  0.01 & -0.02 $\pm$  0.12 & -0.07 $\pm$  0.03 &  0.00 $\pm$  0.01 &  0.04 $\pm$  0.02 & -0.06 $\pm$  0.04 &  0.01 $\pm$  0.01  &-0.00 $\pm$  0.02 & -0.01 $\pm$  0.01 & -0.12 $\pm$  0.02 & -0.11 $\pm$  0.01 & -0.06 $\pm$  0.01 \\
    HD  8004   &      T  & -0.11 $\pm$  0.02 & -0.04 $\pm$  0.01 & -0.09 $\pm$  0.01 & -0.04 $\pm$  0.01 &  0.03 $\pm$  0.02 & -0.07 $\pm$  0.01 &  0.00 $\pm$  0.01  &-0.01 $\pm$  0.01 & -0.00 $\pm$  0.01 & -0.18 $\pm$  0.08 &  0.00 $\pm$  0.03 & -0.03 $\pm$  0.01 \\
    HD 13829   &      T  & -0.10 $\pm$  0.05 & -0.14 $\pm$  0.01 &  0.07 $\pm$  0.17 & -0.06 $\pm$  0.01 &  0.00 $\pm$  0.02 &  0.04 $\pm$  0.13 &  0.05 $\pm$  0.02  &-0.13 $\pm$  0.05 & -0.02 $\pm$  0.02 & -0.04 $\pm$  0.08 & -0.05 $\pm$  0.02 &  0.00 $\pm$  0.01 \\
\end{tabular}
\tablefoot{
\tablefoottext{a}{Instruments employed to acquire the data: HERMES (H), FOCES (F), and TLS (T).}\\
}
\end{table}
\end{landscape}

\begin{table*}
\scriptsize
\caption{[X/Fe] ratios for the s-process elements: Cu, Zn, Y, Zr, Ba, Ce, and Nd. Their 
uncertainties are the line-to-line dispersion. }
\label{galtab3}
\centering
\begin{tabular}{llccccccc}     
\hline
\hline
\noalign{\smallskip}
{\small Name} & {\small Instrument\tablefootmark{a}} & {\small [Cu/Fe]} & {\small [Zn/Fe]} & {\small [Y/Fe]} & {\small [Zr/Fe]} & {\small [Ba/Fe]} & {\small [Ce/Fe]} & {\small [Nd/Fe]} \\
\noalign{\smallskip}
\hline
\noalign{\smallskip}
HD 115043 &      H, T & -0.10 $\pm$  0.11 & -0.06 $\pm$  0.05 &  0.06 $\pm$  0.05 &  0.08 $\pm$  0.10 &  0.29 $\pm$  0.05 &  0.09 $\pm$  0.04 &  0.15 $\pm$  0.03 \\
\noalign{\smallskip}
\hline
\noalign{\smallskip}
    HD  4048 &      T  & -0.16 $\pm$  0.12 & -0.23 $\pm$  0.08 &  0.22 $\pm$  0.01 &  0.20 $\pm$  ---  &  0.16 $\pm$  0.17 &  0.23 $\pm$  0.14 &  0.17 $\pm$  0.01 \\
  V445   And &     F, T & -0.10 $\pm$  0.04 & -0.14 $\pm$  0.04 &  0.14 $\pm$  0.05 & -0.14 $\pm$  ---  &  0.25 $\pm$  0.05 &  0.16 $\pm$  0.06 &  0.12 $\pm$  0.01 \\
    HD  8004 &      T  & -0.05 $\pm$  0.03 & -0.08 $\pm$  0.04 &  0.12 $\pm$  0.01 & -0.06 $\pm$  ---  &  0.16 $\pm$  0.01 & -0.03 $\pm$  0.13 &  0.18 $\pm$  0.01 \\
    HD 13829 &      T  & -0.14 $\pm$  0.01 & -0.16 $\pm$  0.07 &  0.05 $\pm$  0.07 & -0.16 $\pm$  ---  &  0.16 $\pm$  0.04 & -0.05 $\pm$  0.10 &  0.18 $\pm$  0.01 \\
    HD 20367 &      T  & -0.05 $\pm$  0.08 & -0.04 $\pm$  0.12 &  0.09 $\pm$  0.05 & -0.11 $\pm$  ---  &  0.17 $\pm$  0.04 & -0.05 $\pm$  0.07 &  0.21 $\pm$  0.05 \\
$\gamma$  Lep A &    H & -0.07 $\pm$  0.01 & -0.03 $\pm$  0.02 &  0.10 $\pm$  0.04 & -0.06 $\pm$  ---  &  0.15 $\pm$  0.06 &  0.18 $\pm$  0.09 &  0.25 $\pm$  0.16 \\
\end{tabular}
\tablefoot{
\tablefoottext{a}{Instruments employed to acquire the data: HERMES (H), FOCES (F), and TLS (T).}\\
}
\end{table*}

\begin{landscape}
\begin{table}
\scriptsize
\caption{Differential abundances ($\Delta$[X/H]) with respect to HD~115043 for the  $\alpha$-, Fe-peak, and odd-Z elements: Na, Mg, Al, Si, Ca, Sc, Ti, V, Cr, Mn, Co, and Ni. Their uncertainties are the line-to-line dispersion.} 
\label{diftable2}
\centering
\begin{tabular}{llcccccccccccc}     
\hline
\hline
\noalign{\smallskip}
{\small Name} & {\small Instrument\tablefootmark{a}} & {\small $\Delta$[Na/H]} & {\small $\Delta$[Mg/H]} & {\small $\Delta$[Al/H]} &{\small  $\Delta$[Si/H]} & {\small $\Delta$[Ca/H]} &  {\small $\Delta$[Sc/H]} & {\small $\Delta$[Ti/H]} & {\small $\Delta$[V/H]}  & {\small $\Delta$[Cr/H]} & {\small $\Delta$[Mn/H]} & {\small $\Delta$[Co/H]} & {\small $\Delta$[Ni/H]} \\
\noalign{\smallskip}
\hline
\noalign{\smallskip}
    HD  4048 &      T  &  0.01 $\pm$  0.06 &  0.23 $\pm$  ---  &  0.13 $\pm$  ---  &  0.16 $\pm$  0.02 &  0.02 $\pm$  0.02 &  0.26 $\pm$  0.37 &  0.22 $\pm$  0.02  & 0.20 $\pm$  0.11 &  0.07 $\pm$  0.08 &  0.07 $\pm$  0.05 &  0.18 $\pm$  0.04 &  0.16 $\pm$  0.01 \\
  V445   And &     F, T &  0.00 $\pm$  0.01 &  0.01 $\pm$  0.12 &  0.00 $\pm$  0.04 &  0.01 $\pm$  0.01 & -0.05 $\pm$  0.01 & -0.03 $\pm$  0.16 & -0.01 $\pm$  0.01  &-0.01 $\pm$  0.02 & -0.03 $\pm$  0.01 &  0.00 $\pm$  0.07 & -0.03 $\pm$  0.02 & -0.01 $\pm$  0.01 \\
    HD  8004 &      T  & -0.01 $\pm$  0.01 &  0.10 $\pm$  0.05 &  0.03 $\pm$  0.01 &  0.03 $\pm$  0.01 & -0.07 $\pm$  0.02 &  0.27 $\pm$  0.02 &  0.03 $\pm$  0.01  &-0.00 $\pm$  0.05 &  0.05 $\pm$  0.01 & -0.03 $\pm$  0.07 &  0.07 $\pm$  0.05 &  0.08 $\pm$  0.01 \\
    HD 13829 &      T  &  0.01 $\pm$  0.04 & -0.02 $\pm$  0.05 &  0.21 $\pm$  0.18 &  0.04 $\pm$  0.02 & -0.07 $\pm$  0.03 &  0.17 $\pm$  0.12 &  0.11 $\pm$  0.01  & 0.04 $\pm$  0.11 &  0.08 $\pm$  0.02 &  0.18 $\pm$  0.15 &  0.14 $\pm$  0.02 &  0.12 $\pm$  0.01 \\
    HD 20367 &      T  &  0.10 $\pm$  0.01 &  0.12 $\pm$  0.10 &  0.14 $\pm$  0.04 &  0.12 $\pm$  0.01 &  0.06 $\pm$  0.01 &  0.14 $\pm$  0.17 &  0.12 $\pm$  0.01  & 0.12 $\pm$  0.07 &  0.12 $\pm$  0.02 &  0.05 $\pm$  0.12 &  0.09 $\pm$  0.03 &  0.10 $\pm$  0.01 \\
\end{tabular}
\tablefoot{
\tablefoottext{a}{Instruments  employed to acquire the data: HERMES (H), FOCES (F), and TLS (T).}\\
}
\end{table}
\end{landscape}

\begin{table*}
\scriptsize
\caption{Differential abundances ($\Delta$[X/H]) with respect to HD~115043 for the s-process elements: Cu, Zn, Y, Zr, Ba, Ce, and Nd.
Their uncertainties are the line-to-line dispersion.} 
\label{diftable3}
\centering
\begin{tabular}{llccccccc}     
\hline
\hline
\noalign{\smallskip}
{\small Name} & {\small Instrument\tablefootmark{a}}  & {\small $\Delta$[Cu/H]} & {\small $\Delta$[Zn/H]} & {\small $\Delta$[Y/H]} & {\small $\Delta$[Zr/H]} & {\small $\Delta$[Ba/H]} & {\small $\Delta$[Ce/H]} & {\small $\Delta$[Nd/H]} \\
\noalign{\smallskip}
\hline
\noalign{\smallskip}
    HD  4048 &      T  &  0.09 $\pm$  0.12 & -0.12 $\pm$  0.08 &  0.22 $\pm$  0.01 &  0.34 $\pm$  ---  & -0.00 $\pm$  0.17 &  0.34 $\pm$  0.14 &  0.29 $\pm$  0.01 \\
  V445   And &     F, T &  0.04 $\pm$  0.04 & -0.07 $\pm$  0.03 &  0.05 $\pm$  0.10 & -0.10 $\pm$  ---  & -0.03 $\pm$  0.03 &  0.04 $\pm$  0.03 & -0.13 $\pm$  0.01 \\
    HD  8004 &      T  &  0.15 $\pm$  0.03 & -0.02 $\pm$  0.04 &  0.04 $\pm$  0.01 &  0.03 $\pm$  ---  & -0.06 $\pm$  0.01 &  0.03 $\pm$  0.13 &  0.10 $\pm$  0.01 \\
    HD 13829 &      T  &  0.10 $\pm$  0.01 & -0.09 $\pm$  0.07 &  0.04 $\pm$  0.07 & -0.06 $\pm$  ---  & -0.04 $\pm$  0.04 &  0.03 $\pm$  0.10 &  0.04 $\pm$  0.01 \\
    HD 20367 &      T  &  0.21 $\pm$  0.03 &  0.10 $\pm$  0.03 &  0.12 $\pm$  0.08 &  0.06 $\pm$  ---  &  0.03 $\pm$  0.10 &  0.09 $\pm$  0.04 &  0.20 $\pm$  0.01 \\
$\gamma$  Lep A  &   H & -0.05 $\pm$  0.03 &  0.04 $\pm$  0.11 &  0.04 $\pm$  0.08 & -0.10 $\pm$  ---  & -0.21 $\pm$  0.04 & -0.05 $\pm$  0.02 &  0.05 $\pm$  0.01 \\
$\gamma$  Lep B  &   H & -0.05 $\pm$  0.12 & -0.05 $\pm$  0.07 & -0.02 $\pm$  0.07 & -0.06 $\pm$  ---  & -0.17 $\pm$  0.07 & -0.01 $\pm$  0.07 & -0.08 $\pm$  0.09 \\
\end{tabular}
\tablefoot{
\tablefoottext{a}{Instruments employed to acquire the data: HERMES (H), FOCES (F), and TLS (T).}\\
}
\end{table*}

\begin{landscape}
\begin{table}
\scriptsize
\caption{ Differential abundance average sensitivities for Fe, Na, Mg, Al, Si, Ca, Sc, Ti, V, Cr, Mn, Co, and Ni.}
\label{tablasense}
\begin{tabular}{lccccccccccccc}
\hline
\hline
\noalign{\smallskip}

\noalign{\smallskip}
{\small Name} &  {\small $\Delta$ [Fe/H]} &{\small $\Delta$[Na/H]} & {\small $\Delta$[Mg/H]} & {\small $\Delta$[Al/H]} & {\small $\Delta$[Si/H]} & {\small $\Delta$[Ca/H]} & {\small $\Delta$[Sc/H]} & {\small $\Delta$[Ti/H]} & {\small $\Delta$[V/H]} & {\small $\Delta$[Cr/H]} & {\small $\Delta$[Mn/H]} & {\small $\Delta$[Co/H]} & {\small  $\Delta$[Ni/H]} \\
\hline
\noalign{\smallskip}
$\Delta T_{\rm eff}$ = $\pm$ 100 K    &   &      &      &      &      &      &      &      &      &      &      &      &     \\
\hline
\noalign{\smallskip}

HD 64942       & 0.07 & 0.08 & 0.07 & 0.08 & 0.03 & 0.10 & 0.09 & 0.14 & 0.15 & 0.11 & 0.12 & 0.10 & 0.08 \\
HD 184385     & 0.06 & 0.09 & 0.07 & 0.08 & 0.00 & 0.11 & 0.10 & 0.16 & 0.18 & 0.12 & 0.12 & 0.07 & 0.06 \\
HD 56168 & 0.05 & 0.11 & 0.06 & 0.09 & 0.04 & 0.14 & 0.11 & 0.18 & 0.21 & 0.13 & 0.11 & 0.04 &0.02 \\
\hline
\noalign{\smallskip}

Average & 0.06 & 0.09 & 0.07 & 0.08 & 0.02 & 0.12 & 0.10 & 0.16 & 0.18 & 0.12 & 0.12 & 0.07 &0.05 \\
\hline
\noalign{\smallskip}
$\Delta$$\log{g}$ = $\pm$ 0.30 dex  & &      &      &      &      &      &      &      &      &      &      &      &     \\
\hline
\noalign{\smallskip}
HD 64942 & 0.03 & 0.04 & 0.04 & 0.02 & 0.00 & 0.10 & 0.01 & 0.01 & 0.01 & 0.03 & 0.08 & 0.01 & 0.01 \\
\hline
\end{tabular}

\tablefoot{Sensitivities to changes of 100 K in $T_{\rm eff}$, 0.30 dex in $\log{g}$, 0.50 km s$^{-1}$  in $\xi$, and 0.30 dex in [Fe/H]. We verified these sensitivities for the stars HD~64942 (5869 K, 4.63 dex, 1.11 km s$^{-1}$, 0.01 dex), HD~184385 (5511 K, 4.48 dex, 0.94 km s$^{-1}$, 0.05 dex), and HD~56168 (5044 K, 4.53 dex, 0.81 km s$^{-1}$, -0.18 dex).}
\end{table}
 
\begin{table}
\scriptsize
\caption{ Differential abundance sensitivities for Cu, Zn, Y, Zr, Ba, Ce, and Nd.}
\label{tablasense2}
\begin{tabular}{lccccccc}
\hline
\hline
\noalign{\smallskip}

\noalign{\smallskip}
{\small Name} & {\small $\Delta$[Cu/H]} &  {\small $\Delta$[Zn/H]} & {\small $\Delta$[Y/H]} & {\small $\Delta$[Zr/H]} & {\small $\Delta$[Ba/H]} & {\small $\Delta$[Ce/H]} & {\small  $\Delta$[Nd/H]} \\
\hline
\noalign{\smallskip}
$\Delta T_{\rm eff}$ = $\pm$ 100 K       &      &      &     &    &   &    \\
\hline
\noalign{\smallskip}
HD 64942      &  0.06   &     0.01   &     0.01    &    0.11   &     0.01    &    0.03  &      0.02 \\
HD 184385     &  0.06   &     0.01   &     0.01    &    0.12   &     0.01    &    0.02  &      0.02 \\
HD 56168      &  0.04   &     0.03   &     0.02    &    0.14   &     0.02    &    0.02  &      0.03 \\
\hline
\noalign{\smallskip}
$\Delta$$\log{g}$ = $\pm$ 0.30 dex    &      &      &     &    &   &    \\
\hline
\noalign{\smallskip}
HD 64942     &   0.01  &      0.04   &     0.12   &      0.01   &     0.04  &      0.13  &      0.13 \\
HD 184385    &   0.02  &      0.03   &     0.11   &     0.01    &    0.03   &     0.13   &     0.13 \\
\hline

\end{tabular}
\end{table}
\end{landscape}

\begin{landscape} 
\begin{table}
\scriptsize
\caption{ Differential abundance average sensitivities for Cu, Zn, Y, Zr, Ba, Ce, and Nd.}
\label{tablasense2}
\begin{tabular}{lccccccc}
\hline
\hline
\noalign{\smallskip}

\noalign{\smallskip}
{\small Name} & {\small $\Delta$[Cu/H]} &  {\small $\Delta$[Zn/H]} & {\small $\Delta$[Y/H]} & {\small $\Delta$[Zr/H]} & {\small $\Delta$[Ba/H]} & {\small $\Delta$[Ce/H]} & {\small  $\Delta$[Nd/H]} \\
\hline
\noalign{\smallskip}
$\Delta T_{\rm eff}$ = $\pm$ 100 K       &      &      &     &    &   &    \\
\hline
\noalign{\smallskip}
HD 64942      &  0.06   &     0.01   &     0.01    &    0.11   &     0.01    &    0.03  &      0.02 \\
HD 184385     &  0.06   &     0.01   &     0.01    &    0.12   &     0.01    &    0.02  &      0.02 \\
HD 56168      &  0.04   &     0.03   &     0.02    &    0.14   &     0.02    &    0.02  &      0.03 \\
\noalign{\smallskip}
\hline
\noalign{\smallskip}
Average & 0.05 & 0.02 & 0.01 & 0.12 & 0.01 & 0.02 & 0.02 \\
\noalign{\smallskip}
\hline
\noalign{\smallskip}
$\Delta$$\log{g}$ = $\pm$ 0.30 dex    &      &      &     &    &   &    \\
\hline
\noalign{\smallskip}
HD 64942     &   0.01  &      0.04   &     0.12   &      0.01   &     0.04  &      0.13  &      0.13 \\
\hline
\end{tabular}
\tablefoot{Sensitivities to changes of 100 K in $T_{\rm eff}$, 0.30 dex in $\log{g}$, 0.50 km s$^{-1}$  in $\xi$, and 
	0.30 dex in [Fe/H]. We verified these sensitivities for the stars HD~64942 (5869 K, 4.63 dex, 1.11 km s$^{-1}$, 0.01 dex), 
HD~184385 (5511 K, 4.48 dex, 0.94 km s$^{-1}$, 0.05 dex), and HD~56168 (5044 K, 4.53 dex, 0.81 km s$^{-1}$, -0.18 dex).}
\end{table}

\end{landscape}

\end{appendix}
\end{document}